\documentclass[]{aa} 

\newcommand{\artemis}{ArT{\'e}MiS }
\newcommand{\Herschel}{{\it Herschel }}

\usepackage{graphicx}
\usepackage{txfonts}

\begin{document} 

\title{Filaments in the OMC-3 cloud and uncertainties in estimates of filament
profiles
\thanks{{\it Herschel} is an ESA space observatory with science instruments provided by
European-led Principal Investigator consortia and with important participation from
NASA.}}


\author{M. Juvela \inst{1},  E. Mannfors  \inst{1}}

\institute{
Department of Physics, P.O.Box 64, FI-00014, University of Helsinki,
Finland, {\email mika.juvela@helsinki.fi}
}

\date{Received September 15, 1996; accepted March 16, 1997}

\abstract
{Filamentary structures are an important part of star-forming interstellar
clouds. The properties of filaments hold clues to their formation mechanisms
and their role in the star-formation process. }
{We compare the properties of filaments in the Orion Molecular Cloud 3 (OMC-3), as seen 
in mid-infrared (MIR) absorption and far-infrared (FIR) dust emission. We also wish to
characterise some potential sources of systematic errors in filament studies. }
{We calculated optical depth maps of the OMC-3 filaments based on the MIR
absorption seen in \textit{Spitzer} data and FIR dust emission observed with
\Herschel and the \artemis instrument. We then compared the filament
properties extracted from the data. Potential sources of error were
investigated more generally with the help of radiative transfer models.}
{The widths of the selected OMC-3 filament segments are in the range
0.03-0.1\,pc, with similar average values seen in both MIR and FIR analyses.
Compared to the widths, the individual parameters of the fitted Plummer
functions are much more uncertain. The asymptotic power-law index has
typically values $p\sim3$ but with a large scatter.
Modelling shows that the FIR observations can systematically overestimate the
filament widths.  The effect is potentially tens of per cent at column
densities above $N({\rm H}_2)\sim 10^{22}$\,cm$^{-2}$ but is reduced in more
intense radiation fields, such as the Orion region. Spatial variations in dust
properties could cause errors of similar magnitude. In the MIR analysis, dust
scattering should generally not be a significant factor, unless there are
high-mass stars nearby or the dust MIR scattering efficiency is higher than in
the tested dust models. Thermal MIR dust emission can be a more significant
source of error, especially close to embedded sources.}
{The analysis of interstellar filaments can be affected by several sources of
systematic error, but mainly at high column densities and, in the case of FIR
observations, in weak radiation fields. The widths of the OMC-3 filaments were
consistent between the MIR and FIR analyses and did not reveal any systematic
dependence on the angular resolution of the observations. }


\maketitle

\section{Introduction} \label{sect:introduction}

Filaments are an important structural element of the interstellar medium (ISM)
and are intimately linked to the formation of new stars \citep{Andre2010,
Hacar2022}.  Star-forming filaments have been investigated with observations
of thermal dust emission, recently with data from the {\it Herschel} Space
Observatory in particular \citep{Pilbratt2010}. The measurements have been
used to estimate the typical column-density and mass-per-length values of
star-forming filaments, the width of filaments, and parametric representations
for the filament profiles. The questions of the universality of filament
widths and the reliability of filament-property estimates are still topical
\citep{Howard2021,Panopoulou2022}. In \Herschel studies, for clouds within
distances of 0.5\,kpc, the typical filament full-width at half maximum (FWHM)
widths are of the order of 0.1\,pc, which could point to a common formation
mechanism, either through a single event or as a more dynamical accretion
process \citep{Arzoumanian2019,Hacar2022}. However, some higher FWHM values
have also been found in \Herschel studies, often correlated with larger source
distances and thus lower linear resolution \citep{Hennemann2012,
Panopoulou2022}. The estimated filament widths appear to be tied to the range
of angular scales that are probed by the observations, which is qualitatively
in agreement with the idea of a hierarchical structure of the ISM, even in
filaments. Indeed, \Herschel\ filaments themselves often reside in even larger
and often elongated structures. On the other hand, observations at higher
angular resolution, such as with the Atacama Large Millimeter/submillimeter
Array (ALMA) interferometer, have resulted in much smaller values and,
especially in case of line observations, in the detection of `fibres' with
sizes a factor of several below the canonical 0.1\,pc value \citep{Hacar2018,
Schmiedeke2021}. The full picture of ISM structure can only be obtained by
covering a large range of size scales, which may also require the combination
of different tracers and observational methods.

The mass-per-length values and the shapes of the filament profiles are central observational
parameters. Star formation is concentrated in the most massive filaments, and a clear majority of young
stellar objects are found in super-critical filaments \citep{Andre2010, Konyves2015}. The critical
value is typically calculated as $M_{\rm crit}= 2 c_{\rm s}^2/G$, according to the idea of filaments as
isothermal infinite cylinders \citep{Stodolkiewicz1963, Ostriker1964}. The same model predicts for the
filament profiles an asymptotic behaviour $\rho(r)\propto r^{-p}$ with $p=4$. The observed profiles
tend to be shallower, $p\sim$2 \citep{Arzoumanian2011}, which might be explained by deviations from
isothermal conditions and the influence of magnetic fields, external pressure, or the dynamic growth of
filaments \citep{Fischera2012, KashiwagiTomisaka2021}. To arrive at any firm conclusions, the exponent
$p$ should be measured to a precision of a few tenths. So far, most estimates have been based on
observations of dust emission, especially from the large \Herschel surveys, but might suffer from some
bias caused by radial variations in dust temperature and opacity. 

At high column densities, the mid-infrared (MIR) absorption provides an
interesting alternative for estimating column densities
\citep{ButlerTan2012,KainulainenTan2013}. First, it is independent from dust
temperature variations, which can influence the analysis of dust emission in
different ways depending on the presence of internal and external radiation
sources. Second, the MIR absorption should be less affected by dust evolution
when compared to the sub-millimetre and far-infrared (FIR) dust emission
\citep{Roy2013,GCC-V}. Third, the existing MIR observations, especially with
the \textit{Spitzer} \citep{Werner2004} and Wide-field Infrared Survey
Explorer (WISE) \citep{Wright2010} satellites, can provide an angular
resolution ($\sim 2-6\arcsec$) that is better than that of \Herschel ($\sim
18\arcsec$ at 250\,$\mu$m) or even large ground-based single-dish radio
telescopes. However, MIR observations have their own complications, such as
the unknown level of foreground emission and the potential effects of nearby
or embedded radiation sources.

In this paper we examine Orion Molecular Cloud 3 (OMC-3) and
its filaments using FIR and MIR observations. OMC-3 is located in the Orion A
cloud, at the northern end of the integral-shaped filament
\citep[e.g.][]{Stutz2015}. We adopt for the cloud a distance of $d=400$\,pc,
which is within the uncertainties of the estimates that were given by
\citet{Grossschedl2018} based on \textit{Gaia} observations. Their Table 1 summarises
earlier distance estimates for different parts of the Orion A cloud. 

Orion A was mapped by the \textit{Herschel} satellite at 70-500\,$\mu$m wavelengths.
These data have been used in many studies of the dust properties, structure,
and star formation in the region \citep[e.g.][]{Roy2013, Lombardi2014,
Stutz2015, Furlan2016}. \citet{Sadavoy2016} estimated values
$\beta\sim1.7-1.8$ for the dust opacity spectral index in an area that
included OMC-3. Individual cores exhibited a larger dispersion in $\beta$
values, which could be related to dust evolution or to observational
effects from temperature gradients \citep[cf.][]{Juvela2012_Tmix,GCC-VI},
while $\beta$ appears to be systematically lower at millimetre wavelengths
\citep{Schnee2014,Lowe2022}.

The filamentary structure of Orion A has been studied extensively with
observations of the dust continuum emission, dust polarisation, and molecular
and atomic lines, from extended structures down to milliparsec scales
\citep[e.g.][]{Kainulainen2017,Pattle2017,Wu2018_OMC_NH3,Kong2018,Tanabe2019,Goicoechea2020,Salas2021}.
\citet{Hacar2018} identified in combined ALMA and IRAM N$_2$H$^{+}$ line data
narrow fibres with $FWHM \sim 0.035$\,pc. These structures were thus narrower
than typical $\sim$0.1\,pc filaments seen in \textit{Herschel} studies with
lower-resolution ($\sim 20-40\arcsec$) continuum observations
\citep[e.g.][]{Arzoumanian2011,GCC-VII}.
\citet{Schuller2021} combined \textit{Herschel} data with ground-based
continuum observations from the \artemis
instrument\footnote{\url{http://www.apex-telescope.org/instruments/pi/artemis/}
ARchitectures de bolomètres pour des TElescopes à grand champ de vue dans le
domaine sub-MIllimétrique au Sol.} at the Atacama Pathfinder Experiment (APEX)
telescope \citep{ARTEMIS,APEX}. Although the angular resolution of the
combined data was $\sim 8\arcsec$, the estimated filament widths in Orion A
were 0.06-0.11\,pc, close to the earlier \textit{Herschel} results and larger
than the fibres. For the filaments in the OMC-3 region in particular (in both
its eastern and western parts), the widths were $\sim$0.06\,pc. The OMC-3
filaments are dense, with line masses $\sim 200 M_{\sun}\,{\rm pc}^-1$
\citep{Schuller2021}. Based on dust polarisation data from the HAWC+
instrument of the Stratospheric Observatory for Infrared Astronomy (SOFIA),
\citet{Li2022_OMC3} concluded that the filaments are also magnetically
supercritical. 

The OMC-3 region was analysed further by \citet{Mannfors2022} using a
combination of \textit{Herschel} data and an independent set of \artemis observations.
In this paper we compare these FIR data with a new analysis of the MIR
absorption seen in \textit{Spitzer} data. With the help of radiative transfer models,
we also more generally investigate potential sources of systematic errors that
could bias our estimates of the filament masses and profiles, in the case of
both MIR and FIR observations.

The contents of the paper are as follows. In Sect.~\ref{sect:methods} we
present the methods that are used to derive column densities from observations
of dust extinction and emission, and in Sect.~\ref{sect:methods_fit} we
describe the procedures we use to fit the filament profiles with an analytical
function. Section~\ref{sect:methods_RT} explains the radiative transfer
modelling that is used to study potential sources of bias in the filament
analysis. In Sect.~\ref{sect:results} we present the main results, including
the analysis of the OMC-3 filaments and, in Sect.~\ref{sect:bias}, the
analysis of synthetic filament observations. In Sect.~\ref{sect:discussion}
we discuss the observational results and the systematic errors that may affect
the OMC-3 data and, more generally, observations of similar
high-column-density filaments. The final conclusions are listed in
Sect.~\ref{sect:conclusions}.

\section{Methods} \label{sect:methods}

\subsection{Mid-infrared absorption} \label{sect:methods_MIR}

Mid-infrared observations provide a way to measure cloud mass distribution, provided that there is sufficient
background surface brightness and the column density is high enough to result in measurable MIR
extinction. The total observed surface brightness, $I^{\rm obs}$, towards the cloud depends on its
optical depth, $\tau$, as
\begin{equation}
I^{\rm obs} = I^{\rm true} + \Delta I =  I^{\rm fg, true}  + (I^{\rm ext, true}-I^{\rm fg,true}) \times
e^{-\tau} + \Delta I.
\end{equation}
We have explicitly included a correction, $\Delta I$, which is needed if one does
not have absolute surface brightness measurements and the zero point of the intensity
scale is thus uncertain. In the equation, $I^{\rm fg, true}$ is the amount of true emission
that originates in front of the source, and $I^{\rm ext, true}$ is true value of the
extended background, against which the cloud is seen in absorption. The corresponding
observed value of $I^{\rm ext, obs}$ can be estimated by interpolating the observed
surface brightness over the target. 
The optical-depth can be then calculated as
\begin{equation}
\tau = - \ln \left(\frac{I^{\rm true} - I^{\rm fg, true}}
{I^{\rm ext, true} - I^{\rm fg, true}} \right)
= - \ln \left(\frac{I^{\rm obs} - I^{\rm fg, obs}}{I^{\rm ext, obs} -
I^{\rm fg, obs}} \right).
\label{eq:MIRext}
\end{equation}
This requires a separate estimate of the foreground component. In
\citet{ButlerTan2012}, $I^{\rm fg}$ was estimated by assuming that parts of the
target cloud are so optically thick that none of the background radiation comes
through. The minimum surface brightness is then also an estimate for $I^{\rm
fg}$. Because $I^{\rm obs}$ and $I^{\rm fg, obs}$ are affected by the same
correction $\Delta I$, this correction term disappears, and the optical depth
can be estimated even with an arbitrary zero-point offset in the data. However,
unless peak optical depths are very high, the minimum surface brightness
provides only an upper limit of the true foreground component. In real
observations, the non-constant background and the effect of local radiation
sources add to the overall uncertainty.

\subsection{Far-infrared dust emission}  \label{sect:methods:FIR}

The observed FIR emission from the target cloud is
\begin{equation}
I_{\nu} = \int B_{\nu}(T) e^{-\tau_{\nu}} d\tau_{\nu} 
\label{eq:FIR1}
\end{equation}
and thus depends via the Planck function, $B_{\nu}$, on the dust temperature, $T$,
and optical depth, $\tau_{\nu}$, along the line of sight. In most practical work,
the medium is assumed to be homogeneous and optically thin, which gives the
simple relationship
\begin{equation}
I_{\nu} =   B_{\nu}(T) (1-e^{\tau_{\nu}}) \approx B_{\nu}(T) \tau_{\nu}.
\label{eq:FIR2}
\end{equation}
The optical depth can be calculated from the modified blackbody
(MBB) function above
and with further assumptions of the dust absorption coefficient $\kappa_{\nu}$, optical depths can be
converted to column density. Optical depth is obtained from Eq.~(\ref{eq:FIR2}), once the dust
temperature $T$ has been first calculated by fitting the same Eq.~(\ref{eq:FIR2}) to multi-frequency
observations. In the optically thin case, the temperature determination does not depend on the absolute
value of the opacity $\kappa_{\nu}$, only on its frequency dependence. At FIR wavelengths, this is
usually written assuming a power-law dependence,
\begin{equation}
\kappa(\nu) = \kappa(\nu_0) \,\, (  \nu/\nu_0)^{\beta}.
\end{equation}
We used either $\beta=1.8$ (OMC-3 observations) or
$\beta=2.0$ (radiative transfer models). 

Column densities are estimated based on \Herschel 160-500\,$\mu$m data. In OMC-3, with the assumed dust
opacity, the column density reaches maximum values above $N({\rm H}_2)= 10^{23}$\,cm$^{-2}$,
which corresponds to a 160\,$\mu$m optical depth of $\tau=0.14$. The assumption of optically thin
emission therefore seems valid. There can be some optically thick regions, but only at scales below the
\Herschel resolution. Saturation of short-wavelength emission could lower the estimated colour
temperatures and lead to higher column-density estimates. However, even when optical depths are not
negligible, the use of the approximate form of Eq.~(\ref{eq:FIR2}) may still be preferred over the full
equation.

We calculated a low-resolution map (LR map) by convolving the \Herschel data to
a common 41$\arcsec$ resolution and deriving a column-density
map at the same
resolution. An alternative high-resolution map (HR map) is calculated
following \citet{Palmeirim2013}. We use the convolution kernels presented in
\citet{Aniano2011}, and the nominal resolution of the resulting column-density
map is 20$\arcsec$.

\citet{Mannfors2022} presented \artemis observations of the OMC-3 field. The
350\,$\mu$m map is centred at RA=5$^{\rm h}$35$^{\rm m}$20$^{\rm s}$ 
Dec=-5$\degr$1$\arcmin$31$\arcsec$ (J2000), and it covers an area slightly
larger than $9.3\arcmin \times 11.7\arcmin$ at an angular resolution of
$8.5\arcsec$. The noise level is $\sim$0.2\,Jy\,beam$^{-1}$, and the
signal-to-noise ratio exceeds 100 in the brightest part of the filament. The
main filament is equally well visible in the simultaneously observed \artemis
450\,$\mu$m map. However, because the 450\,$\mu$m map is slightly smaller (not
covering the filament segment A) and there are no \textit{Herschel} observations at
this wavelength, the 450\,$\mu$m data are not used in this paper.
The third column-density map is based on combined
(feathered\footnote{\url{https://github.com/radio-astro-tools/uvcombine}})
\Herschel and \artemis 350\,$\mu$m surface brightness map. The map has a
nominal resolution of 10$\arcsec$, although the temperature information is
available only at a lower resolution. In the following, we refer to this as
the AR map.

The analysis with a single MBB is common, but an observed spectrum is never
going to precisely match any single MBB function, because $\beta$ and $T$ are
not constant in the clouds. The effects of the line-of-sight temperature
variations are well known \citep{Shetty2009a, Malinen2011, Juvela2012_Tmix},
and we return to these questions in Sect.~\ref{sect:bias_FIR} and in a future
paper. With a sufficient number of frequency channels and data with high
signal-to-noise ratio, the observations can be modelled as a sum of several
temperature components, thus reducing the bias associated with the
single-temperature assumption. Such an analysis could be more sensitive to
other assumptions, such as the dust opacity spectral indices. Nevertheless,
methods such as point process mapping (PPMAP) \citep{Marsh2015} and inverse
Abel transform \citep{Roy2014, Bracco2017} have been successfully applied to
many dust continuum observations. Of these, PPMAP sets fewer requirements on
the symmetry of the modelled object but is computationally more expensive.
\citet{Howard2019} compared the MBB and PPMAP methods for filaments in the
Taurus molecular cloud. The PPMAP method resulted in some 30\% decrease in the
estimated filament FWHM values (Gaussian fits) and, perhaps surprisingly, in a
significant reduction in the estimated line masses. However, that analysis
also made use of the longer-wavelength SCUBA-2 observations, in addition to
the \Herschel 160-500\,$\mu$m data.

\subsection{Filament profile fitting} \label{sect:methods_fit}

We fit filament profiles with a Plummer-type function:
\begin{equation}
P(r; N_0, R, p, \Delta r) = N_0 [ 1 + ((r-\Delta r)/R)^2 ]^{(1-p)/2}
\label{eq:Plummer}
\end{equation}
\citep{Whitworth2001, Arzoumanian2011}. Here $r$ is the distance from the filament centre, in the
direction perpendicular to the filament path. In the case of OMC-3 data, the paths are defined by 
parametric splines through a number of hand-picked points. The free parameters of the fit are the peak
column density, $N_0$, the size of the central flat part of the profile, $R$, the power-law index, $p$,
and the sideways adjustment, $\Delta r$. The parameter $N_0$ is related to the central density of the
filament but also depends on the values of $R$ and $p$ as well as the scaling between column density
and mass \citep{Arzoumanian2011}. The parameter $\Delta r$ allows a shift if the local filament centre
does not perfectly align with the spline description of the filament.  By allowing the shift $\Delta r$
in individual profiles, one also avoids the artificial widening that would result from imperfect
alignment of the profiles.

The actual model fitted to OMC-3 data includes a linear background and the
final convolution of the model to the resolution of fitted column-density data,
\begin{equation}
F(r; N_0, R, p, \Delta r, A, B) = \mathrm{Con}(P(r; N_0, R, p, \Delta r)) + A + B\cdot r.
\label{eq:fit}
\end{equation}
The fit was done independently for each extracted 1D profile. This includes the replacement of the
2D-convolution of an image with a 1D-convolution of individual profiles. This is exact only if the 2D
filament does not change along its length, at the scale of a single beam. This also assumes that the
filament shape is close to Gaussian, because only in the case of a Gaussian filament convolved with a
Gaussian beam are the 1D and 2D results identical. Ideally, one would build a 2D model of the entire
region (with a global model for the background as well) that would be convolved in 2D during the fitting.
However, the 1D approximation of Eq.~(\ref{eq:fit}) is sufficiently accurate and, of course, is much
faster to calculate \citep[cf.][]{Mannfors2022}.

The fitting of Eq.~(\ref{eq:fit}) with six free parameters was done both with a normal least-squares
routine and with a Markov chain Monte Carlo (MCMC) routine, the latter providing the full posterior
probability distributions for the estimated parameters. The MCMC fitting uses uninformative priors,
except for forcing positive values for $N_0$ and $R$. We also required $p>1$, since $p=1$ corresponds
to a flat column-density profile. Appendix~\ref{sect:app_MCMC_test} shows one test on the expected
accuracy of the parameter estimates as a function of the noise level.

\subsection{Radiative transfer models} \label{sect:methods_RT}

We used a simple cloud model to study how the extracted filament parameters
might be affected by different error sources. In the case of FIR emission,
these include especially the line-of-sight temperature variations that cause
radially varying bias in the column-density maps. In the case of MIR analysis,
errors may be introduced by dust scattering and local thermal dust emission.

The cloud model was discretised onto a Cartesian grid of 200$^3$ cells, with a
cell size of 0.0116\,pc. The size of the cloud model is thus 2.32\,pc or
20$\arcmin$ for a distance of 400\,pc. The model consists of a single linear
filament, with a density profile matching the Plummer function with
$R$=0.0696\,pc and $p$=3. These correspond to a filament FWHM value of
0.14\,pc or 72$\arcsec$ at the 400\,pc distance. Tests are carried out with
different values of the filament column density, and observing the filament
mostly in a direction perpendicular to its main axis. 

We used different dust models for the emission and scattering calculations.
FIR emission is calculated using the core-mantle-mantle (CMM) dust that is
included in the the Heterogeneous dust Evolution Model for Interstellar Solids
(THEMIS) model \citep{Jones2013, Kohler2015, Ysard2016}. Although the dust
model does not contain large aggregates or ice mantles, it is appropriate for
dense environments. As an example of a more evolved dust populations, we use
in some tests the THEMIS AMMI model, aggregate grains with ice mantles. 

Because the CMM model does not include small grains, the calculations of MIR
thermal dust emission were performed with the \citet{Compiegne2011} dust model
(in the following, COM), which is appropriate for a diffuse medium. The
thermal emission should come mainly from outer filament layers that consist of
relatively pristine material. This is not necessarily true for the MIR
scattering, which originates in regions where the optical depth at MIR
wavelengths, rather than at optical wavelengths reaches unity. Dust scattering
is therefore also calculated using the CMM dust model. Differences in the
scattering properties of different dust models are discussed, for example, in
\citet{Ysard2016} and \citet{Juvela_L1642}.

The model filament is illuminated by an isotropic background according to the
\citet{Mathis1983} model of the interstellar radiation field (ISRF). In part
of the calculations, we included an additional point source, which was
modelled as a $T=15 700$\,K blackbody with a total luminosity of 590 solar
luminosities (similar to a B5V star). The source was placed at distance of
0.23\,pc from the top end of the filament (in the orientation used in the
figures) and at distance of 0.93\,pc from the filament axis. The viewing angle
was varied such that the source is in front of the filament, behind the
filament, or to the left of the filament. Since the amount of scattered light
is directly proportional to the illumination, these results can be easily
scaled for any source luminosity.

The radiative transfer calculations were performed with the SOC program
\citep{Juvela2019_SOC}, which gives 200$\times$200 pixel maps of the dust
emission at 8, 160, 250, 350, and 500\,$\mu$m and maps of scattered light at
8\,$\mu$m. The pixel size matches the model resolution and is 0.0116\,pc or
6$\arcsec$ for the assumed 400\,pc distance. Since the effect of stochastic
grain heating is small at long wavelengths, the 160-500\,$\mu$m dust emission
was calculated assuming equilibrium between the radiation field and the grain
temperatures. The stochastic nature of the grain temperatures was naturally
taken into account in the calculations of the MIR dust emission. The SOC
program is based on Monte Carlo simulations. The number of simulated photon
packages was selected so that the noise is a fraction of one per cent in the
FIR maps and $\sim 1\%$ or less in the computed maps of scattered light
(pixel-to-pixel). 

The FIR surface brightness maps were further convolved with Gaussian beams. In
most tests, this was set to $FWHM$=24$\arcsec$ and the column-density maps
were calculated at the same resolution. The simulated MIR data were used at
the full model resolution, because the model pixels are already larger than,
for example, the \textit{Spitzer} resolution at distances below 1.2\,kpc.

\section{Results from OMC-3 observations} \label{sect:results}

\subsection{OMC-3 filament in MIR absorption} \label{sect:results_MIR}

\begin{figure}
\sidecaption
\centering
\includegraphics[width=9cm]{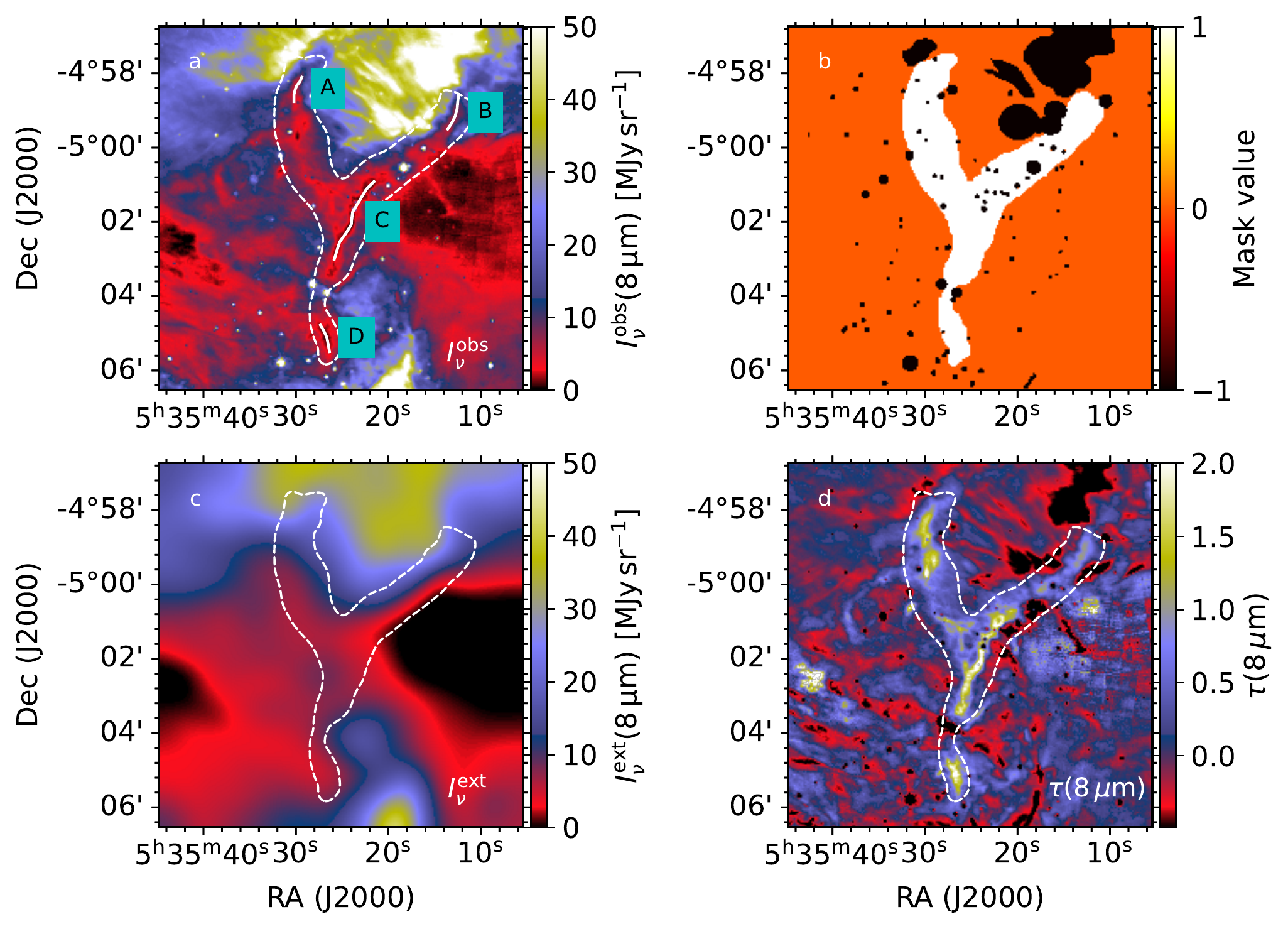}
\caption{
MIR observations of the OMC-3 field. Frame (a) shows the observed 8\,$\mu$m
surface brightness. Frame (b) shows the masks used, where the value 1
corresponds to the chosen filament region (shown with dashed contours in the
other frames) and the value -1 corresponds to masked stars. Frame (c) shows
the estimated extended emission, $I^{\rm ext}$, and frame (d) the resulting
8\,$\mu$m optical depth. Frame (a) also indicates the four filament fragments
that were chosen for further analysis (solid white lines with labels A-D).
}
\label{fig:G208_MIR_maps}
\end{figure}

We used Eq.(\ref{eq:MIRext}) and \textit{Spitzer} data to calculate an 8\,$\mu$m
optical-depth map for the OMC-3 field. We started by creating a mask for those
filament regions that are clearly visible in absorption
(Fig.~\ref{fig:G208_MIR_maps}a). The extended component $I^{\rm ext}$ was
calculated by replacing the masked pixels with interpolated values. In
practice, this was done by convolving the map with a Gaussian beam with
$FWHM$=40$\arcsec$, where the convolution ignored as inputs all pixels inside
the filament mask or inside the manually created masks for point sources
(Fig.~\ref{fig:G208_MIR_maps}b-c). Comparison of MIR and \Herschel data showed
that the MIR surface brightness rises towards the north, and the MIR data are
affected by local radiation sources, whose effect changes rapidly as a
function of the sky position. In Fig.~\ref{fig:G208_MIR_maps}b, the masks have
already been reduced to an area, where the relationship between the MIR
absorption and the FIR-based column densities appeared consistent.
The increased MIR emission towards the equatorial north is caused mainly by
the NGC~1977 open cluster (its northern sub-cluster), but the closest B stars
are still some 4 arcmin or 0.5\,pc (projected distance) north of the area
included in Fig.~\ref{fig:G208_MIR_maps} \citep{Getman2019, Megeath2022}.

The level of foreground emission is not known. \citet{ButlerTan2012} estimated the quantity
corresponding to $I^{\rm fg}+\Delta I$ statistically as
\begin{equation}
I^{\rm fg}+\Delta I =  
\langle  I^{\rm obs}(I^{\rm obs}<I^{\rm obs, min}+2\sigma) \rangle - 2\sigma,
\label{eq:BT2012}
\end{equation}
where $I^{\rm obs, min}$ is the minimum observed surface brightness and
$\sigma$ is the noise. In the OMC-3 region, the extended surface brightness,
and therefore probably also the foreground component, increases strongly
towards both north and south. We set $I^{\rm fg}$ equal to the minimum
observed surface brightness, which is found close to the centre of the area
shown in Fig.~\ref{fig:G208_MIR_maps}. This may lead to some overestimation of
the $\tau$ values at that position (and a few undefined values), which needs
to be taken into account in the subsequent analysis. On the other hand, the
selected value may underestimate the foregrounds further in the south and the
north.

The calculated map of the 8\,$\mu$m optical depth is shown in
Fig.~\ref{fig:G208_MIR_maps}d. Our main interest is in the shape of the
filaments, and scaling of $\tau$ to column density is not needed. However,
with a 8-$\mu$m dust opacity of 7.5\,cm$^2$\,g$^{-1}$, the values obtained for
the clean parts of the filaments (without strong point-source contamination)
are within a factor of $\sim$2 of those derived from the \Herschel data (when
compared at 20$\arcsec$ resolution). This scaling is used in some subsequent
plots where column-density units are used.

\begin{figure}
\sidecaption
\centering
\includegraphics[width=9cm]{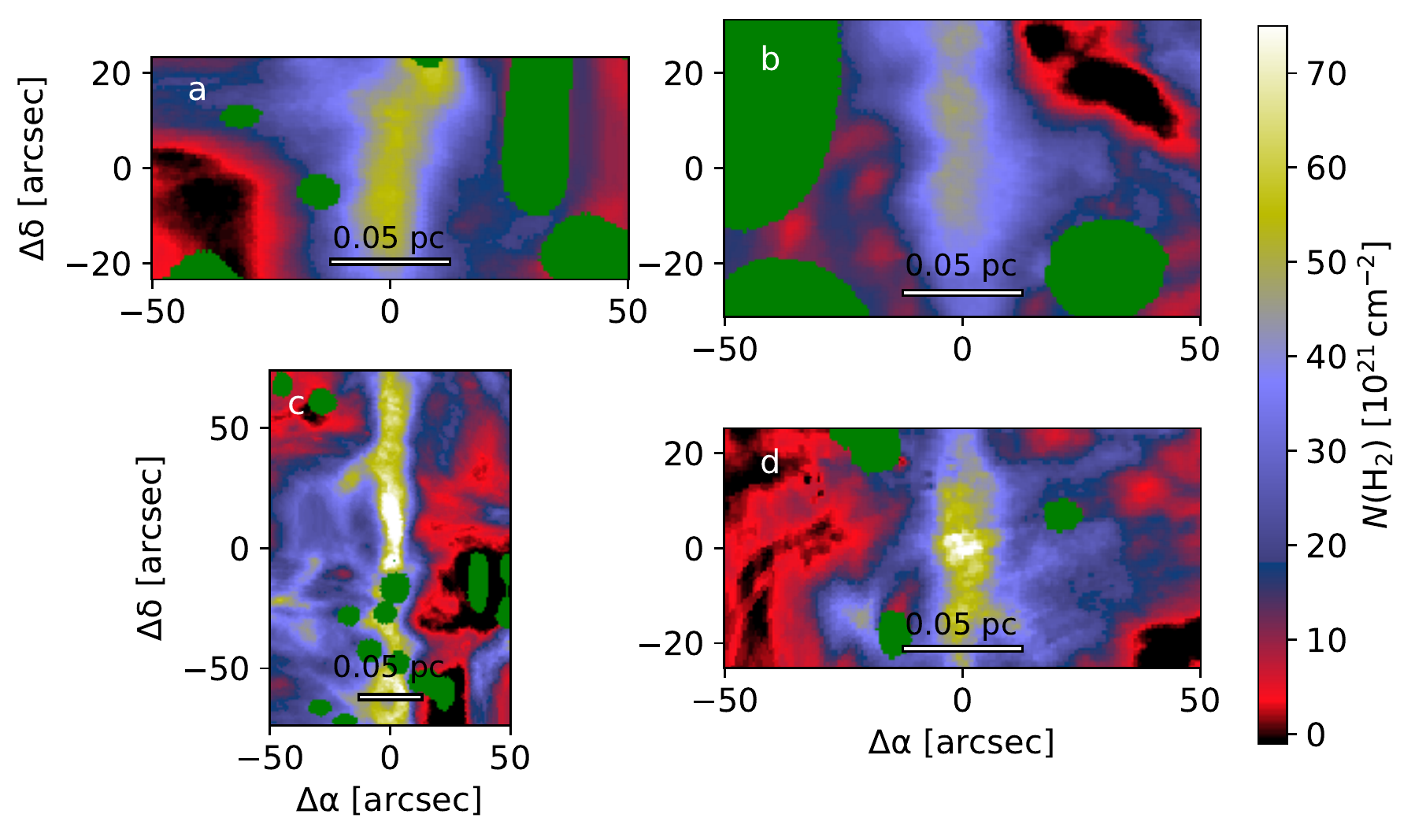}
\caption{
Two-dimensional images of the four OMC-3 filament segments that are marked in
Fig.~\ref{fig:G208_MIR_maps}a. Each frame shows one segment that is extracted
from the column-density map so that the filament runs vertically at the centre of
each frame. The green areas correspond to pixels that are masked because of point
sources.
}
\label{fig:MIR_filaments_W50as}
\end{figure}

Figure~\ref{fig:G208_MIR_maps} shows the MIR data and the derived 8\,$\mu$m
optical depth map. The strongest MIR absorption is found within a y-shaped
region, which can be interpreted as one main filament on the west side and one
side filament extending towards the north-east. For further analysis, we
selected from this area four isolated filament segments that show the deepest
MIR absorption (i.e. correspond to the highest column densities) and are not
significantly contaminated by emission from embedded or nearby stars. The
segments are labelled with letters A-D in Fig~\ref{fig:MIR_filaments_W50as}a.

\begin{figure}
\sidecaption
\centering
\includegraphics[width=9cm]{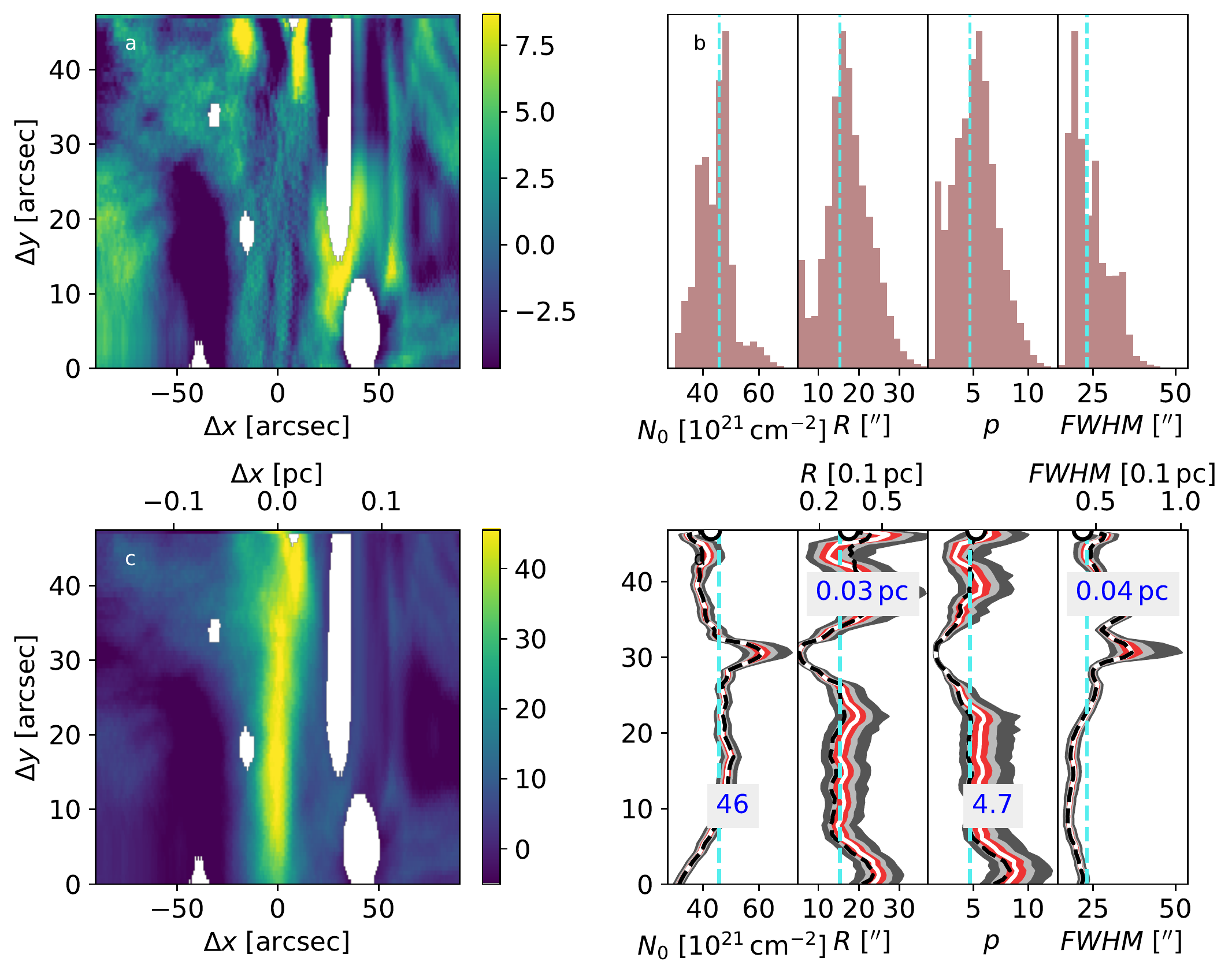}
\caption{
Results of the Plummer fits to OMC-3 filament segment A that was measured
using MIR extinction. Frame (c) shows the filament column densities (units
$10^{21}$\,cm$^{-2}$) as a 2D image, the filament running vertically in the plot.
The top row contains the median profile. Frame (a) shows the fit residuals. Frame (d)
shows the parameter estimates along the filament. The shaded regions correspond
to different percentile ranges of MCMC samples: [1, 99]\% in dark grey, [10,
90]\% in light grey, and [25, 75]\% in red. The MCMC median values are plotted as
solid white curves. The values from separate $\chi^2$ minimisation are plotted
with dashed black curves. The vertical dashed light green lines (frame b and
frame d) show the median of the parameter values in the individual least-squares
fits, while the white half-circles at the top of frame (d) correspond to the
fit to the median profile. Frame (b) shows histograms of the parameter
distributions based on MCMC samples over all profiles.
}
\label{fig:MIR0}
\end{figure}

The column-density maps of the segments were then extracted as 2D images,
where the filaments are aligned vertically. We fitted each individual row as
well as the median profile of each filament segment with Eq.~(\ref{eq:fit}).
The model includes a linear background, the Plummer profile with a possible
shift in the direction perpendicular to the filament length, and convolution
with a Gaussian beam with FWHM=2.0$\arcsec$ (approximating the
\textit{Spitzer} beam size). The results are shown in Fig.~\ref{fig:MIR0} for
the filament segment A, based on data within [-90$\arcsec$, +90$\arcsec$] of
the filament centre. Corresponding plots for the three other segments can be
found in Appendix ~\ref{sect:app_OMC3_MIR}.

The figures also show the filament FWHM values that were calculated from the
parameters of the Plummer fits. The $FWHM$ values are usually much more robust
than the values of the individual parameters $R$ and $p_0$ \citep{Suri2019}.
There is no significant correlation between the fitted filament column density
(parameter $N$) and the FWHM. If the filament is not well defined (or well
matched by the assumed functional form), the parameter $N_0$ does not
accurately represent the central column density, because of partial degeneracy
with the fitted linear background. The median FWHM value of the filament is
$\sim$0.05\,pc.

\subsection{OMC-3 filaments in FIR emission} \label{sect:results_FIR}

For comparison with the MIR results, we analysed the column-density maps
obtained from \Herschel and the combined \Herschel and \artemis data. We
extracted column densities for the same short filament segments as in the MIR
analysis and fitted these with the model of Eq.~(\ref{eq:fit}).

\begin{figure}
\sidecaption
\centering
\includegraphics[width=9cm]{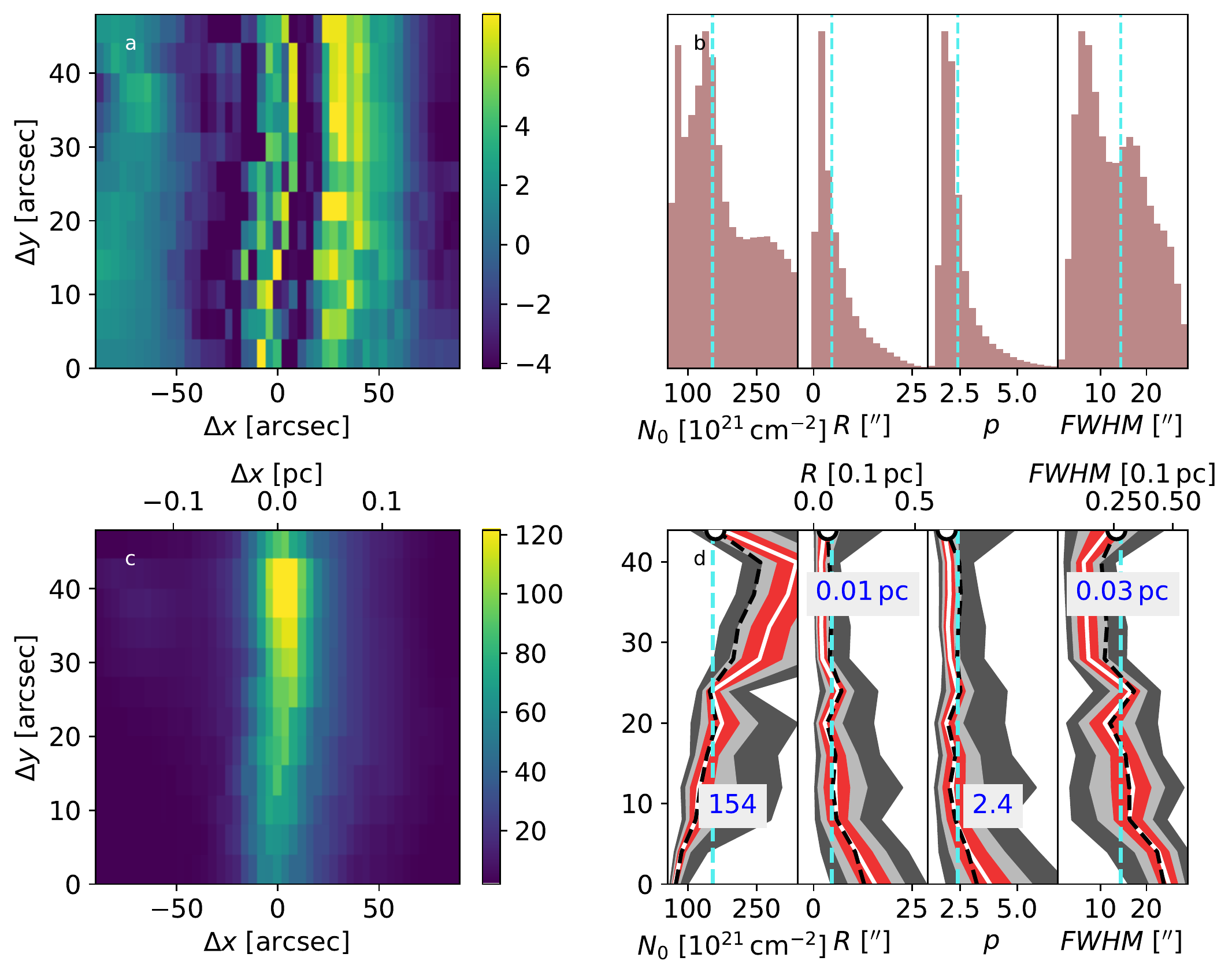}
\caption{
Plummer fits of OMC-3 filament segment A, using the HR column-density map (angular
resolution 20$\arcsec$). Frame (c) shows the filament segment as a 2D image (the
top row containing the median profile), frame (a) the fit residuals, frame (d) the
parameter estimates along the filament, and frame (b) the parameter histograms (cf.
description in Fig.~\ref{fig:MIR0}). The column-density map has an angular
resolution of 20$\arcsec$, and the fitted area is [-90$\arcsec$, +90$\arcsec$] in
the cross-filament direction.
}
\label{fig:G208_NH2_Palmeirim_NH2_MCMC_A0}
\end{figure}

\begin{figure}
\sidecaption
\centering
\includegraphics[width=9cm]{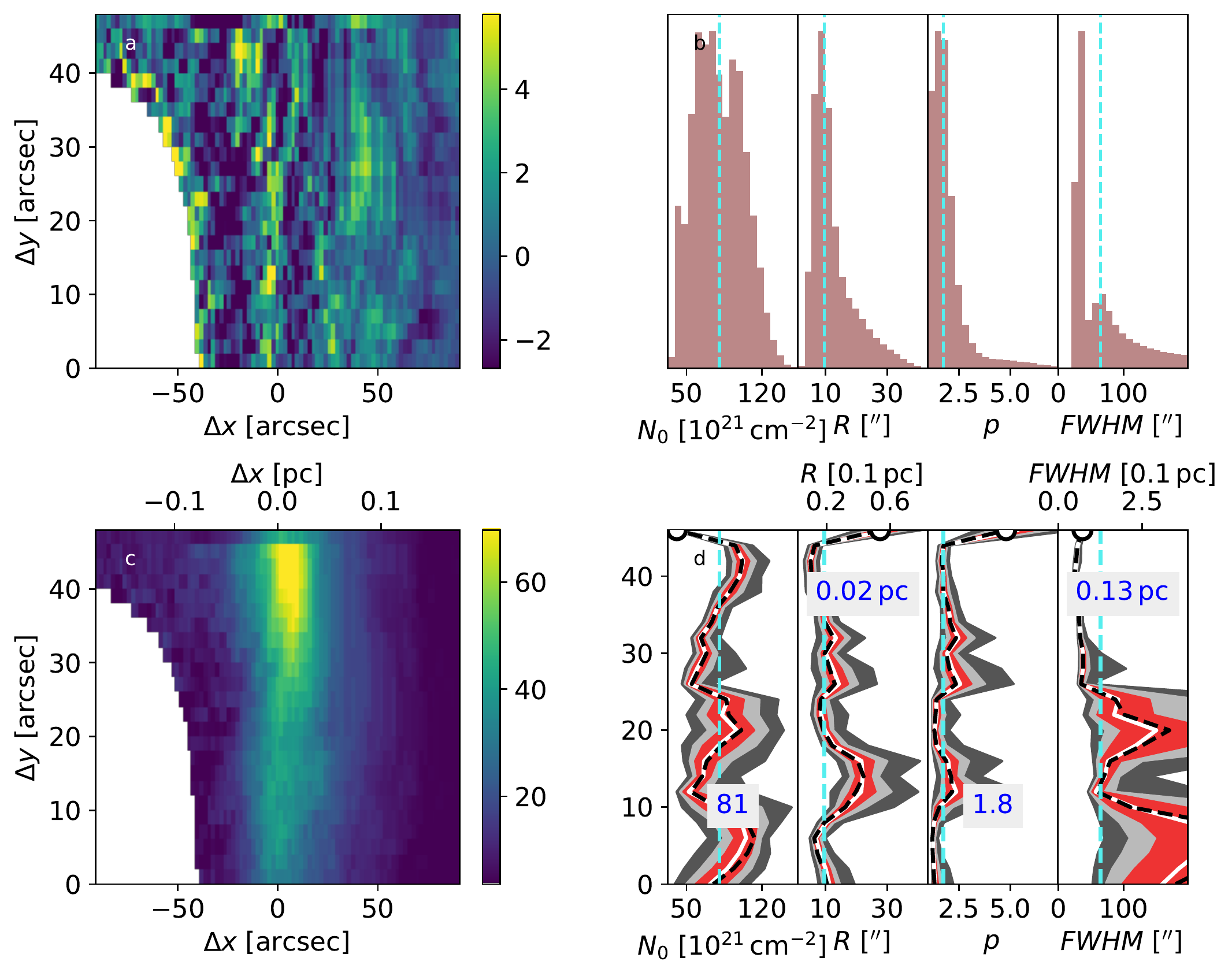}
\caption{
Plummer fits of OMC-3 filament segment A. The figure is the same as
Fig.~\ref{fig:G208_NH2_Palmeirim_NH2_MCMC_A0} but uses the AR column-density map
that is based on combined \Herschel and \artemis data and has an angular
resolution of 10$\arcsec$.
}
\label{fig:G208_NH2_feathered_MCMC_A0}
\end{figure}

We used the LR and HR column-density maps estimated with \Herschel
160-500\,$\mu$m data and the AR map that also makes use of \artemis data.
Figures~\ref{fig:G208_NH2_Palmeirim_NH2_MCMC_A0} and
\ref{fig:G208_NH2_feathered_MCMC_A0} show the results for the first filament
segment A, using the HR column-density map (angular resolution 20$\arcsec$)
and the AR map (angular resolution 10$\arcsec$). The corresponding plots for
the other filament segments are shown in Appendix~\ref{sect:app_OMC3_FIR}. 
The values of $p$ are of the order of $\sim$3, but with significant scatter.
The higher-resolution maps result in lower FWHM values, with the exception of
segment A. 

\begin{figure*}
\sidecaption
\centering
\includegraphics[width=12cm]{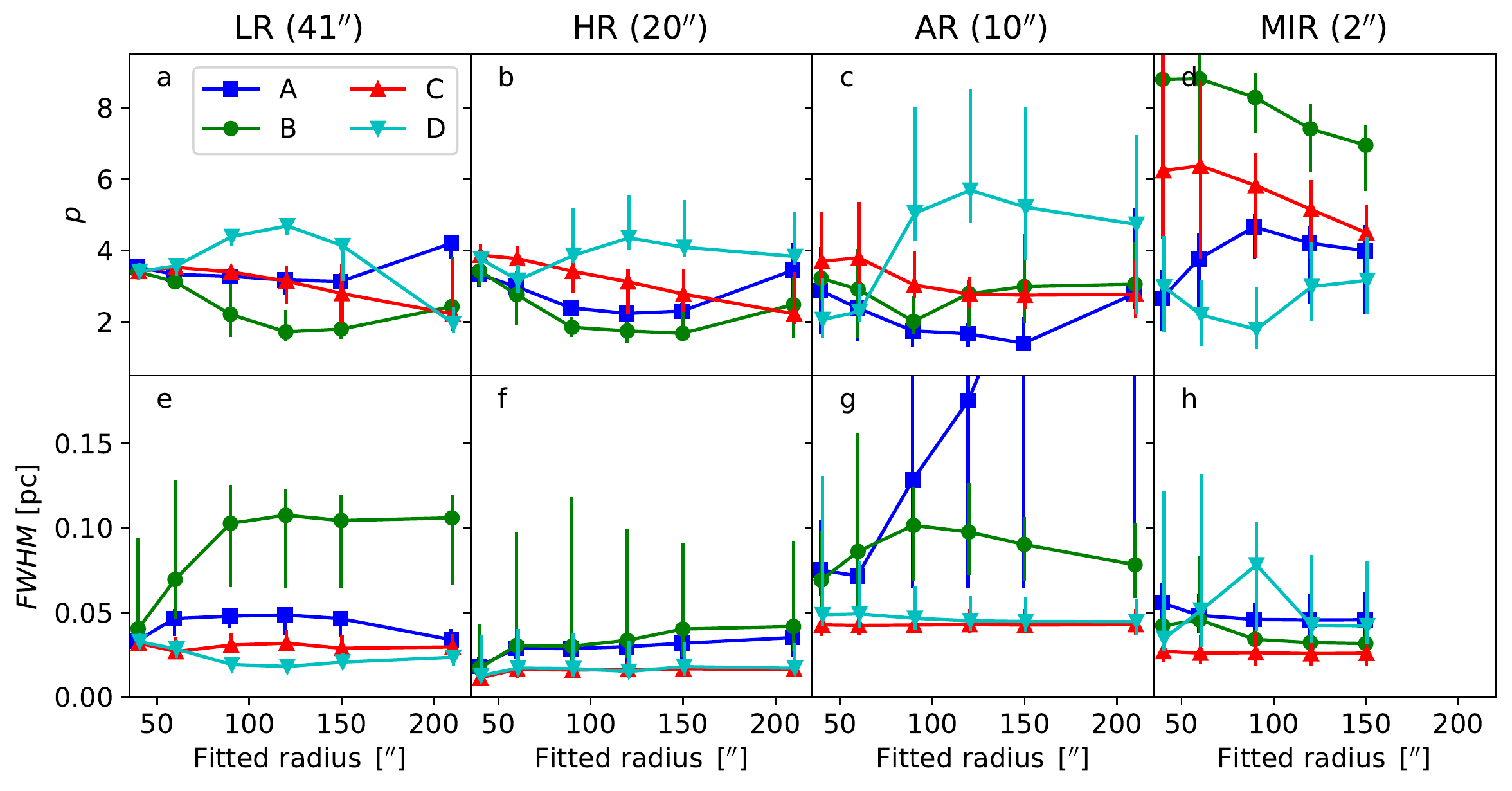}
\caption{
Parameters $p$ (upper frames) and FWHM (lower frames) in fits to OMC-3 LR, HR, and AR maps (based on
dust emission) and the MIR maps. Each frame shows results for the four filament segments, A-D, the
symbols corresponding to median and the inter-quartile range of the least-squares parameters over the
filament length.
}
\label{fig:plot_FIR_para}
\end{figure*}

The above results apply to fits to data within $|r|<90 \arcsec$ of the filament centre. To test the
sensitivity to the extent of the fitted region, the analysis was repeated by varying the maximum
distance between $r_{\rm max}=60\arcsec$ to $r_{\rm max}=210\arcsec$. Figure~\ref{fig:plot_FIR_para}
shows the resulting median parameter values for the four filament segments. In addition to the HR and
AR data (20$\arcsec$ and 10$\arcsec$ resolutions), we include here the results using the LR
(41$\arcsec$ resolution) column-density map.

The FWHM values obtained from different column-density maps are relatively
consistent. The results are similar for the LR and AR maps, and HR data result
in only slightly lower values. We also repeated the LR and HR analysis using
the column-density maps provided by the Gould Belt Survey, where the maps have
resolutions of 36.3$\arcsec$ \citep{Roy2013} and 18.2$\arcsec$
\citep{Polychroni2013}. The 36.3$\arcsec$ resolution maps showed no noticeable
differences to our results with the LR map. The 18.2$\arcsec$ maps resulted in
slightly higher FWHM values that match our LR and AR results more closely.
These small differences could be caused by differences in the background
subtraction (which in the case of the HR map was done close to the filament),
the convolution kernels, and even the assumed $\beta$ values \citep[$\beta=2$
in][]{Polychroni2013}.

One clear outlier is segment A in the AR map (Fig.~\ref{fig:plot_FIR_para}g), where the $FWHM$
value increases with increasing $r_{\rm max}$. However, the values appears to be affected by the masked
area, which corresponds to the edge of the \artemis coverage
(Fig.~\ref{fig:G208_NH2_feathered_MCMC_A0}). In the better defined end of the segment, the FWHM
values also drop in segment A close to the general $\sim 0.05$\,pc level. Segment B also shows
larger FWHMs in the LR and AR maps than in the HR map. The widths are smaller in the high-density end
and larger in the low-density end, the median in this case picking the larger value.

The MIR data result in FWHM estimates that are even surprisingly close to the values derived from
dust emission. However, while all emission maps give values $p\sim 3$ for the power-law index, the MIR
results show a much larger scatter. The values are high ($p>4$) for the B and C segments, and, as shown
in Figs.~\ref{fig:MIR1}-~\ref{fig:MIR2}, the values are consistently high along the entire filament
segments.

\section{Analysis of synthetic filament observations} \label{sect:bias}

In this section we use the cloud model of Sect.~\ref{sect:methods_RT} to
examine sources of potential bias in the dust emission and extinction
observations. With the parameters used in the simulations (0.0116\,pc pixels,
$R=6$ pixels, and $p=3.0$), the model filament has a FWHM size of 0.14\,pc or
some 72$\arcsec$ at a distance of 400\,pc (Sect.~\ref{sect:methods_RT}).
The synthetic observations were made using a Gaussian beam with
$FWHM=24\arcsec$. The beam size and filament properties are roughly similar to
those found in some \textit{Herschel} studies
\citep[e.g.][]{Arzoumanian2011,GCC-VII,Panopoulou2022} but the simulations are
not intended to directly replicate the OMC-3 observations. In particular, we
examine a wider range of models with maximum column densities ranging from
$N({\rm H}_2)=10^{21}$\,cm$^{-2}$ to $N({\rm H}_2)=10^{24}$\,cm$^{-2}$.  

\subsection{Bias in FIR observations} \label{sect:bias_FIR}

\begin{figure}
\sidecaption
\centering
\includegraphics[width=9.2cm]{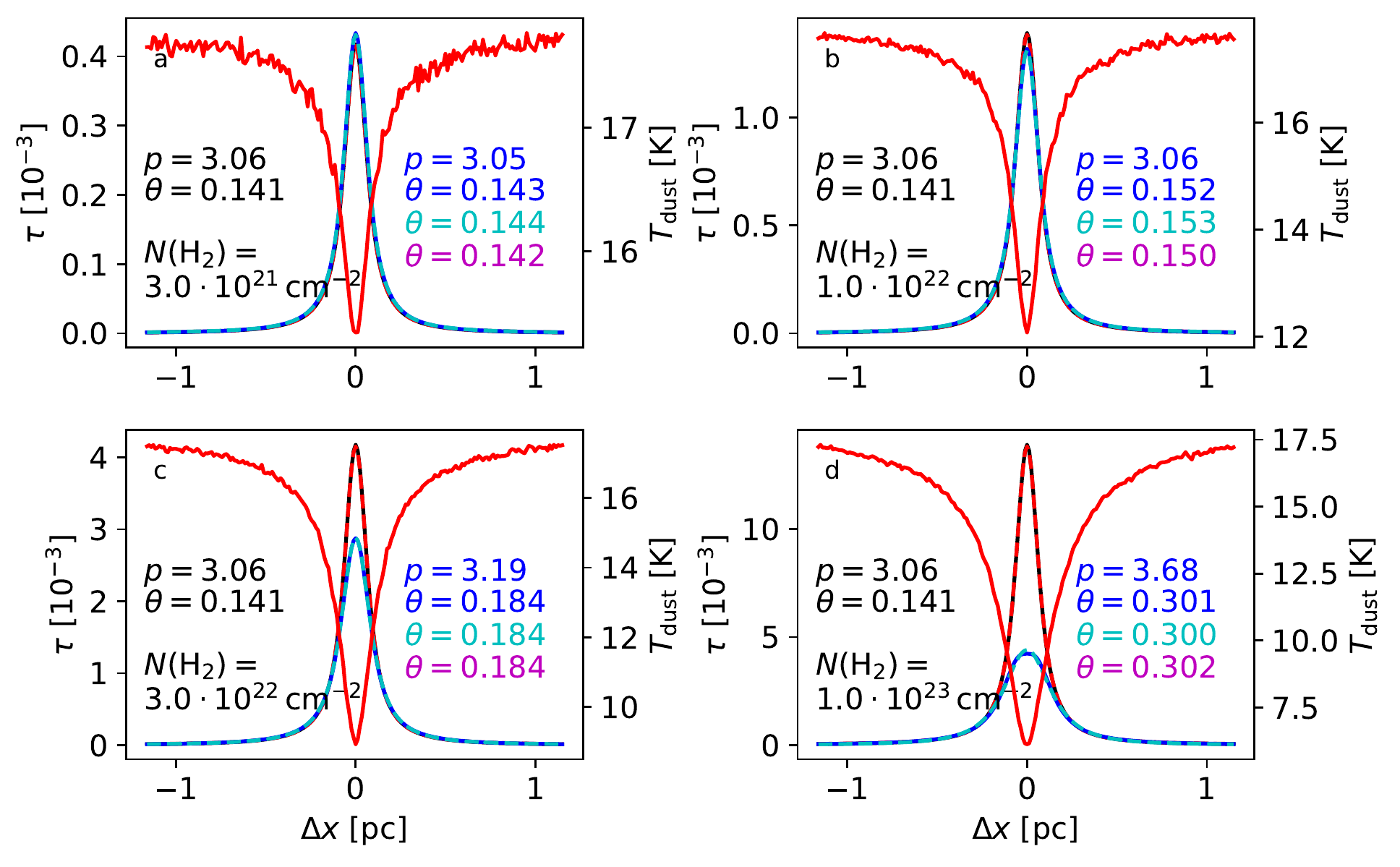}
\caption{
Profiles of optical depth, $\tau$, and dust temperature in the case of model
filaments illuminated by an isotropic external field. The black curves show
the true optical-depth profiles and the dashed red curves (fully overlapping
the black curves) the Plummer functions fitted to those profiles. The blue and
dashed cyan lines are, respectively, the optical depth profile derived from
synthetic 160-500\,$\mu$m surface brightness observations and the Plummer fit
to those data. The values of $p$ and FWHM (in parsecs, in the plot marked as
$\theta$) are shown for both the true profile (left side, black font) and the
estimated (right side, blue font) $\tau$ profiles. The analysis assumes
$\beta=2.0$, but FWHM values for $\beta=1.8$ and $\beta=2.2$ are also shown
in cyan and magenta, respectively. The assumed beam size is 24$\arcsec$. The
dust temperature profiles (solid red curves) are cross-sections of the 3D
model and are shown at full model resolution.
}
\label{fig:SED_eq_bg_fwhm4}
\end{figure}

We look first at FIR observations of a filament illuminated by an isotropic
external radiation field. Figure~\ref{fig:SED_eq_bg_fwhm4} shows the true
optical depth profiles and the profiles estimated from synthetic surface
brightness maps for filaments of different column density.

The results are almost correct up to $N({\rm H}_2) \sim 10^{22}$\,cm$^{-2}$,
although $\tau$ is increasingly underestimated. When the column density
reaches $N({\rm H}_2)=3\cdot 10^{22}$\,cm$^{-2}$, the estimated peak $\tau$ is
some 75\% of the correct value and $p$ is overestimated by $\sim$4\%. The
filament FWHM is overestimated by $\sim$30\%, and the effect increases with
increasing column density. 
The results were not sensitive to the assumed beam size.

The MBB fits were made with $\beta=2.0$, which is close to the actual value in
the dust model. At $N({\rm H}_2)= 10^{22}$\,cm$^{-2}$, the use of $\beta=1.8$
or $\beta=2.2$ would change the column-density estimates by -22\% or +27\%,
respectively, while the FWHM is affected only at the $\sim$1\% level.

\begin{figure}
\sidecaption
\centering
\includegraphics[width=9cm]{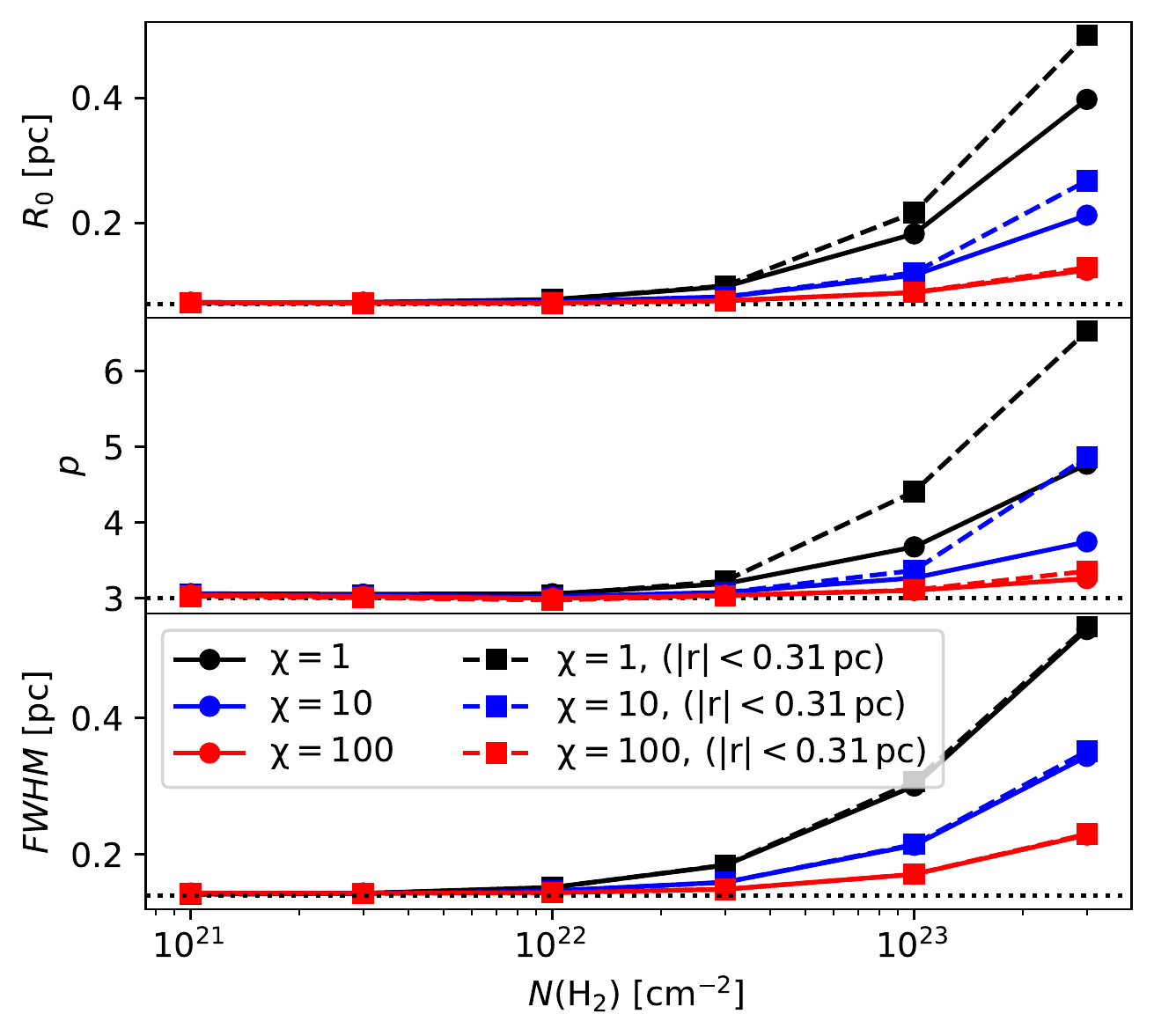}
\caption{
Estimated $p$ and $R$ values and the corresponding FWHM as a function of the
peak column density of isotropically illuminated model filaments. The black,
blue, and red curves correspond, respectively, to observations of model
filaments in a normal ISRF, $\chi=1$, and in stronger fields with $\chi=10$
and $\chi=100$. The solid lines with circles are fits to the model profile at
$|r|<0.58$\,pc, while the dashed lines with squares are fits to a narrower
region, $|r|<0.35$\,pc. The horizontal dotted lines indicate the true parameter
values in the models.
}
\label{fig:SED_eq_vs_NH2}
\end{figure}

Figure~\ref{fig:SED_eq_vs_NH2} shows how the bias in the $p$, $R$, and FWHM
estimates increases with column density. For the normal ISRF and fits to the
$|r|<0.58$\,pc area, the FWHM is at $N({\rm H}_2)=10^{23}$\,cm$^{-2}$
overestimated by a factor of two.
The bias in $p$ is 20\% at $N({\rm H}_2)=10^{23}$\,cm$^{-2}$ and more than
50\% at $N({\rm H}_2)=3 \cdot 10^{23}$\,cm$^{-2}$.
The fractional errors in $R$ are larger, but even there become significant
only beyond $N({\rm H}_2)=3\cdot 10^{22}$\,cm$^{-2}$. 

Figure~\ref{fig:SED_eq_vs_NH2} also shows results for a radiation field that
is a factor of $\chi=10$ or $\chi=100$ stronger at all
frequencies.\footnote{More luminous sources would intrinsically have higher UV
luminosity, but the UV-to-IR ratio of the radiation field at the filament
location can still be much lower, depending on the intervening extinction.}
This increases the temperature contrast in the filament but also the average
temperature ($\sim$7\,K for the $N({\rm H}_2)=10^{22}$\,cm$^{-2}$ and
$\chi=10$ case). Therefore, a change from $\chi=1$ to $\chi=10$ decreases the
systematic errors by a factor of two, and for $\chi=100$ the FWHM estimates
remain accurate up to $N({\rm H}_2)=10^{23}$\,cm$^{-2}$.

In many observations, fits can only be done using a limited sky area. 
Figure~\ref{fig:SED_eq_vs_NH2} also shows results for fits within the smaller
$r_{\rm max}=0.35$\,pc area. This has no effect on the FWHM values, but the
errors in $p$ increase. While the observed column-density profile can be
fitted with a Plummer function (with parameters different from the true
values), it does not match the Plummer profile exactly, since the result
depends on $r_{\rm max}$. In the case of noiseless synthetic observations, a
wider area always results in more accurate estimates. The situation can be
different in real observations if the signal in the filament wings is
dominated by noise and emission from unrelated structures.

A change in inclination changes the line-of-sight optical depths but, unlike a true
increase of the filament column density, does not affect the temperatures.
Appendix~\ref{sect:app_inclination} confirms that the inclination has only a
minor effect on the extracted filament parameters.

\begin{figure}
\sidecaption
\centering
\includegraphics[width=9cm]{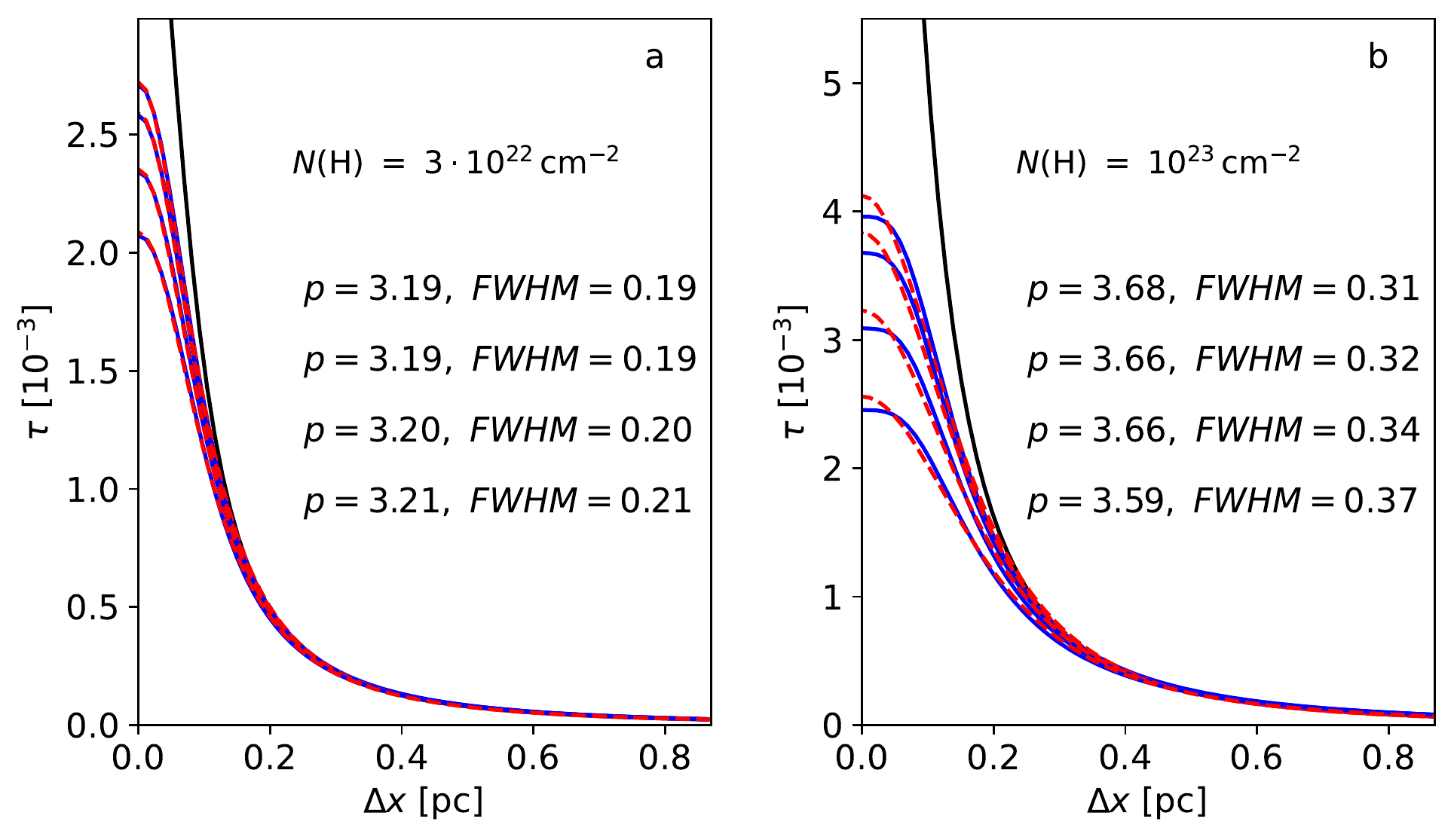}
\caption{
Selected optical-depth profiles along model filaments that are illuminated by
a point source. The filament column density is $N({\rm H}_2)=3\cdot
10^{22}$\,cm$^{-2}$ (frame a) or $N({\rm H}_2)=1\cdot 10^{23}$\,cm$^{-2}$
(frame b). The black curves show the true optical depths, the blue curves the
optical depths estimated from FIR observations, and the dashed red curves the
Plummer fits to those optical depths. The cross-sections are selected from positions
$\Delta y=$0.46, 0.93, 1.39, and 1.86\,pc along the filament, when the point
source is located at the position $\Delta y$=2.09\,pc along the filament and a
distance 0.93\,pc behind the filament. The curves from top to bottom are in
order of decreasing distance to the point source (increasing order of $\Delta
y$), and the parameters are listed in the same order (FWHM in units of
parsec).
}
\label{fig:SED_eq_PS}
\end{figure}

When the model includes a discrete radiation source, the parameter estimates
depend on the distance to the source. Figure~\ref{fig:SED_eq_PS} shows the
true and the recovered profiles at four positions along a model filament. The
point source is located behind the filament but, because the filament is
optically thin for FIR emission, the results are similar if the source were in
front of the filament. At $N({\rm H}_2)=10^{22}$\,cm$^{-2}$ and $N({\rm
H}_2)=3\cdot 10^{22}$\,cm$^{-2}$, the FWHM errors increase towards the 
point source location but the parameter $p$ is much less affected.
Figure~\ref{fig:SED_eq_PS}b demonstrates the central flattening of the
recovered $\tau$ profiles, and, as suggested by Fig.~\ref{fig:SED_eq_vs_NH2},
the Plummer fits do not follow the actual profile. \cite{Schuller2021} reached
qualitatively similar results, although in their case the effects were smaller
because of the stronger radiation field ($G_0$=1000).

\begin{figure}
\sidecaption
\centering
\includegraphics[width=9cm]{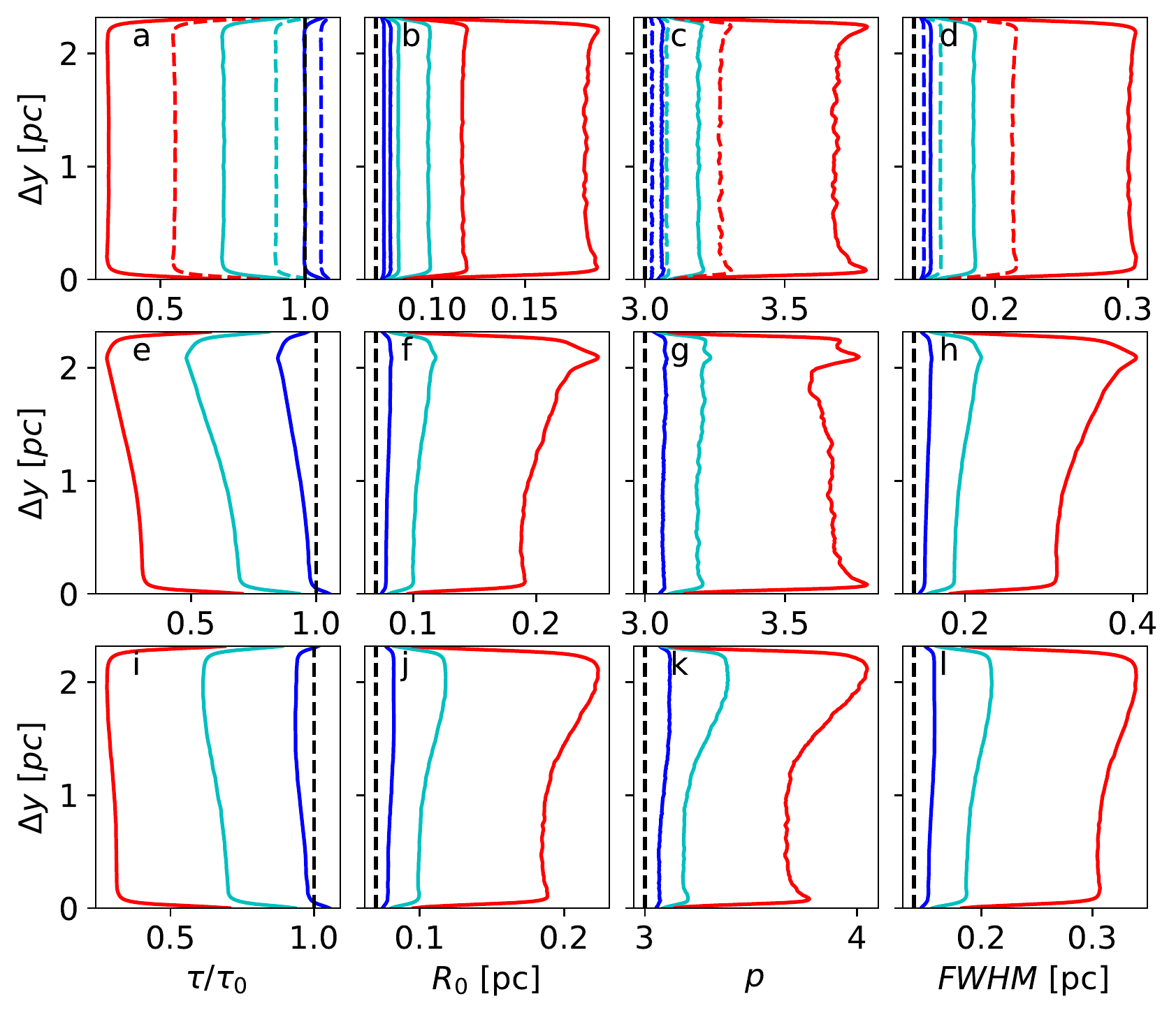}
\caption{
Variation in FIR-estimated parameters along model filaments in cases of:
isotropic illumination (frames a-d); a 590\,$L_{\sun}$ point source at
$\Delta y=2.09$\,pc and 0.93\,pc behind the filament (frames e-h); and to one
side of the filament (frames i-l). The plotted quantities are the ratio
between the estimated and true FIR optical depths ($\tau/\tau_0$), the Plummer
parameters ($R$ and $p$), and the filament FWHM calculated based on these. Each
frame shows results for three model filaments with peak column densities
$N({\rm H}_2)=10^{22}$\,cm$^{-2}$ (blue lines), $N({\rm H}_2)=3\cdot
10^{22}$\,cm$^{-2}$ (cyan lines), and $N({\rm H}_2)= 10^{23}$\,cm$^{-2}$ (red
lines). The dashed lines in frames (a)-(d) correspond to cases with a higher
isotropic radiation field ($\chi=10$); all other cases include an isotropic field
with $\chi=1$. True values are plotted with dashed black lines.
}
\label{fig:SED_eq_trace_3}
\end{figure}

Figure~\ref{fig:SED_eq_trace_3} shows how the parameter estimates vary along the
filament in the cases of isotropic illumination ($\chi=1$ or $\chi=10$) and
the sum of an isotropic field ($\chi=1$) and a point source. The figure
confirms the rapid increase of bias at column densities above $N({\rm
H}_2)=10^{22}$\,cm$^{-2}$. The $p$ values are more sensitive to point-source
illumination from one side, while FWHM is more affected when the source is
along the line of sight towards the filament. The small bias at the filament ends is
caused by these being subjected to the full unattenuated external field.

\begin{figure}
\sidecaption
\centering
\includegraphics[width=9cm]{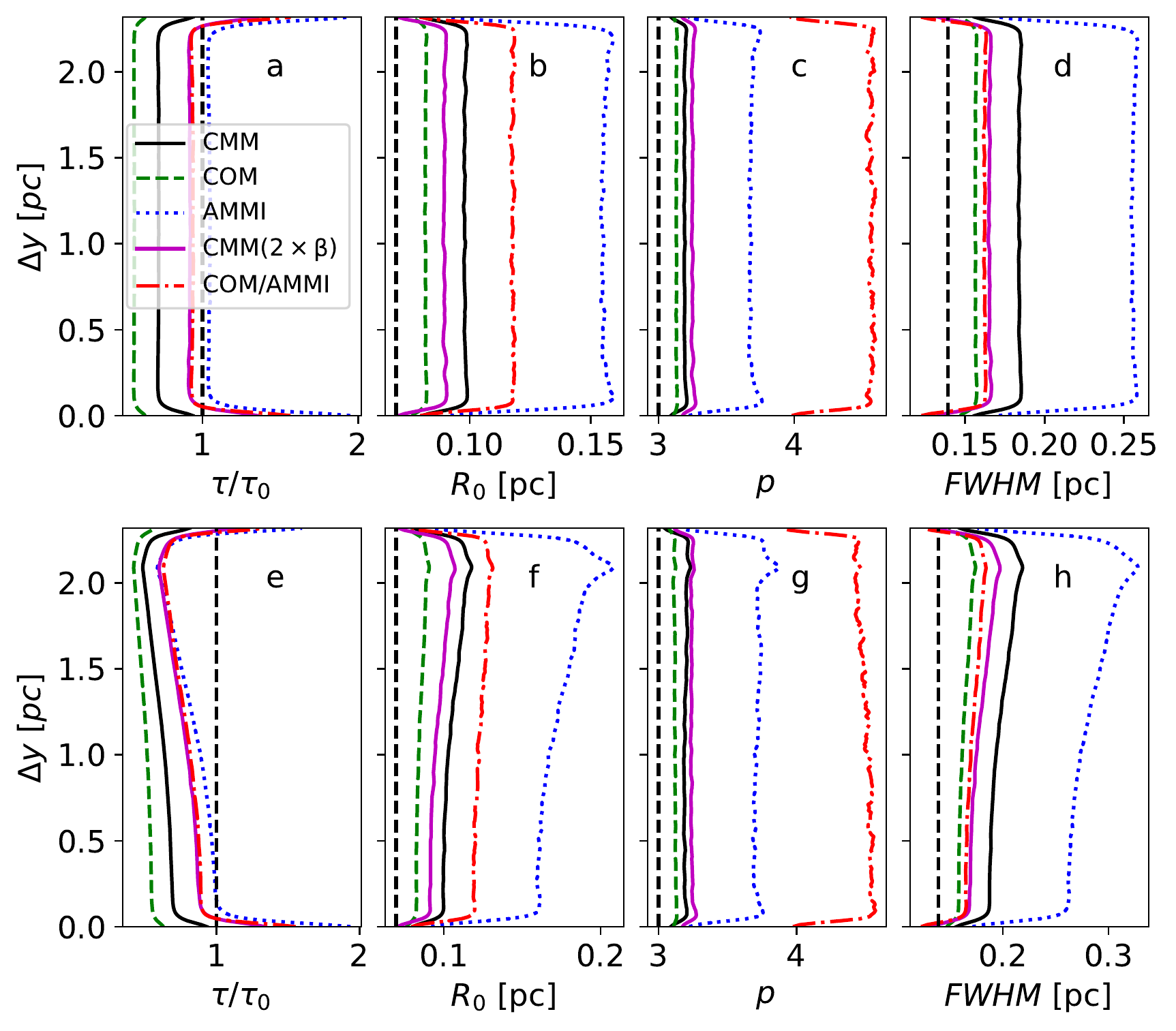}
\caption{
FIR-estimated filament parameters for single-dust models (COM, CMM, or AMMI) and two models with
spatial dust-property variations. In $\rm CMM(2\time \beta)$, the transition is from normal CMM dust in
the outer parts to modified dust with $\beta=1.5$ in the inner part. For COM/AMMI, the transition is
from COM dust to AMMI dust.
The filament column density is $N({\rm H}_2)=3 \cdot 10^{22}$\,cm$^{-2}$, and the
filament is illuminated by an isotropic radiation field (frames a-d) or with a
590\,$L_{\sun}$ point source at $\Delta y$=2.09\,pc and 0.93\,pc to one side of
the filament (frames e-h). The plotted parameters are the estimated optical depth
relative to its true value ($\tau/\tau_0$), the parameters $R$ and $p$ of the
fitted Plummer functions, and the resulting filament FWHM estimate. 
}
\label{fig:SED_eq_trace_TWIN}
\end{figure}

The FIR results can also be affected by changes in dust properties. These
could be related to changes in the grain sizes and optical properties,
following the formation of larger aggregates and ice mantles
\citep{Ossenkopf1994,Ormel2011,Jones2016}.
Figure~\ref{fig:SED_eq_trace_TWIN} compares calculations with uniform dust
properties to with two two-component models. The single-component models
consist of COM, CMM, or AMMI dust. The AMMI model tends to result in
the largest systematic errors, especially in the FWHM. The 250\,$\mu$m
opacity of AMMI is five times higher than for COM, and a similar difference
also exists at shorter wavelengths, where dust absorbs energy. The differences
are thus caused mainly by changes in the optical depth, since spectral index
of AMMI dust ($\beta\sim2.02$) is similar to the value $\beta=2.0$ that was
used in the MBB analysis.

We tested two cases with spatially varying dust properties. The first
two-component model consists of CMM dust and a modified CMM, where $\beta$ is
decreased from the original 160-500\,$\mu$m spectral index $\beta \sim 1.97$
down to $\beta=1.5$. The abundance of the modified dust is calculated as
$\tanh(n/[10^5\,{\rm cm}^{-3}])$, so that the filament centre consists
entirely of the modified dust. This shows the effect of a change in the
spectral index, without a net change in the opacity. The results show
systematic but relatively minor variations in the filament parameters
(Fig.~\ref{fig:SED_eq_trace_TWIN}).

The second two-component model consists of COM dust in the outer parts and
AMMI dust in the inner part. The relative abundance of the AMMI component
follows the same density dependence as above. The spectral indices are
different ($\sim$2.02 and $\sim$1.83 for AMMI and COM, respectively) but there
is a larger difference in the absolute opacities. While the pure AMMI model
led to the largest $R$ and FWHM estimates, the COM+AMMI combination leads to
the largest $p$ values. This illustrates qualitatively the potential effects
from spatial dust property variations. However, the quantitative results will
also be sensitive to the radial position and steepness of the transition in
dust properties.

\subsection{Bias in MIR observations} \label{sect:bias_MIR}

The filament models of Sect.~\ref{sect:bias_FIR} were also used to examine how
the MIR analysis is affected by in situ dust scattering and emission.
Radiative-transfer calculations provide the surface brightness due to
8\,$\mu$m dust scattering with the CMM dust model and the thermal emission
from stochastically heated grains with the COM dust model. We concentrate here
on the systematic effects. Appendix~\ref{sect:app_MIR} examines further some
effects related to observational noise.

\subsubsection{Effect of MIR scattering} \label{sect:bias_SCA}

For a filament with $N({\rm H}_2)= 3 \cdot 10^{23}$\,cm$^{-2}$, the scattering
results in errors of less than $\sim$1\% in the estimated $\tau(8\,\mu{\rm m})$.
This remains true even if the isotropic radiation field is increased to
$\chi=10$ or the default point-source luminosity is increased by a factor of
50. 
The calculations assume an intensity $I^{\rm bg}$=10\,MJy\,sr$^{-1}$ for the
background sky. This is similar to the OMC-3 field but still a relatively low
value. If the background is higher, the effects from scattering in the cloud
would be further reduced. 

\begin{figure}
\sidecaption
\centering
\includegraphics[width=9cm]{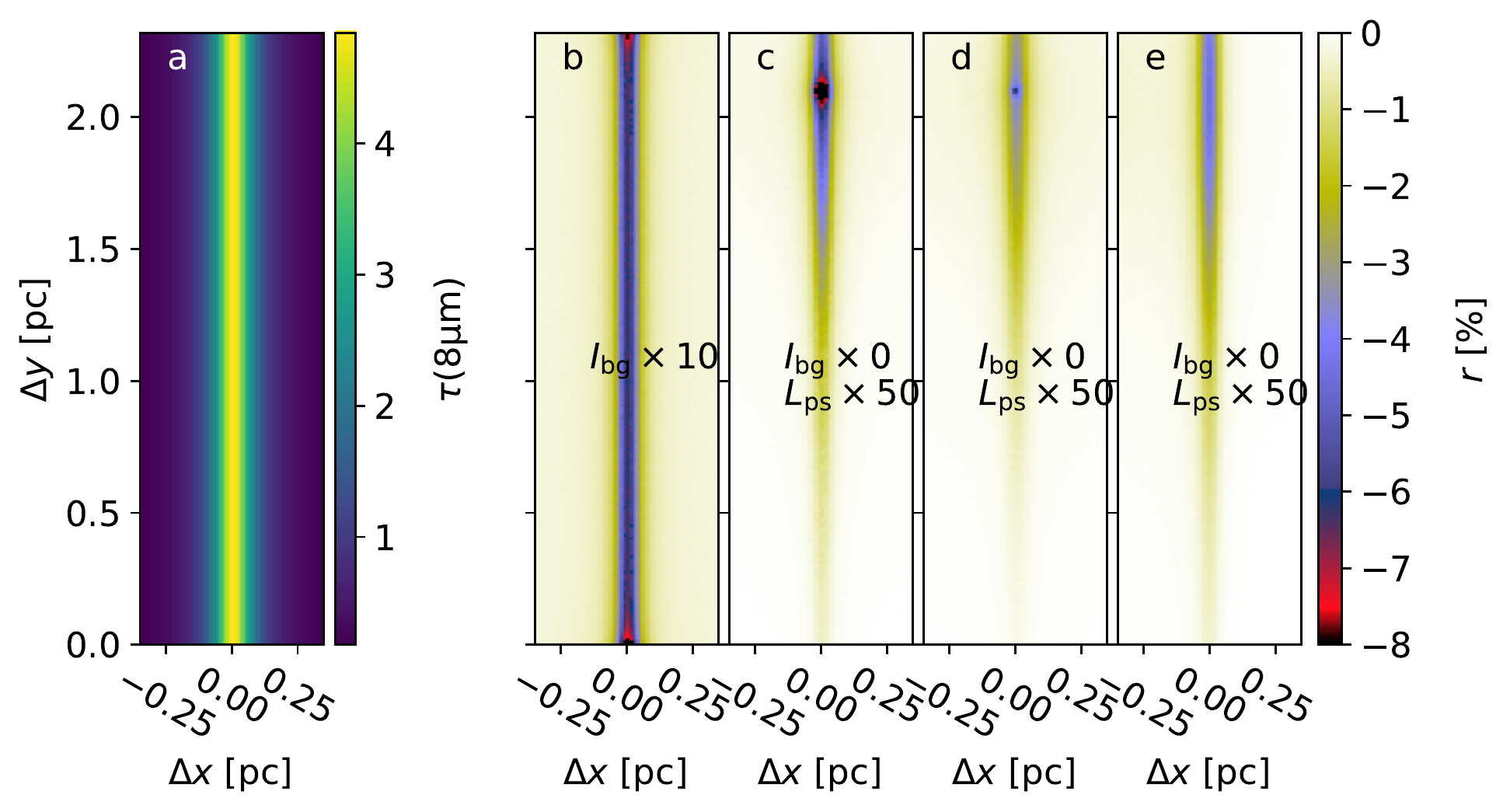}
\caption{
Modelled effect of scattering on the MIR optical-depth estimates towards filament centre. Frame (a) shows
the model optical depth, which corresponds to a peak column density of $N({\rm H}_2)=
10^{24}$\,cm$^{-2}$ with the CMM dust model. The other frames show the relative error in $\tau(8\mu{\rm
m})$ estimates. Frame (b) includes only an isotropic radiation field with $\chi=10$. Frames (c)-(e) show the
errors, when a point source, with luminosity 50 times the default value, is included at $\Delta
y=2.09$\,pc and 0.93\,pc behind the filament (frame c), in front of the filament (frame d), or to one
side. 
}
\label{fig:SED_SCA_1_HI1}
\end{figure}

The importance of scattering increases with increasing column density. For
$\tau(8\,\mu{\rm m})<1$, the intensity of the scattered light follows the
column-density profile. If the filament is optically thick, the scattered
light will peak on either side of the column-density peak, with potentially
larger impact of the parameter estimates. However,
Fig.~\ref{fig:SED_SCA_1_HI1} shows that for a filament with $N({\rm H}_2)=
10^{24}$\,cm$^{-2}$, the maximum optical-depth errors remain below 10\%, both
for an isotropic field $\chi=10$ or a point source with luminosity 50 times
the default value.  
If the isotropic radiation field is increased to $\chi=50$, the errors exceed
20\% for a $N({\rm H}_2)= 10^{24}$\,cm$^{-2}$ filament. If the column density
is increased further by a factor of three (to rather extreme values), the
errors would exceed 60\%. The bias is determined mainly by the optical depth.
The optical depth depends on the column density but also the dust properties
and would be more than two times higher for the AMMI dust than for the CMM
dust was used in Fig.~\ref{fig:SED_SCA_1_HI1}.

\begin{figure}
\sidecaption
\centering
\includegraphics[width=9cm]{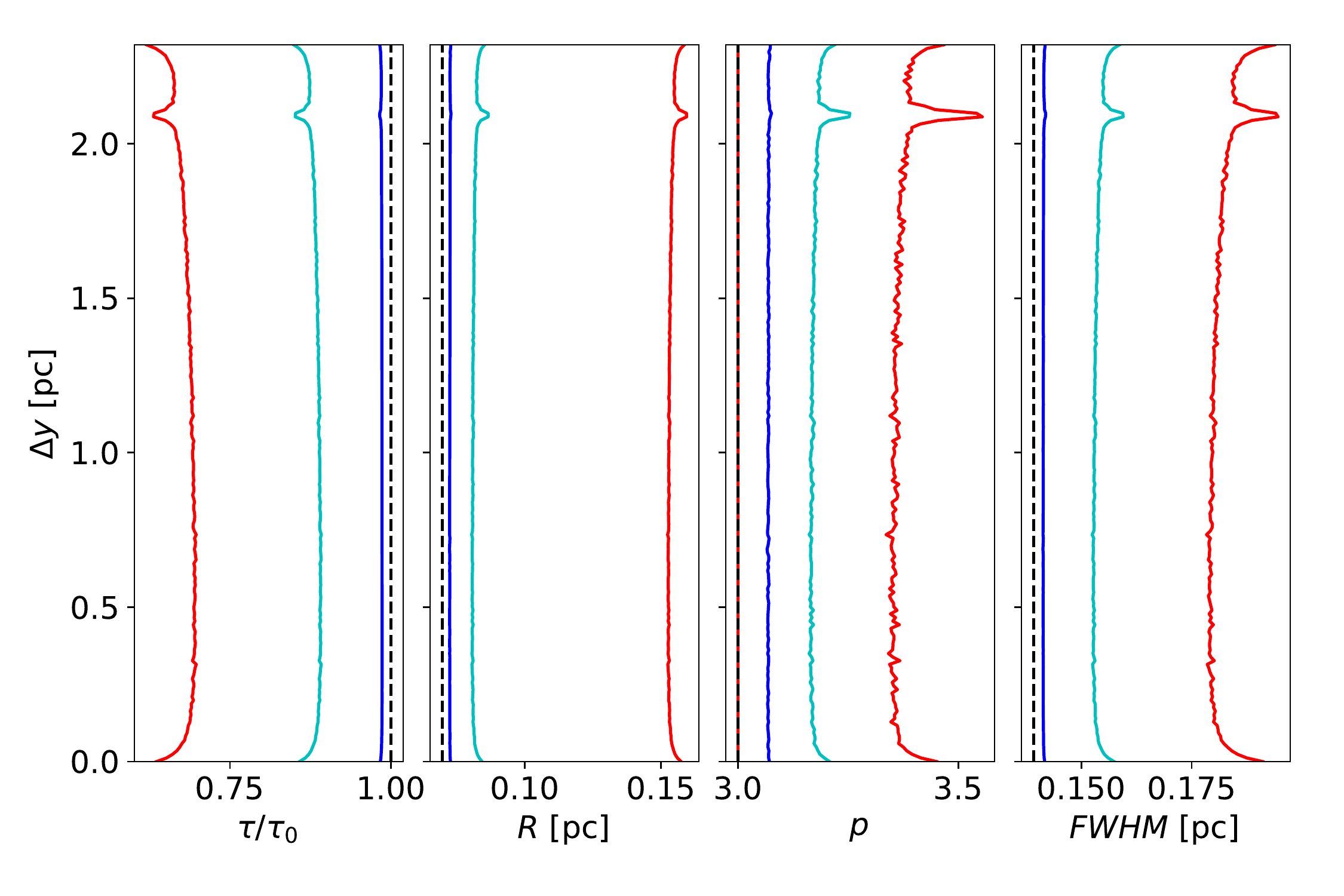}
\caption{
Modelled effect of scattered light on the filament parameters derived from MIR observations. The
filament column density is $N({\rm H}_2) = 10^{24}$\,cm$^{-2}$, and it is illuminated by an isotropic
radiation field and a foreground point source. The blue curves correspond to the case of $\chi=1$ for
the isotropic component and the point source with the nominal luminosity. The cyan and red curves show,
respectively, the results when both radiation-field components are scaled by a factor of 20 or 100. The
dashed black lines show the correct values of the parameters.
}
\label{fig:SED_sca_trace_PS1}
\end{figure}

Figure~\ref{fig:SED_sca_trace_PS1} shows the effect of scattering on the
filament parameter in the case of a $N({\rm H}_2) = 10^{24}$\,cm$^{-2}$
filament illuminated by an isotropic field and a foreground point source. For
the default radiation-field values, the bias caused by light scattering is
$\sim$1\% or less. When the isotropic field is increased to $\chi=20$, the
errors in $p$ and FWHM rise to a few per cent. If the point source is made
50 times stronger, the errors exceed $\sim$15\%, but only in a small area
closest to the point source. 

In summary, the scattering in the filament has only a minor effect on the
parameter estimation. The effect become visible only if the local radiation
field is very strong, the filament has a very high column density, and the
background surface brightness is low.

\subsubsection{Effects of MIR dust emission} \label{sect:bias_SHG}

\begin{figure}
\sidecaption
\centering
\includegraphics[width=9cm]{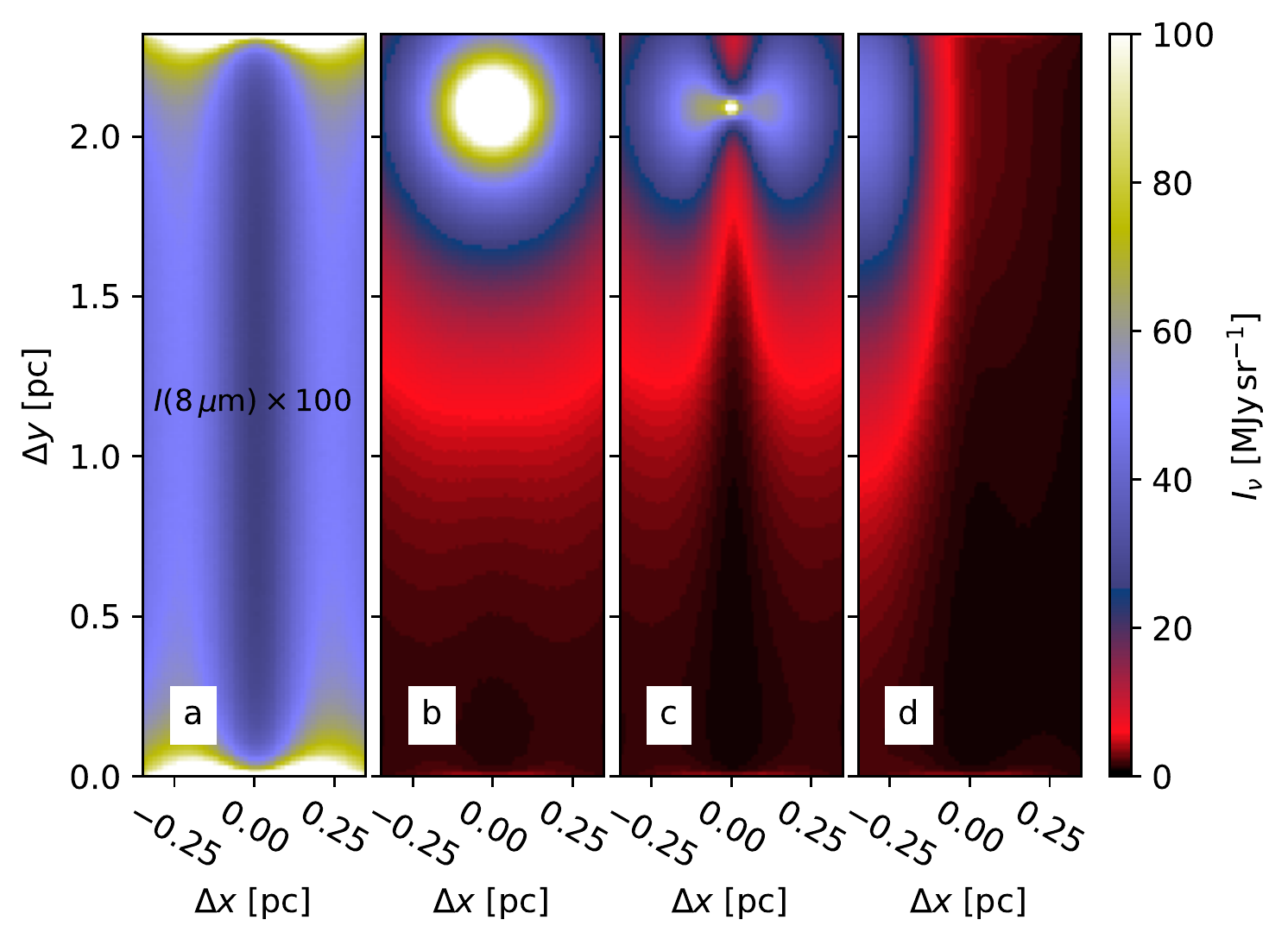}
\caption{
Maps of thermal dust emission $I^{\rm shg}$ calculated for the $N({\rm H}_2) =
3 \cdot 10^{23}$\,cm$^{-2}$ filament model. Frame (a) corresponds to isotropic
illumination with $\chi=1$, but the surface-brightness values are multiplied
by 100 just for plotting (the same colour bar applies to all frames). The other
frames include both the isotropic field and a point source located at $\Delta
y=2.09$\,pc and at a distance of 0.93\,pc in front, behind, or to the left of
the filament centre axis (frames b, c, and d, respectively).
}
\label{fig:SED_SHG_maps}
\end{figure}

The effects of the 8 $\mu$m thermal emission from stochastically heated grains
($I^{\rm shg}$) was examined using the COM dust model.
Figure~\ref{fig:SED_SHG_maps} shows maps of the dust emission for a $N({\rm
H}_2) = 3 \cdot 10^{23}$\,cm$^{-2}$ filament. An isotropic radiation field
($\chi$=1) results in emission at a level of $I_{\nu}(8\,\mu{\rm m})\sim
0.5$\,MJy\,sr$^{-1}$. This is not completely negligible if the background sky
brightness is low. The 590\,$L_{\sun}$ point source has a larger effect, which
ranges from less than 1\,MJy\,sr$^{-1}$ far from the source to
$\sim$100\,MJy\,sr$^{-1}$ close to the source. For the column density of
$N({\rm H}_2) = 10^{23}$\,cm$^{-2}$, the surface brightness remains at a
similar level, but the morphology is different. The surface brightness follows
more closely the column-density distribution, and, for a source behind the
filament (Fig.~\ref{fig:SED_SHG_maps} frame c), the intensity peaks towards
the centre of the filament, with only a minor dip at $|\Delta x| <0.05$\,pc.

Because the thermal emission is stronger and more extended than the scattered
light, it could even affect the observer's estimate of $I^{\rm bg}$. We
calculated alternative $\tau$ maps, where the median value of  the thermal
emission $I^{\rm shg}$ from stochastically heated grains (in the area visible
in the Fig.~\ref{fig:SED_SHG_maps}) was added to the original $I^{\rm bg}$.
The added component is not truly part of the sky background, because it
originates within the source itself and preferentially on the observer's side
of the source. 

\begin{figure}
\sidecaption
\centering
\includegraphics[width=9cm]{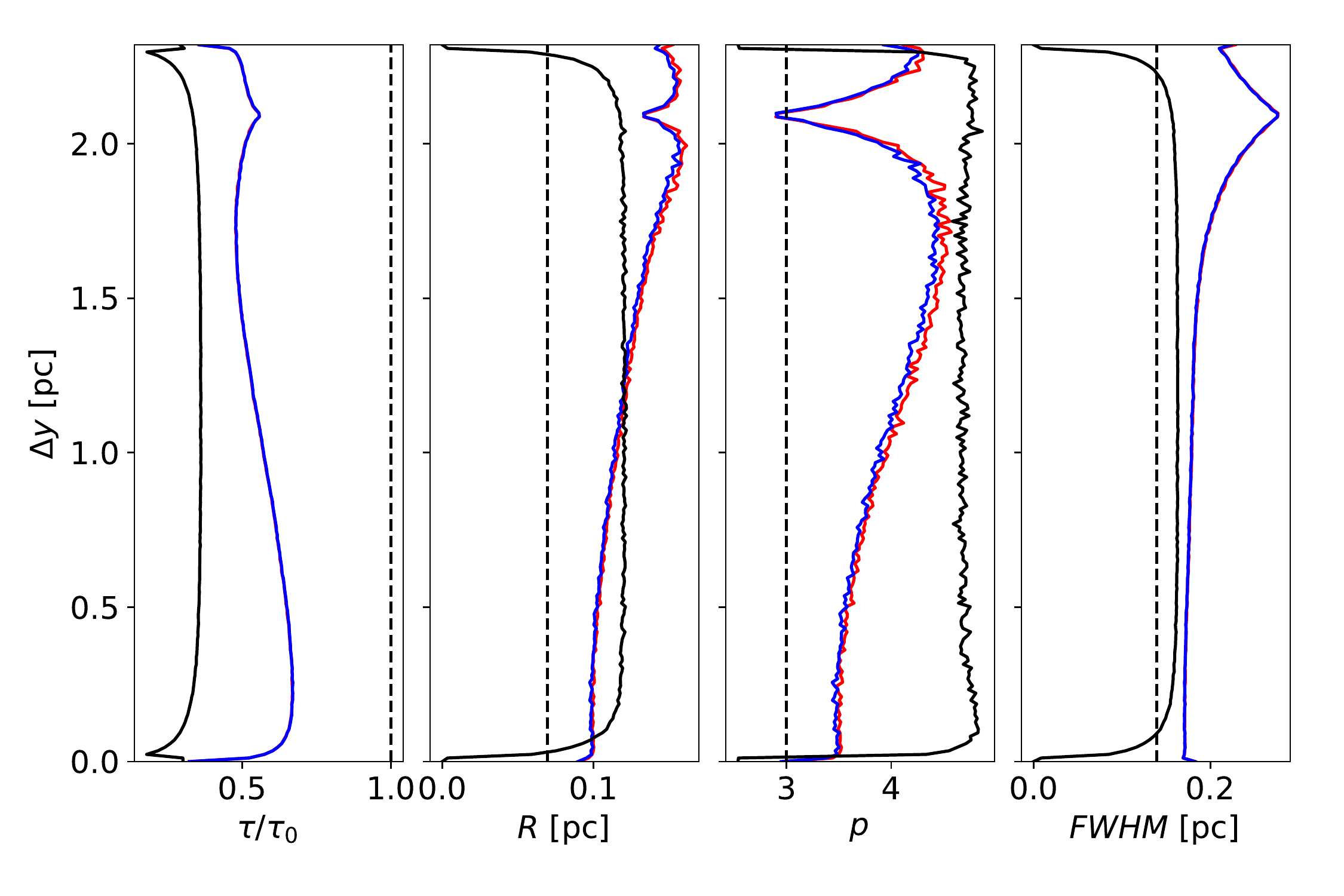}
\caption{
Estimated filament parameters along the $N({\rm H}_2) = 3 \cdot
10^{23}$\,cm$^{-2}$ model filament, when the surface brightness includes
8\,$\mu$m thermal dust emission. The filament is illuminated by an isotropic
background ($\chi=1$) and a point source at $\Delta y= 2.09$\,pc and a
distance 0.94\,pc to one side of the filament (cf.
Fig.~\ref{fig:SED_SHG_maps}d).  Red curves correspond to calculations with the
true value of $I^{\rm bg}=10$\,MJy\,sr$^{-1}$ and blue curves to a case where
the median surface brightness from Fig.~\ref{fig:SED_SHG_maps}d is added to
the estimate of $I^{\rm bg}$. The black curves are for the case of a $\chi=10$
isotropic radiation field (no point source), with a similarly adjusted $I^{\rm
bg}$ estimate.
}
\label{fig:SED_SHG_trace_PS3_HI1}
\end{figure}

Figure~\ref{fig:SED_SHG_trace_PS3_HI1} shows the results for $N({\rm H}_2) = 3
\cdot 10^{23}$\,cm$^{-2}$, when the point source is towards one side of the
filament. Because the observed sky brightness varies along the $\Delta y$
coordinate, the $\tau$ profiles do not drop to zero in the filament wings.
This effect is mostly eliminated by the linear background component that is
part of the fitted profile function (Eq.~(\ref{eq:fit})). Nevertheless, the
parameter estimates vary by up to 50\% with the distance to the point source.
Both $p$ and FWHM are more overestimated near the point source, although $p$
drops sharply at the position closest to the point source.

Figure~\ref{fig:SED_SHG_trace_PS3_HI1} also shows results for a stronger
isotropic field with $\chi=10$, where $I^{\rm bg}$ has to take into account
the extended emission. The optical depths are now underestimated more, and the
especially $p$ shows large bias. Nevertheless, the filament FWHM is
overestimated only by some 25\%.

Other cases with $N({\rm H}_2) = 3 \cdot 10^{23}$\,cm$^{-2}$ but different
radiation fields are shown in Appendix~\ref{sect:app_bias_emission}. If the
point source source is directly in front of or behind the filament, its effect
is amplified, as more line-of-sight material is heated. The parameter values are then
lower at $\Delta y \ga 1$, and the filament disappears close to the projected
point-source location, when the dip caused by the background extinction is
filled by thermal emission. This happens even earlier for models of lower
column density, because the lower opacity reduces the background absorption
more than it reduced the thermal emission.

In Appendix~\ref{sect:app_bias_emission} we examine a model with lower column
density, $N({\rm H}_2) = 3 \cdot 10^{22}$\,cm$^{-2}$, and higher background
intensity, $I^{\rm bg}=100$\,MJy\,sr$^{-1}$.  Filament parameters are there
generally accurately recovered. However, the errors in $p$ and FWHM still
reach 30\% close to the point source and, if the source is along the line of sight, the
filament disappears as an absorption feature.

\section{Discussion} \label{sect:discussion}

We have examined the estimation of the filament properties with observations
of MIR extinction and FIR dust emission. In Sect.~\ref{sect:disc_OMC3}, we
discuss the observational results on the OMC-3 field. In 
Sect.~\ref{sect:disc_bias}, we concentrate on the radiative transfer models
and, based on the models, the systematic errors that may affect the filament
observations.

\subsection{OMC-3 filament parameters} \label{sect:disc_OMC3}

We analysed four filament segments, named A-D, in the OMC-3 cloud. Based on
MIR absorption, the median FWHM widths are $\sim$0.04\,pc, with little
dependence on the fitted cross-filament extent $r_{\rm max}$. The values were
on average consistent between the four segments, but there were also
significant variations along the filaments. These are discussed further in
Sect.~\ref{sect:reliability}. The analysis of FIR emission also gave median
FWHM$\sim$ 0.03-0.05\,pc, but, in the case of segment B, up to $\sim$0.1\,pc.
FIR emission was analysed using column-density
maps with angular resolutions
from 10$\arcsec$ to 41$\arcsec$. There were only small differences between the
different map versions, the HR version (20$\arcsec$ resolution) resulting in
the smallest values, with median values of 0.02-0.04\,pc in
Fig.~\ref{fig:plot_FIR_para}. Fits to the median profiles gave values that
were similar to the median parameter values along the filaments.

Previous \Herschel studies have typically found filament widths of
$\sim$0.1\,pc. \citet{Arzoumanian2011} reported a narrow distribution of
$0.10\pm0.03$\,pc in the cloud IC5146, at a distance of 460\,pc. These
(deconvolved) widths were based on Gaussian fits, and column-density
maps at
$\sim 37\arcsec$ resolution and 250\,$\mu$m surface brightness maps at $\sim
18\arcsec$ resolution gave similar results. \citet{GCC-VII} analysed 29
filaments in 13 separate fields at $d=100-500$\,pc distances, the analysis of
column-density maps of 41$\arcsec$ resolution resulting in widths of
$0.13\pm0.05$\,pc. In the above papers, the fits were done to the mean (or
median) radial profile of an entire filament that was detected with automated
methods, using DisPerSE \citep{Sousbie2011} in the case of
\citet{Arzoumanian2011} and getfilaments \citep{Menshchikov2013} in the case
of \citet{GCC-VII}. \citet{Arzoumanian2011} extended \Herschel studies to 599
filaments in eight regions at 140-460\,pc distances, the distribution again
peaking at $\sim 0.1$\,pc with an interquartile range of 0.07\,pc. However, 
based on \Herschel data in \citet{Arzoumanian2019}, \citet{Panopoulou2022}
concluded that the filament width estimates also depend on the source distance
and appear to be 4-5 times the beam size. Thus, the widths would increase from
less than 0.1\,pc in the closest fields (e.g. Taurus and Ophiuchus) to almost
0.3\,pc at $\sim$800\,pc (the IC5146 cloud). This scaling appeared to hold for
regions of very different types, from the low-density Polaris field to the
Orion-B cloud with active star formation. Some distance dependence was also
noted in \citet{GCC-VII}. 

Although deconvolution by a larger telescope beam increases uncertainties,
with perfect observations of perfect Plummer profiles, the FWHM estimates
should not depend on the resolution of observations. Indeed, in our
analysis, the factor of 20 range in the angular scales is not associated with
corresponding systematic changes in the estimated filament widths. The average
FWHM were at or below 0.05\,pc for all maps, although \textit{Herschel} data could in
some cases lead to higher values FWHM$\sim$0.1\,pc (segments A and B,
Fig.~\ref{fig:plot_FIR_para}).
The values are lower than those reported in \citet{Panopoulou2022} for fields
at similar distances (including Orion B with FWHM$\ga$0.15\,pc), and do
not match distance/resolution dependence reported there. In all the above
studies, the column densities were based on the modelling of dust emission
with a single-component MBB. \citet{Howard2019} and \citet{Howard2021} used
the PPMAP method \citep{Marsh2015} to analyse \Herschel and SCUBA-2
observations of filaments in the Taurus and Ophiuchus clouds. They noted that
the use of the PPMAP method, which takes temperature variations into account,
resulted in a reduction in the estimated filament widths. Our results
were also roughly constant with respect to the size of the fitted
cross-filament area. Therefore, even if the extent of the fitted area
typically increases with the target distance, that should not necessarily lead
to larger FWHM estimates.

Filaments in the OMC-3 region were already investigated by
\citet{Schuller2021}, who used \Herschel 160-500\,$\mu$m and \artemis
350\,$\mu$m and 450\,$\mu$m data (different observations from the \artemis
data analysed in \citet{Mannfors2022}). The \artemis surface-brightness data
and \Herschel temperature information at 18.2$\arcsec$ resolution were
combined to a column-density map with 8$\arcsec$ resolution. This resulted in
filament FWHM estimates of 0.06$\pm$0.02\,pc. These values are similar to
our AR results, where (ignoring the outlying values of segment A) the median
values range from 0.04\,pc to $\sim$0.1\,pc (Fig.~\ref{fig:plot_FIR_para}g).
While \citet{Schuller2021} studied long, automatically detected filaments, our
analysis is limited to short segments at the highest column densities.

In addition to FWHM, the values of the asymptotic power-law index, $p$, of the
fitted Plummer function are of interest. The \Herschel and combined \Herschel
and \artemis data gave typically values $p=2-5$, with some variations
depending on the extent of the fitted area. These suggest that nearby cloud
structures are affecting the fits in the tails of the profile function (cf.
Sect.~\ref{sect:reliability}). Compared to FWHM, the individual $p$ and $R$
values could be measured less precisely. Especially the MIR $p$ values showed
a large scatter, which could be caused in part by changes in the local MIR
radiation field. The MIR $p$ estimates are similar for different values
of $r_{\rm max}$, which suggests that individual point sources within that
area do not have a strong effect on the results.

\subsection{Systematic errors in filament parameters} \label{sect:disc_bias}

We used radiative transfer simulations to study error sources in the analysis
of MIR and FIR observations. In the following we discuss their relative
importance and the potential effects on the OMC-3 results.

\subsubsection{MIR observations}

Dust scattering and thermal dust emission are potential error sources in the
analysis of MIR extinction. They depend on the local radiation field and are
significant at high column densities. We assumed in the tests a background
level of 10\,MJy\,sr$^{-1}$. However, if the background level is higher, the
effects of local emission and scattering will be reduced.

For dust scattering, errors reached tens of per cent only if the column
density is $N({\rm H}_2)\sim 10^{24}$\,cm$^{-2}$ or higher. The intensity of
the isotropic background also had to be a factor of $\chi=100$ above the
normal ISRF or the luminosity of a point source at $\sim$1\,pc distance had to
be of the order of $\sim 10^4$L$_{\sun}$. Thus, scattering should not be
a significant source of errors in the OMC-3 field. Scattering would tend to
decrease the $\tau$ estimates, increase the $p$ estimates, and lead to
overestimations of the filament FWHM (Fig.~\ref{fig:SED_sca_trace_PS1}).
These effects are more pronounced close to point sources, especially if the
source is on the line of sight towards the filament. The presence of such point sources
would be evident in the observed maps, although their effect can extend over
distances of several parsecs.

The MIR scattering depends strongly on the grain properties. The so-called
coreshine, which is observed at 3-4\,$\mu$m wavelengths towards many dense
cores, has indicated surprisingly strong MIR scattering. This requires strong
dust evolution relative to diffuse clouds \citep{Steinacker2010, Pagani2010,
GCC-III, Steinacker2014_L1506, Lefevre2014}. \citet{Lefevre2016} investigated
the 8\,$\mu$m scattering towards the pre-stellar, high-column-density core of
L~183, where the estimated intensity of the scattered light was hundreds of
kJy\,sr$^{-1}$. This corresponds to the scattering in a more or less normal
ISRF, and the high scattering efficiency could be explained by large aggregate
grains. The L~183 observations (and the possibility of very large aggregates)
suggests that the scattered signal could be stronger than in our simulations.
If the scattered light is in L~183 at a level of $\sim 0.1$\,MJy\,sr$^{-1}$,
scattering could cause noticeable errors in MIR observations of high-mass
star-forming regions, where the radiation fields are much stronger.

In our models, the local thermal dust emission was a more significant factor
than the scattering. For the high column density of $N({\rm H}_2)\sim
10^{24}$\,cm$^{-2}$ filament, the emission effects were tens of per cent, both
in the case of the normal ISRF and in the case of the 590\,$L_{\sun}$ point
source. This led to the filament optical depths being underestimated and the
individual filament parameters ($p$, $R$, FWHM) to be overestimated. The
magnitude of the effects depends strongly on the presence of local radiation
sources (Fig.~\ref{fig:SED_sca_trace_PS1}). If there is a point source on the
line of sight towards the filament, the thermal emission can completely mask the MIR
absorption (Appendix Sect.~\ref{sect:app_bias_emission}). Quantitatively,
these effects are sensitive to the filament column density, the level of the
background surface brightness, and the location of the point source relative
to the filament (Fig.~\ref{fig:SED_SHG_maps}). In the comparison to the
models, one must also take into account that observations tend to
underestimate the true column densities.

Scattering is not only a source of errors but can itself be used to study
filaments at high angular resolution. At near-infrared wavelengths the
scattering is stronger (in absolute terms and relative to the in situ thermal
emission) but the larger optical depths complicate the analysis
\citep{Juvela2012_CrA, Malinen2013}. Scattering may also be measurable at MIR
wavelengths, above the MIR absorption \citep[e.g.][]{Steinacker2014_L1506,
Lefevre2014}. However, this requires a low background sky brightness, such as
found at high Galactic latitudes \citep{Steinacker2014}.

\subsubsection{FIR observations}

Far-infrared analysis of Sect.~\ref{sect:results_FIR} is affected especially by the
bias of column-density estimates, which is caused by temperature variations in
the source and the analysis using the single-temperature MBB model.
Appendix~\ref{sect:app_toy} shows that these effects can be easily demonstrated
even without complex modelling. 

In the radiative-transfer simulations of Sect.~\ref{sect:bias}, the
column-density estimates vary depending on the data resolution (beam sizes),
the analysis method, the temperatures, and dust optical properties. It can be
instructive to compare the peak optical depths obtained at different
resolutions and with different methods. Table~\ref{table:1e22} lists values
for the $N({\rm H}_2)=10^{22}\,{\rm cm}^{-2}$ model. Since the filament FWHM
is quite large, $\sim 72\arcsec$, the peak values of the 40$\arcsec$ and
20$\arcsec$ maps differ only little. 
Shorter wavelengths are more sensitive warm dust, tend to bias the
temperatures more upwards, and result in lower $\tau$ values.  In
Table~\ref{table:1e22} the effect is the opposite for $\beta=1.8$, because
the simulation used a dust model with $\beta\sim 1.95$. A modest increase in
the assumed $\beta$ (to a value above the $\beta$ in the simulations) can even
negate the natural tendency to underestimate the column densities. When real
observations are analysed, the precise value of $\beta$ is unknown. However,
because $\beta$ affects all column-density estimates similarly, its effect on
the observed filament profiles is limited - as long as $\beta$ does not
change significantly as a function of the filament radius.

\begin{table}
\caption{Comparison of peak 250 $\mu$m optical depths estimated with different analysis
methods. The estimates are based on synthetic surface-brightness maps of the $N({\rm H}_2)=10^{22}\,{\rm
cm}^{-2}$ filament model.}
\begin{tabular}{lcccc}
\hline
Case  &  Beam           &  $\beta$ & $\tau^{\rm max}(250\,\mu{\rm m})$ &  Rel. error  \\
&  [$\arcsec$]    &          &  $[10^{-3}]$            &  [\%]        \\
\hline   
true &   6  &  --   &  1.49  &   0.00  \\
MBB  &   6  &  1.8  &  1.11  &  -25.77 \\
MBB  &  24  &  1.8  &  1.04  &  -29.99 \\
HR  &  18  &  1.8  &  1.04  &  -30.34 \\
HR  &  18  &  2.0  &  1.33  &  -10.48 \\
HR  &  18  &  2.2  &  1.71  &  14.61 \\
HR (w) &  18  &  1.8  &  1.00  &  -32.54 \\
HR (w) &  18  &  2.0  &  1.32  &  -11.46 \\
HR (w) &  18  &  2.2  &  1.73  &  16.05 \\
LR  &  40  &  1.8  &  0.98  &  -33.98 \\
LR  &  40  &  2.0  &  1.22  &  -17.85 \\
LR  &  40  &  2.2  &  1.51  &   1.64 \\
LR (w) &  40  &  1.8  &  0.96  &  -35.27 \\
LR (w) &  40  &  2.0  &  1.22  &  -18.42 \\
LR (w) &  40  &  2.2  &  1.52  &   2.30  \\
\hline
\end{tabular}
\tablefoot{Relative errors are with respect to the true values that are read from the
3D model and reported on the first row. MBB refers to modified blackbody fits where
all data at the same resolution, `HR' to the higher-resolution versions
\citep{Palmeirim2013}, and `(w)' to alternative fits where the 160\,$\mu$m data are
given a lower weight. The third column refers to the constant value of $\beta$ used
in the SED fits.} 
\label{table:1e22}
\end{table}

The bias of the column-density estimates increase with column density and
exceeded a factor of two at $N({\rm H}_2)=10^{22}\,{\rm cm}^{-2}$
(Fig.~\ref{fig:SED_eq_bg_fwhm4}). Since these errors are correlated with the
column density, they also affect the observed filament profiles. The
parameters $p$, $R$, and FWHM are all biased upwards. Unlike in MIR
observations, a stronger isotropic radiation field decreases the errors by
reducing the amount of very cold dust. In the normal ISRF, the systematic
errors of filament FWHM reach 50\% when the column density exceeds $N({\rm
H}_2)=10^{23}\,{\rm cm}^{-2}$. In a $\chi=10$ field, the errors are less than
half of this, and are they are further halved in a $\chi=100$ field. This
means that the errors can be of similar magnitude in a low-mass star-forming
region (low column density and low radiation field) as in a high-mass
star-forming region (high column density and high radiation field). The errors
in $R$ and $p$ also depend on both the column density and the radiation field,
and can reach 50\% for a $N({\rm H}_2)=10^{23}\,{\rm cm}^{-2}$ in the normal
ISRF (Fig.~\ref{fig:SED_eq_trace_3}). A point source has a similar effect, the
errors increasing close to the source and with only a small dependence on the
source location (on the line of sight vs to one side of the filament).

The maximum column density of the OMC-3 filaments is in MBB analysis $N({\rm
H}_2) \sim 10^{23}\,{\rm cm}^{-2}$, but the true column density could be even
a few times higher. Because the quiescent parts of the filament are
likely to be subjected to a radiation field of at most $\chi\sim 100$
\citep{Mannfors2022}, and the effects of MIR scattering are likely to be 
insignificant.
The models predict a more significant role
for the MIR dust emission. Figure~\ref{fig:SED_SHG_trace_PS3_HI1} indicated a
$\sim$25\% effect for an isotropic radiation field with $\chi=10$. This
requires the extended thermal emission to also be taken into account in the
$I^{\rm bg}$ estimates. In observations this happens automatically (to some
accuracy), and the local thermal emission need not be separated from other
contributions to $I^{\rm bg}$. The fact that the OMC-3 FIR and MIR
observations resulted in similar FWHM estimates also suggests that the
systematic errors of the MIR analysis are unlikely to amount to tens of per
cent.

Filaments are located inside dense clouds, which significantly reduces the UV
flux that reaches the filament. Most of the 8\,$\mu$m emission would then
originate in extended regions, and, unlike in our simple models, the emission
would be mostly uncorrelated with the filament structure. This will reduce the
systematic errors in the filament parameters. A second important factor is the
abundance of polycyclic aromatic hydrocarbons and other very small
grains \citep{Draine2003ARAA}. If these have already partly disappeared within
the filament (e.g. by sticking onto larger grains), the MIR emission would be
suppressed. Based on our models, FIR analysis could overestimate the width of
$N({\rm H}_2) \sim 10^{23}\,{\rm cm}^{-2}$ filaments by tens of per cent in
the normal ISRF. However, a stronger radiation field reduces the errors to
$\sim$10\% level, which is within the uncertainties of the MIR versus  FIR
comparison (Fig.~\ref{fig:plot_FIR_para}). Therefore, although the models show
the possibility of significant systematic errors in all observations, there is
no contradiction in the approximate agreement between the OMC-3 MIR and FIR
estimates.

\subsection{Reliability of profile fits}  \label{sect:reliability}

Based on synthetic observations of magnetohydrodynamic cloud simulations,
\citet{Juvela2012_MHDFIL} concluded that, at the nominal \textit{Herschel} resolution,
the filament parameters could be recovered reliably only up to $\sim$400\,pc
(assuming $\sim$0.1\,pc filaments). OMC-3 is at this limit and the filaments
are partly narrower than 0.1\,pc. Nevertheless, the angular resolution of at
least the MIR and AR maps should be sufficient for the profile analysis.

In addition to systematic errors, the results show significant random
fluctuations, especially in the $p$ and $R$ estimates. Some fits resulted in
values $p\ga5$ that are inconsistent with most physical filament models. 
Individual $p$ and $R$ values are more difficult to measure because a larger
value of $p$ can in Eq.~(\ref{eq:Plummer}) be compensated with a larger value
of $R$, the fit still recovering the same FWHM and generally fitting the
observed profile. The parameter $R$ probes the structure in the inner part of
the filament and is dependent on the data being able to resolve those smaller
scales. Conversely, $p$ probes the asymptotic behaviour at large distances and
is sensitive to nearby cloud structures and the general background
fluctuations. 

The error distributions of the parameters (including FWHM) are often
asymmetric for the individual profiles, with the MCMC estimates showing a long
tail to high values (cf. Fig.~\ref{fig:MCMC_test}). The overall dispersion
along the segments (as shown by the histograms e.g. in Fig.~\ref{fig:MIR0}b)
give an empirical estimate for the total uncertainty.
Our data also contain missing values, which affect the reliability of MIR
profiles (e.g. Fig.~\ref{fig:MIR0}) and the analysis of the AR map of the
filament segment A (Fig.~\ref{fig:G208_NH2_feathered_MCMC_A0}). The missing
data also bias the median profiles. Rather that rejecting all profiles that
contained missing values (which could be almost all of the data), the median
values were calculated over the remaining pixels. Therefore, at a given
distance from the filament centre, the median value is based on different
sections along the length of the filament, leading to random errors in the
median profile.

\begin{figure}
\sidecaption
\centering
\includegraphics[width=9cm]{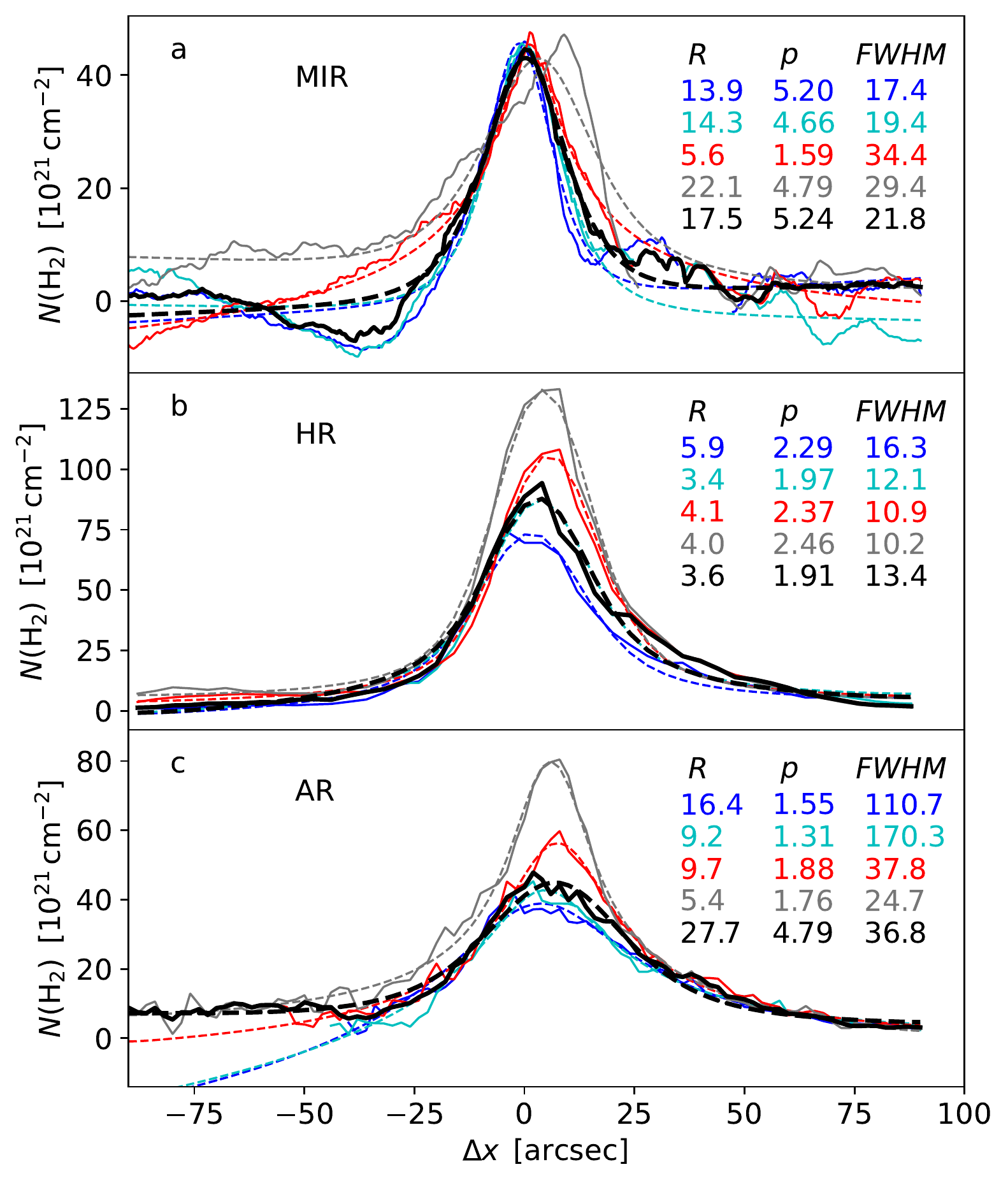}
\caption{
Plummer fits for selected cross-sections of the OMC-3 filament segment A. The
three frames correspond to the MIR, HR, and AR column-density
maps,
respectively. Each frame shows individual profiles for the offsets $\Delta
y$=10, 20, 30, and 40 arcsec (cf.
Fig.~\ref{fig:MIR0}-\ref{fig:G208_NH2_feathered_MCMC_A0}; blue, cyan, red, and
grey lines, respectively) and the median profile (thick black lines). The
best-fit Plummer profiles are plotted with dashed lines of the same colour,
and $R$, $p$, and FWHM values of the fits are listed in the frames.
}
\label{fig:real1}
\end{figure}

Figure~\ref{fig:real1} shows examples of the fits to the OMC-3 filament
segment A, based on the MIR, HR, and AR column-density estimates. Most MIR
profiles do not have data around offset $\Delta x=30\arcsec$ but in this case
this does not have a major effect on the fits. MIR profiles also tend to show
a dip around $\Delta x=-30\arcsec$, which could be caused by imperfections in
the column-density estimation (e.g. dust to nearby sources), or random column-density fluctuations. These fits are associated with high values of $p\sim 5$.
In comparison, the profile at offset $\Delta y=30\arcsec$ along the filament
(red curve) decreases rather than increases towards negative $\Delta x$, and
has $p\sim 1.6$. For the maps HR and AR based on dust FIR emission, the fitted
profiles match better the observed profiles. However, the effect of the
missing data is clear in the AR results, where the observations at $\Delta
y$=10 and 20 arcsec (i.e. the profiles most affected by missing values, cf.
Fig.~\ref{fig:G208_NH2_feathered_MCMC_A0}c) provide only weak constraints at
negative $\Delta x$ values. This leads to degeneracy between the Plummer and
the background parameters, the background is underestimated at negative
$\Delta x$ offsets, and the fit results in abnormally large FWHM values.

\begin{figure}
\sidecaption
\centering
\includegraphics[width=9cm]{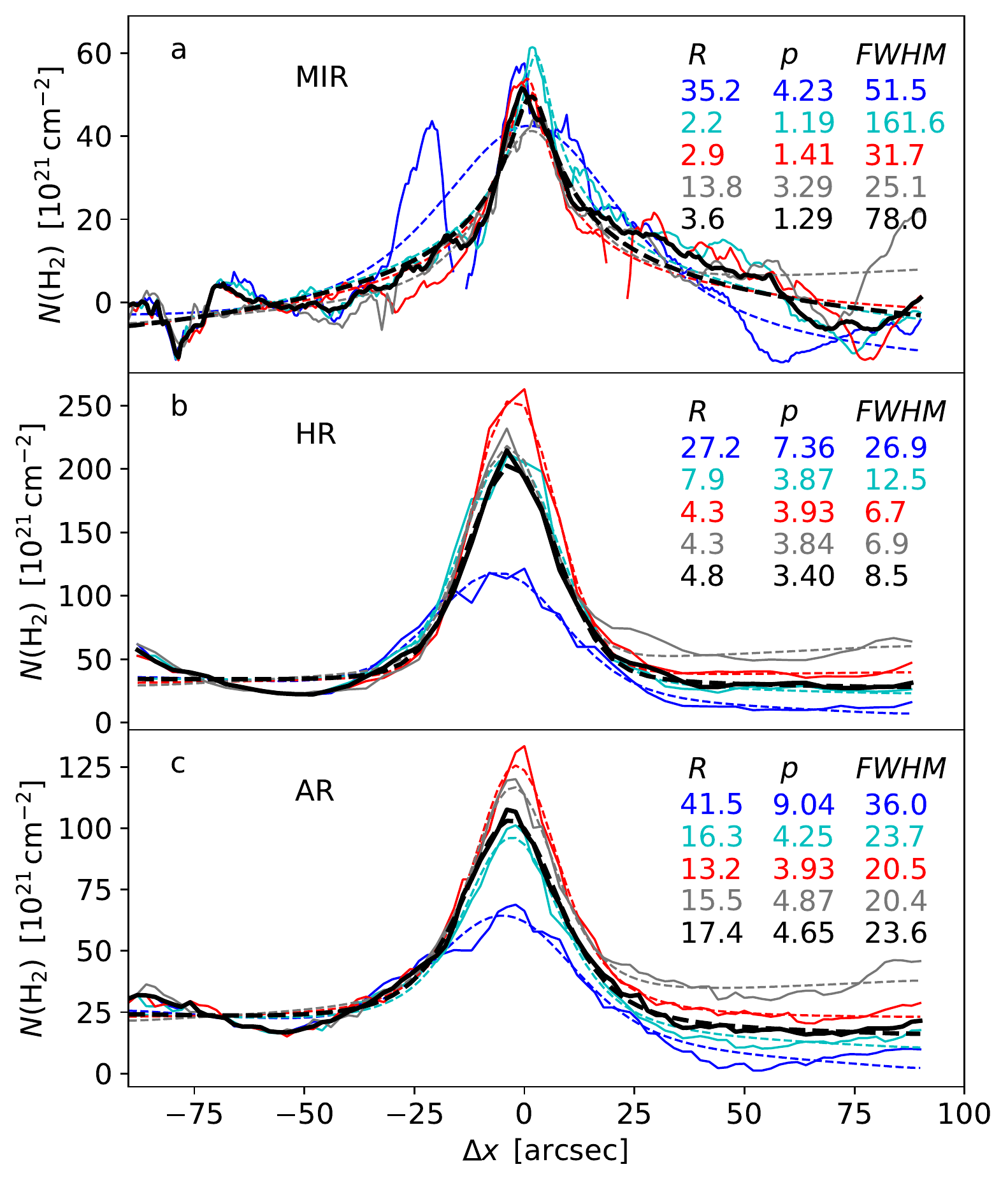}
\caption{
Selected cross-sections of the OMC3-filament segment D.  These correspond to
HR and AR data (blue and red lines, respectively) and the offsets $\Delta
y$=10 and 25 arcsec, where the $\Delta y$=10 arcsec profiles have the lower
column density. The dashed lines are the best-fit Plummer profiles with the
parameters listed in the figure.
}
\label{fig:real2}
\end{figure}

Large $p$ values are also observed for some other dust emission data,
typically in fainter and less clear parts of the filaments. They can be
related to interference from nearby cloud structures, which can be other
filaments or clumps (e.g. northern part of filament B,
Fig.~\ref{fig:G208_NH2_Palmeirim_NH2_MCMC_A1}) or diffuse emission that appear
as an extension of the filament itself (e.g. segment C, $\Delta x \sim
60\arcsec$, Fig.~\ref{fig:G208_NH2_Palmeirim_NH2_MCMC_A2}). However, the
filament segment D is associated with large values $p \ge 4$ over its full
length. Figure~\ref{fig:real2} shows examples of the individual profiles. The
main feature (in HR and AR maps) is a dip at negative $\Delta x$ values. This
is similar to the one seen in Fig.~\ref{fig:real1} and similarly appears to be
the origin of the large $p$ values. The situation is worst for the $\Delta
x=10 \arcsec$ profile, where the background also curves up at positive $\Delta
x$ offsets, raising the $p$ value above nine. This suggest that the background
might need to be modelled using a second order polynomial, although $p$ would
be partly degenerate with an second order background term.

\begin{figure}
\sidecaption
\centering
\includegraphics[width=9cm]{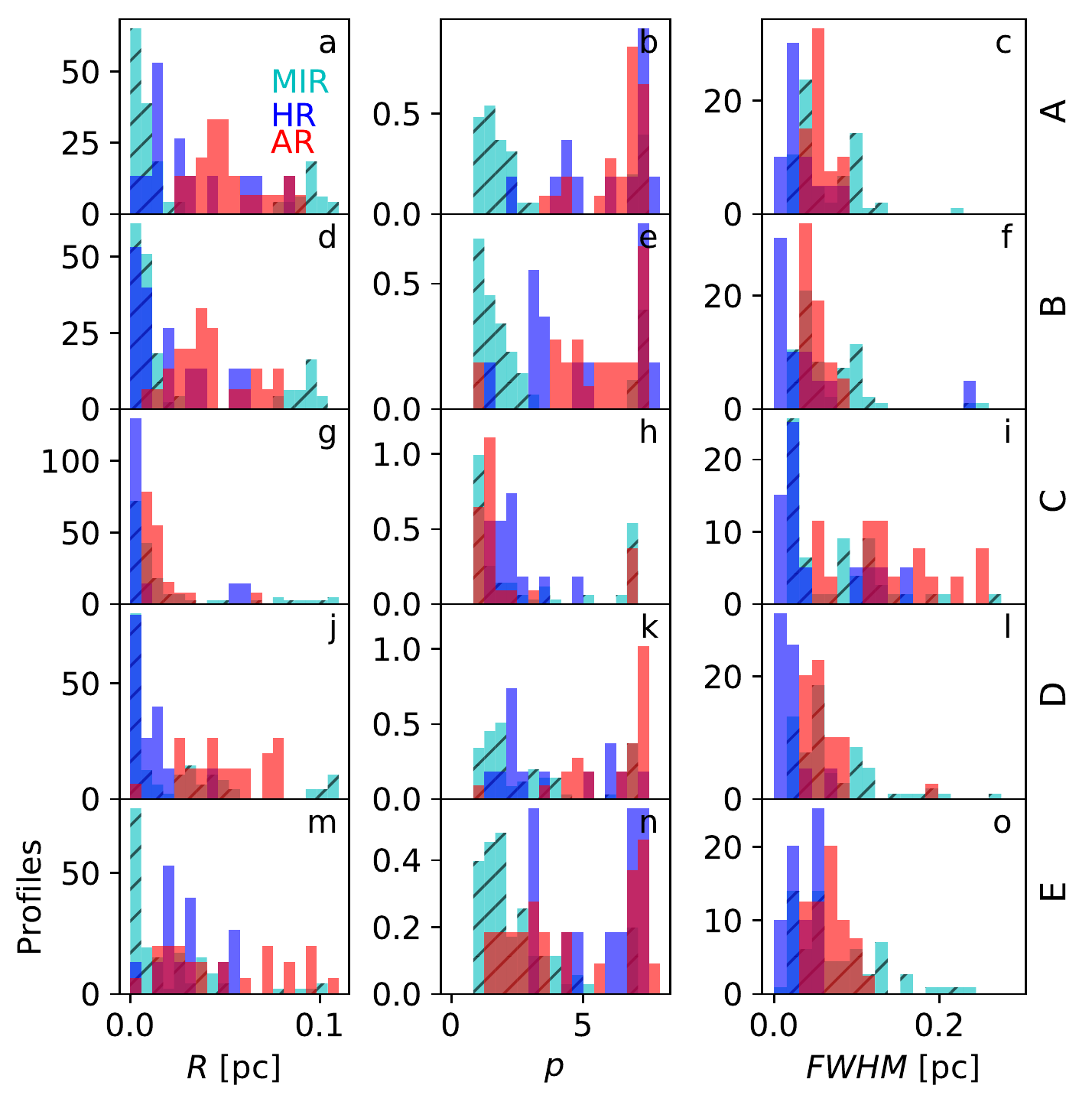}
\caption{
Alternative fits of the filament segment D. The histograms show the
distributions of the $R$, $p$, and FWHM parameters that are estimated from
the MIR, HR, and AR maps (cyan, blue, and red histograms, respectively). The
rows are for alternative models of the profile function, where row B
corresponds to our default model. In the other fits, row A omits the shift of
the filament centre, row C adds a second order term to the background, row D
fits two-sided Plummer functions with linear background, and row E fits 
two-sided Plummer functions with a second order polynomial for the background.
}
\label{fig:APP_plot_2_hist}
\end{figure}

Figure~\ref{fig:APP_plot_2_hist} compares the results that are based on the
MIR, HR, and AR data on the filament segment D and six versions of the fitted
profile function. The row B corresponds to our default model in
Eq.~(\ref{eq:fit}), which has six parameters: three for the Plummer function
itself, one allowing a shift along the $\Delta x$ axis, and two parameters
describing the linear background. In Fig.~\ref{fig:APP_plot_2_hist}, one of
the alternative fits omits the shift (row A) and one adds the second order
term to the background component (row C). So far all fits assume that the
filament profiles are symmetric, which is only approximately true for the
selected segments. In Fig.~\ref{fig:APP_plot_2_hist}, we also show results for
asymmetric Plummer functions (separate $R$ and $p$ parameters on each side of
the peak) that are combined with a background modelled as a linear first order
(row D) or a second order (row E) polynomial.

If the filament shifts in the $\Delta x$ direction at small scales, the
omission of the $\Delta r$ term in Eq.~\ref{eq:Plummer} should lead to larger
FWHM values. No such effect is seen in Fig.~\ref{fig:APP_plot_2_hist},
although this could still play a role in fits of longer and more fragmented
filaments. The addition of the second order background term (figure row C)
reduces the $p$ values significantly from $p\sim 5$ to $p\sim 2.5$. The $R$
values are also smaller, but the mean value and the scatter of the FWHM
values has increased. When the second order background term is included in the
asymmetric Plummer fits, the results are much less affected and especially the
FWHM values remain practically unchanged (row E vs row B). With the
exception of the row C, the FWHM values are thus similar for all the
alternative fits.

In Fig.~\ref{fig:APP_plot_2_hist}, the fits have between five and nine free
parameters. In these calculations, we also added penalties for values
$R<0.005$\,pc and $p>8$. The AR results were somewhat sensitive to the $R$
threshold, because the filament widths are not much larger than the beam size
and unresolved filaments can result in small $R$ values. The FWHM estimates
from the LR maps could thus be even more sensitive to any priors that are
used. Accurate beam models are also important, since they are used to
deconvolve the observations. The LR beam is well defined, while the effective
beams of the HR and AR maps depend on the way these maps are constructed. The
HR column density is based on a combination of intermediate maps that have
different angular resolutions and different sensitivity to temperature
variations. The AR map is the combination of data from two instruments, and
the effective beam could be affected by calibration differences or
imperfections of the feathering procedure.

\section{Conclusions}  \label{sect:conclusions}

We have studied four filament segments in the OMC-3 cloud in Orion, using
observations of MIR extinction and FIR dust emission and with the goal of
measuring the filament widths. The \Herschel FIR data were converted to
column-density maps at 20$\arcsec$ (LR maps) and 41$\arcsec$ (HR maps)
resolution, as well as an additional map of 10$\arcsec$ resolution from combined
\Herschel and \artemis 350\,$\mu m$ observations (AR map). In addition to
using observational results from the OMC-3 filaments, we performed radiative transfer
simulations to investigate sources of systematic errors that could affect the
measurement of filament profiles. The study led to the following conclusions:
\begin{enumerate}
\item The OMC-3 filament segments have FWHM values of 0.03-0.05\,pc. Similar
values are obtained with the three column-density
maps derived from FIR
observations (10-41$\arcsec$ angular resolution) and based on the MIR
extinction map ($\sim 2\arcsec$ resolution). In the LR and AR maps, the
estimated width was only higher in segment B and (partly) in segment A,
with $FWHM\sim 0.1$\,pc.
\item The MIR results showed little dependence on the extent of the fitted area,
which was varied between 60$\arcsec$ and 210$\arcsec$ maximum distances from the filament centre. Some
variation was observed due to map edges and, to a lesser extent, due to the
influence of unrelated background structures. The results are based on Plummer fits, where the model itself includes terms for a linear background.
\item When estimated from FIR data, the values of the asymptotic power-law 
index, $p$, in the Plummer function were $\sim 2-5$, with an average value of $p\sim 3$. The estimates
derived from MIR extinction had a large scatter, and the median values of the four segments ranged from
$p\sim 3$ to as high as $p \sim 8$.
\item The FWHM estimates are quite robust, even when the individual Plummer
parameters, $p$ and $R$, show a large scatter. This applies to both MIR and FIR analysis.
\item Synthetic observations of model filaments were analysed. The MBB 
fits to 160-500\,$\mu$m dust emission led to the expected underestimation of
column densities. The error exceeded $\sim 50$\% above $N({\rm H}_2) = 3
\times 10^{23}$\,cm$^{-2}$ but depended on the dust properties.
\item A similar bias exists in the filament parameters derived from FIR dust
emission. For a $N({\rm H}_2)=10^{23}$\,cm$^{-2}$ filament heated by the normal ISRF, the FWHM is
overestimated by more than 50\%. However, the error decreases by more than a factor of two if the
radiation field is ten times stronger and the dust temperatures correspondingly higher. The errors
also decrease rapidly at lower column densities.
\item The effect of point sources on the FIR analysis is qualitatively similar to
the isotropic field, systematic errors increasing closer to the source. There
is only a minor dependence on the source location: a source along the line of sight versus  a
source to one side of the filament.
\item The accuracy of the MIR extinction analysis is affected by the uncertainty of the
foreground emission and the potential effects of MIR dust scattering and
emission.
\item In the models, MIR scattering shows only minor effects.
The errors in filament parameters reach 10\% only at very high column
densities ($N({\rm H}_2) \sim 10^{24}$\,cm$^{-2}$) and in strong radiation
fields ($\chi > 10$). Errors are larger near luminous point sources, whose
presence should be clearly visible in the maps. Scattering is sensitive to
dust properties, and strong grain growth could increase its effects by a
factor of several.
\item 
Thermal MIR emission is, in our models, a more significant error source than MIR
scattering. In a $N({\rm H}_2) = 3 \times 10^{23}$\,cm$^{-2}$ model filament,
errors already reach the 10\% level in the normal ISRF or within a 1\,pc
distance of a 590\,L$_{\sun}$ point source. Errors are still of the same order
of magnitude even in a stronger isotropic radiation field of $\chi=10$.
However, if the diffuse UV field is attenuated by the surrounding cloud or the
abundance of very small grains is lower inside the filament, the significance
of MIR emission is correspondingly decreased.
\end{enumerate}
The estimated widths of the OMC-3 filaments were roughly equal between
different tracers and observations of different angular resolution. This is
encouraging, considering the many potential sources of systematic errors.
However, it also further highlights the differences between the dust filaments
and the narrower fibres that are observed in spectral lines. High-resolution
comparisons of the dust and gas tracers, within the same sources, are needed
to understand these differences and the exact role of filaments in the
star-formation process.

\bibliographystyle{aa}
\bibliography{my.bib}

\appendix


\section{Test of the fitting routines} \label{sect:app_MCMC_test}

\begin{figure}
\sidecaption
\centering
\includegraphics[width=9cm]{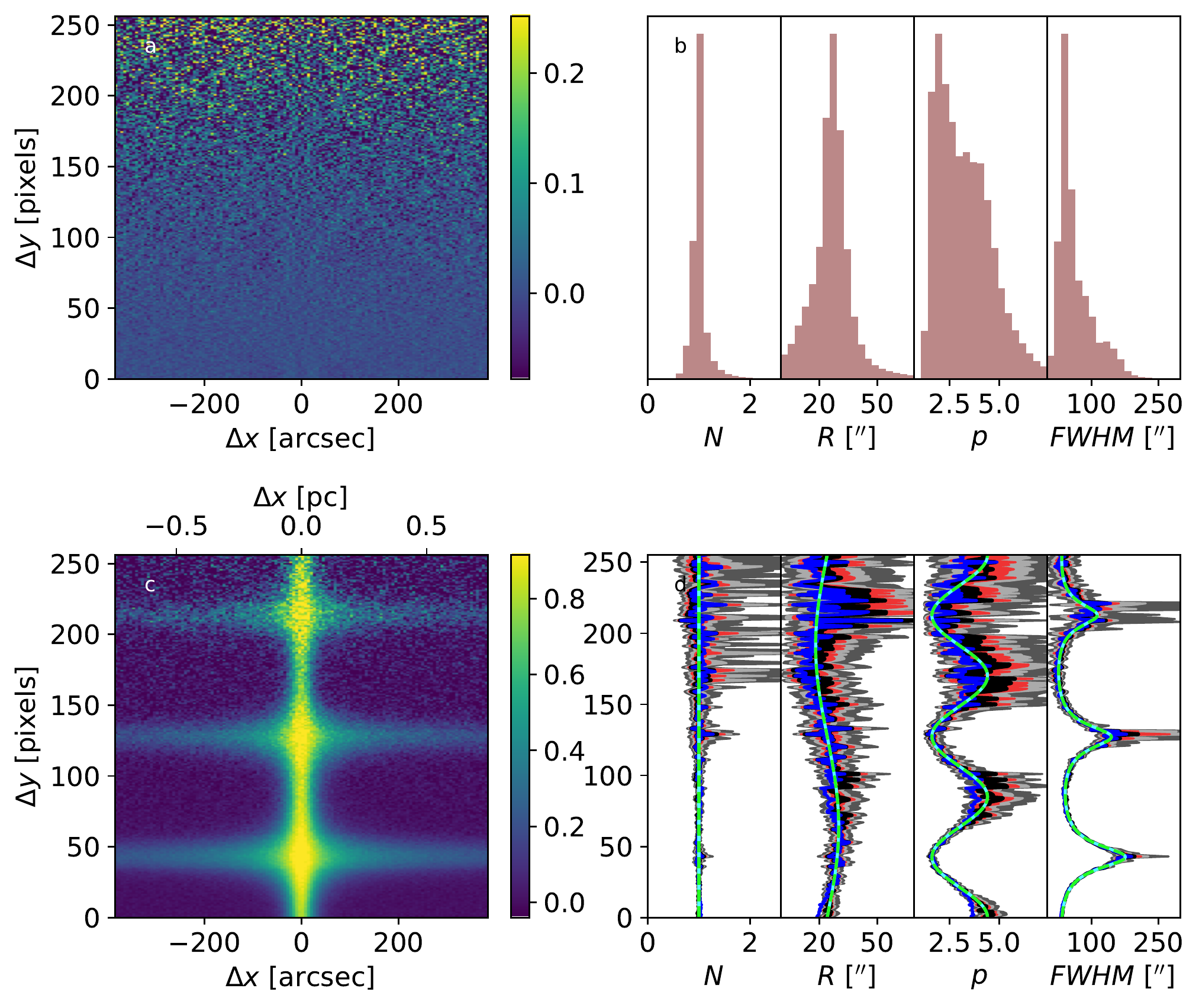}
\caption{
Fits of synthetic filament profiles. Frame (c) shows 256 generated profiles, where the
noise relative to the filament peak increases logarithmically from 1\% at the bottom (at
$\Delta y$=0) to 20\% at the top (at $\Delta y$=255). The residuals of the least-squares
fits are show in frame (a). Frame (d) shows the recovered parameter values, where the
light-green lines correspond to the true input values, and the least-squares estimates
are plotted in blue.  For the MCMC calculations, the black lines show the mean of the
MCMC samples and the red, light-grey, and dark-grey bands correspond to the [25, 75],
[10,90], and [1, 99] percentile intervals, respectively. Frame (b) shows histograms
corresponding to MCMC samples along the full length of the filament.
}
\label{fig:MCMC_test}
\end{figure}

The profile function of Eq.~(\ref{eq:fit}) was fitted with a least-squares
routine and with a MCMC routine. Figure~\ref{fig:MCMC_test} shows results for
the analysis of a synthetic dataset. The inputs consist of a set of ideal
profiles, which match the model setup, and white noise that is varied from 1\%
to 20\% of the filament peak value. For illustration, we fix the pixel size to
6$\arcsec$ and the data resolution (beam size) to 24$\arcsec$. The profile
parameters are varied in the ranges $R=18-30\arcsec$ and $p=1.6-4.4$. The
simulation is not concerned with the way the synthetic column-density
observations are obtained (emission or extinction), only on the noise effects
in the actual profile fitting.

The figure shows that the parameter uncertainty depends not only on the noise
level but also on the values of $R$ and $p$. For example, $p$ is correlated
with $R$, and its uncertainty is much higher when the true value of $p$ is
higher. At high noise levels, the MCMC mean value tends to be above the
least-squares solution, because of the skewed error distribution (and the used
$p \ge 1$ constraint). Compared to the individual parameters, the FWHM
estimate is generally more accurate, but its uncertainty still increases
significantly when the value of $p$ is low. 

The results depend on the size of the analysed region, and fits to a narrower
area around the filament will naturally lead to higher errors (especially for
low $p$). However, the errors in real observations can be dominated by
structured background emission rather than observational noise.

\section{Additional figures of fits to OMC-3 MIR data} \label{sect:app_OMC3_MIR}

\begin{figure}
\sidecaption
\centering
\includegraphics[width=9cm]{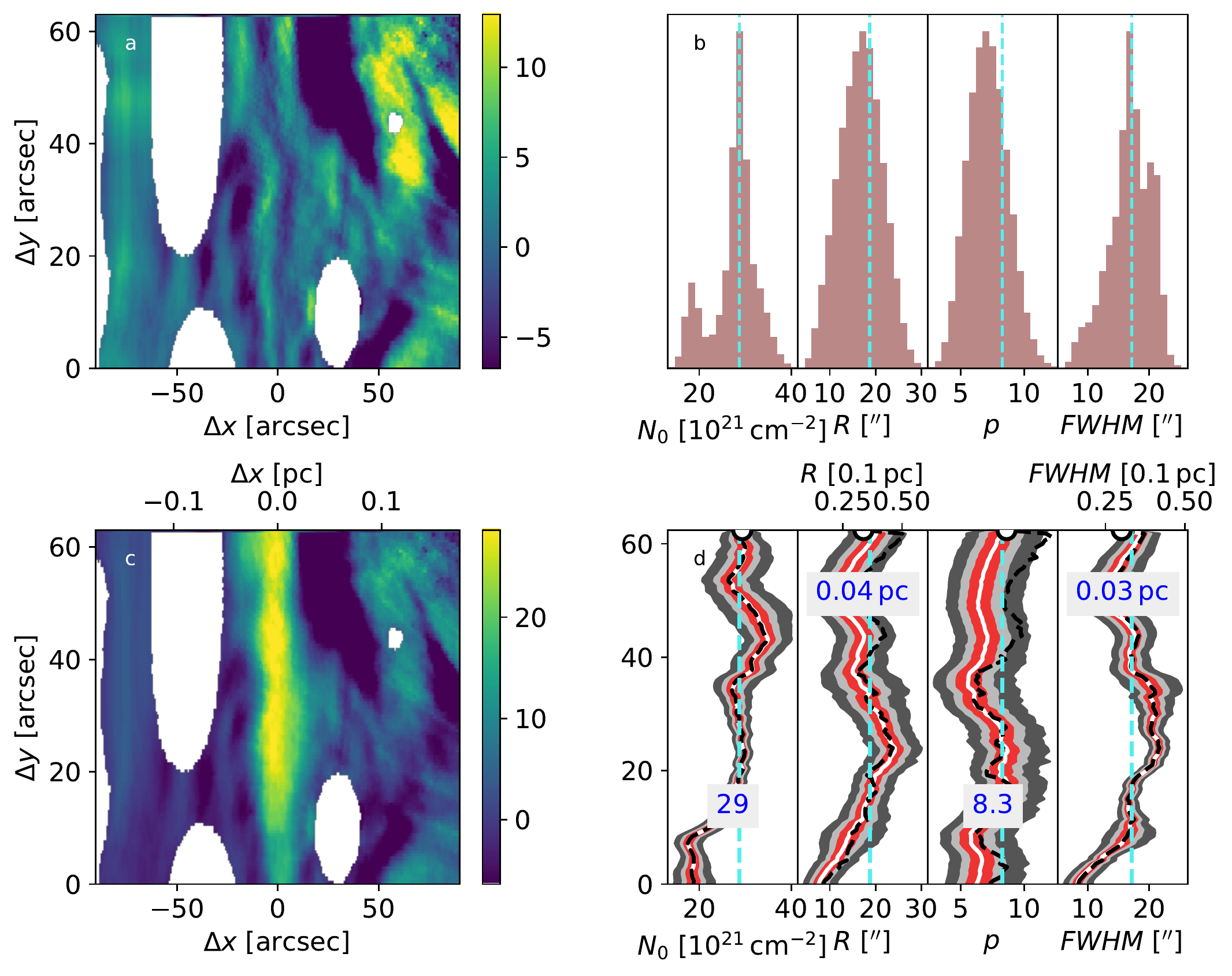}
\caption{
Plummer fits of MIR absorption. As
Fig.~\ref{fig:MIR0} but for OMC-3 filament segment B.
}
\label{fig:MIR1}
\end{figure}

\begin{figure}
\sidecaption
\centering
\includegraphics[width=9cm]{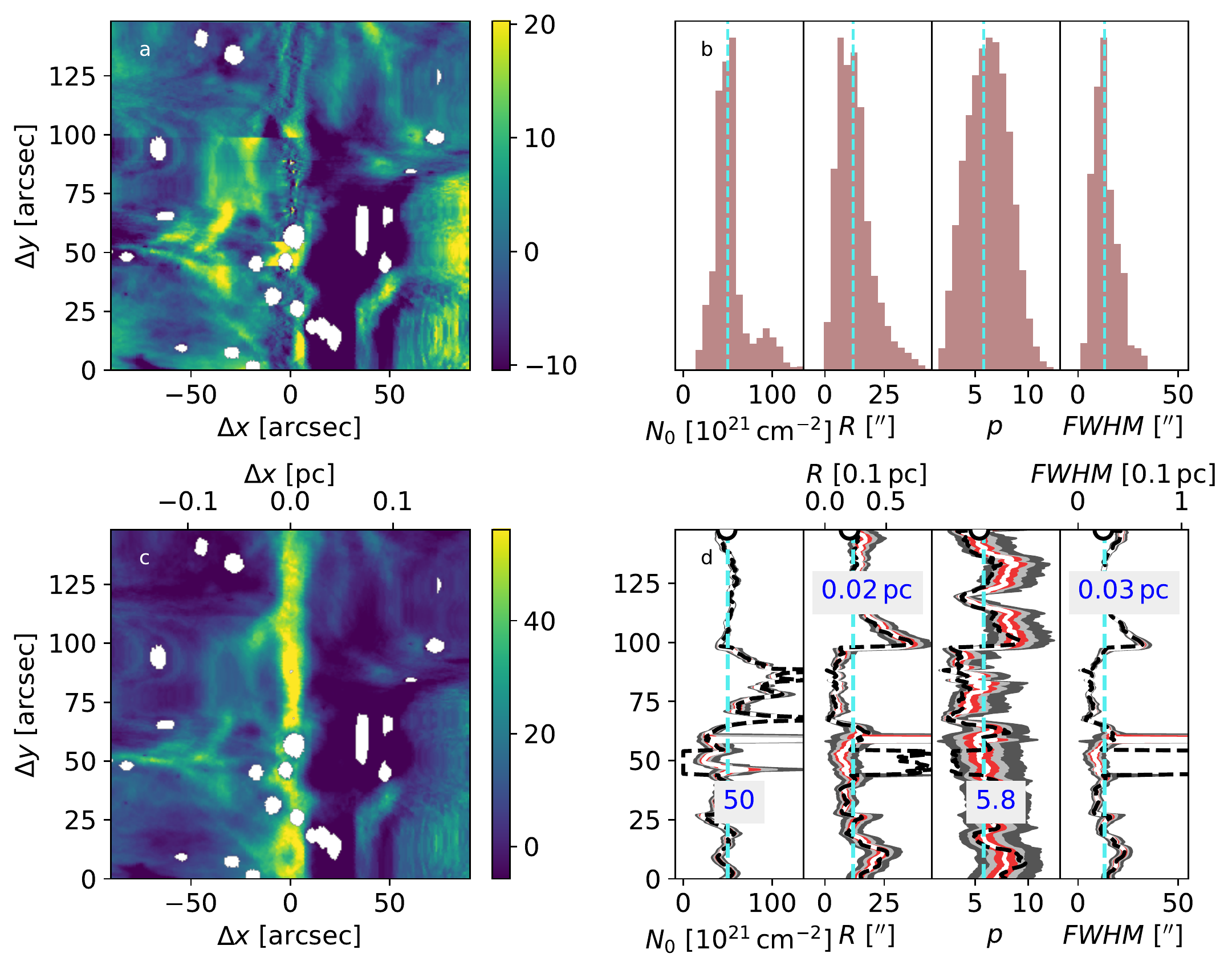}
\caption{
Plummer fits of MIR absorption. As Fig.~\ref{fig:MIR0} but for OMC-3 filament segment C.
}
\label{fig:MIR2}
\end{figure}

\begin{figure}
\sidecaption
\centering
\includegraphics[width=9cm]{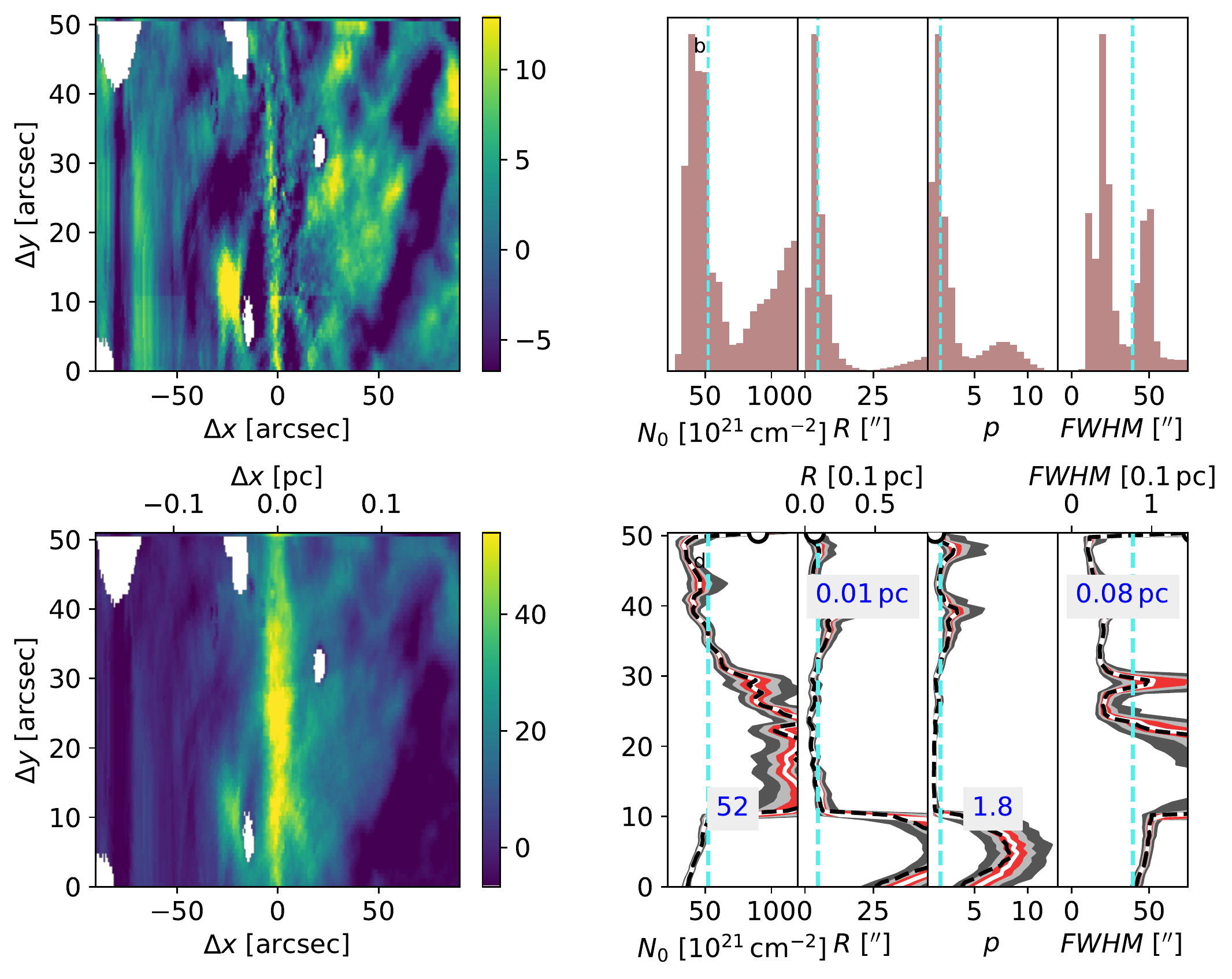}
\caption{
Plummer fits of MIR absorption. As Fig.~\ref{fig:MIR0} but for OMC-3 filament segment D.
}
\label{fig:MIR3}
\end{figure}

Section~\ref{sect:results_MIR} discussed the filament properties based on OMC-3 MIR
observations, and Fig.~\ref{fig:MIR0} showed the estimated parameters for the filament
segment A. Here Figs.~\ref{fig:MIR1}-\ref{fig:MIR3} show the corresponding results for
the other segments B-D.

\section{Additional figures of fits to OMC-3 FIR data} \label{sect:app_OMC3_FIR}

\begin{figure}
\sidecaption
\centering
\includegraphics[width=9cm]{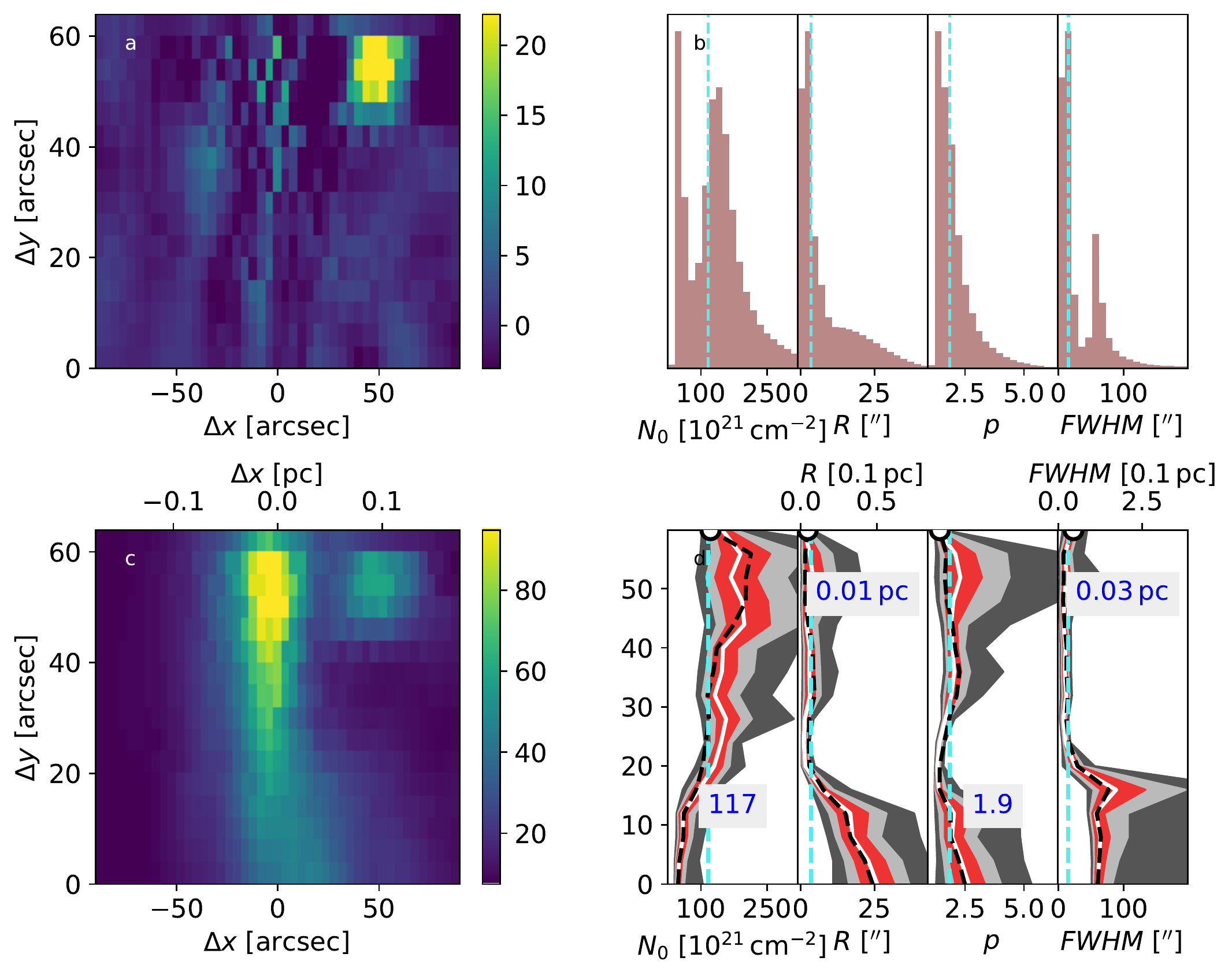}
\caption{
Plummer fits of \Herschel column-density data at 20$\arcsec$ resolution. As
Fig.~\ref{fig:G208_NH2_Palmeirim_NH2_MCMC_A0} but for OMC-3 filament segment B.
}
\label{fig:G208_NH2_Palmeirim_NH2_MCMC_A1}
\end{figure}

\begin{figure}
\sidecaption
\centering
\includegraphics[width=9cm]{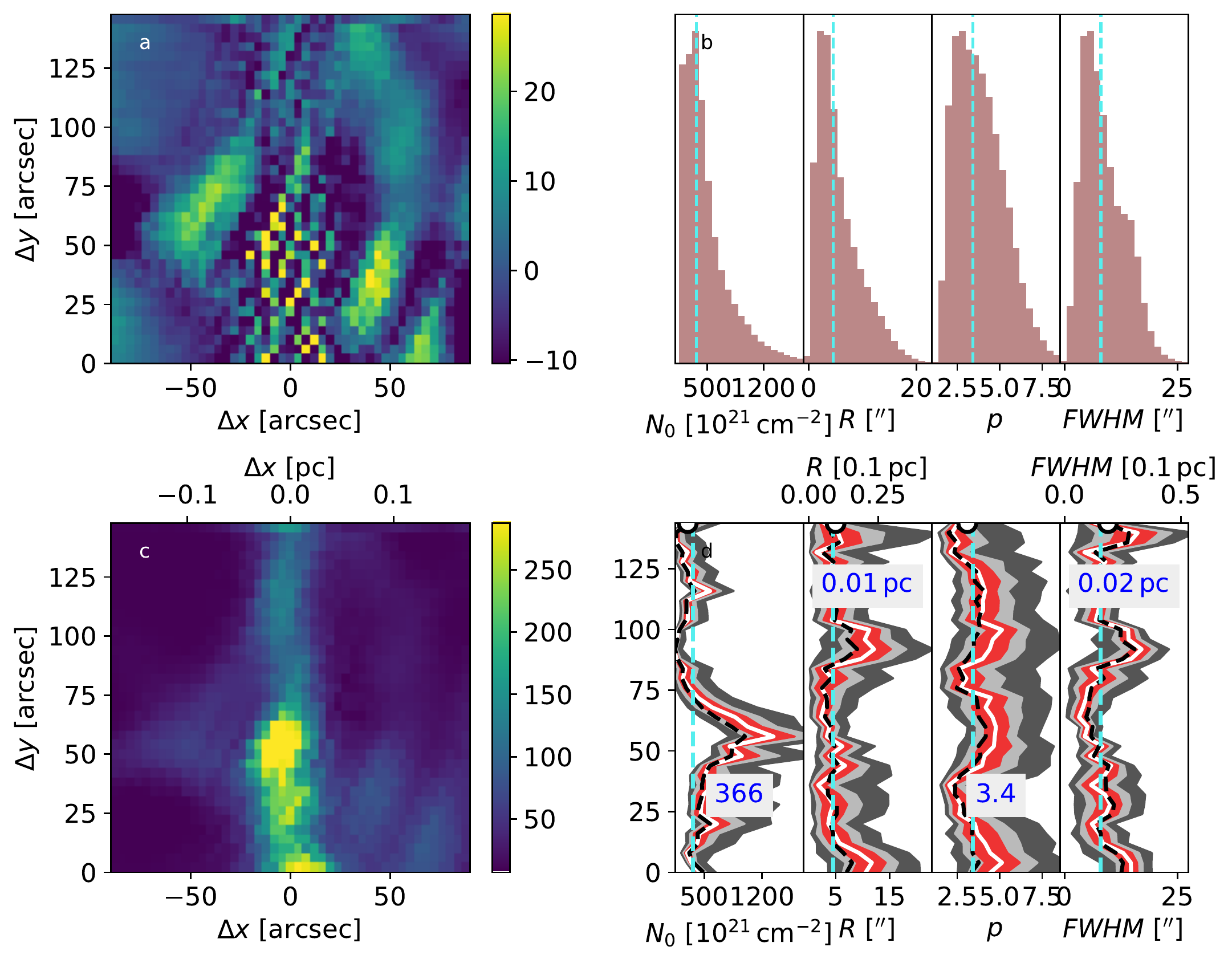}
\caption{
Plummer fits of \Herschel column-density data at 20$\arcsec$ resolution. As
Fig.~\ref{fig:G208_NH2_Palmeirim_NH2_MCMC_A0} but for OMC-3 filament segment C.
}
\label{fig:G208_NH2_Palmeirim_NH2_MCMC_A2}
\end{figure}

\begin{figure}
\sidecaption
\centering
\includegraphics[width=9cm]{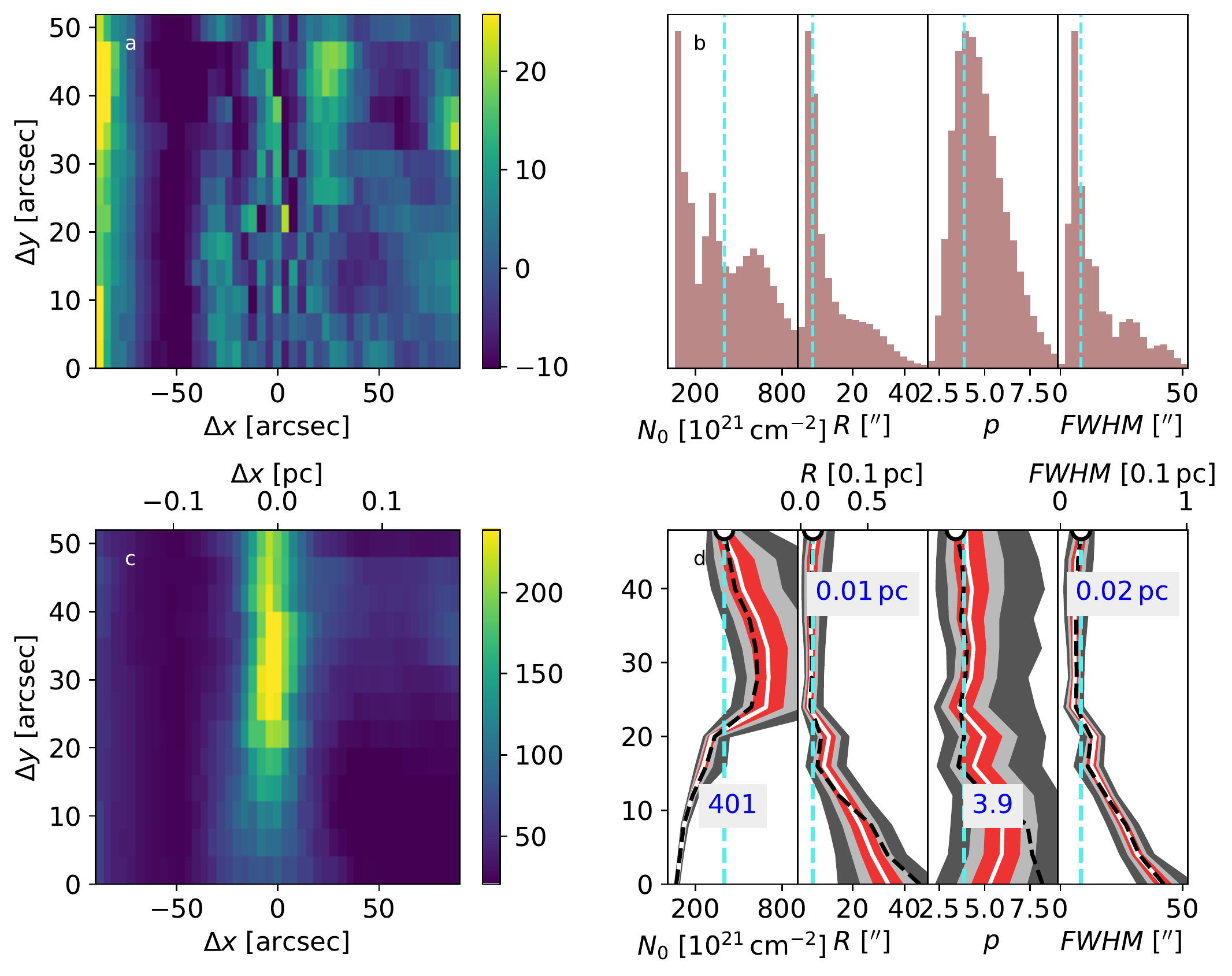}
\caption{
Plummer fits of \Herschel column-density data at 20$\arcsec$ resolution. As
Fig.~\ref{fig:G208_NH2_Palmeirim_NH2_MCMC_A0} but for OMC-3 filament segment D.
}
\label{fig:G208_NH2_Palmeirim_NH2_MCMC_A3}
\end{figure}

\begin{figure}
\sidecaption
\centering
\includegraphics[width=9cm]{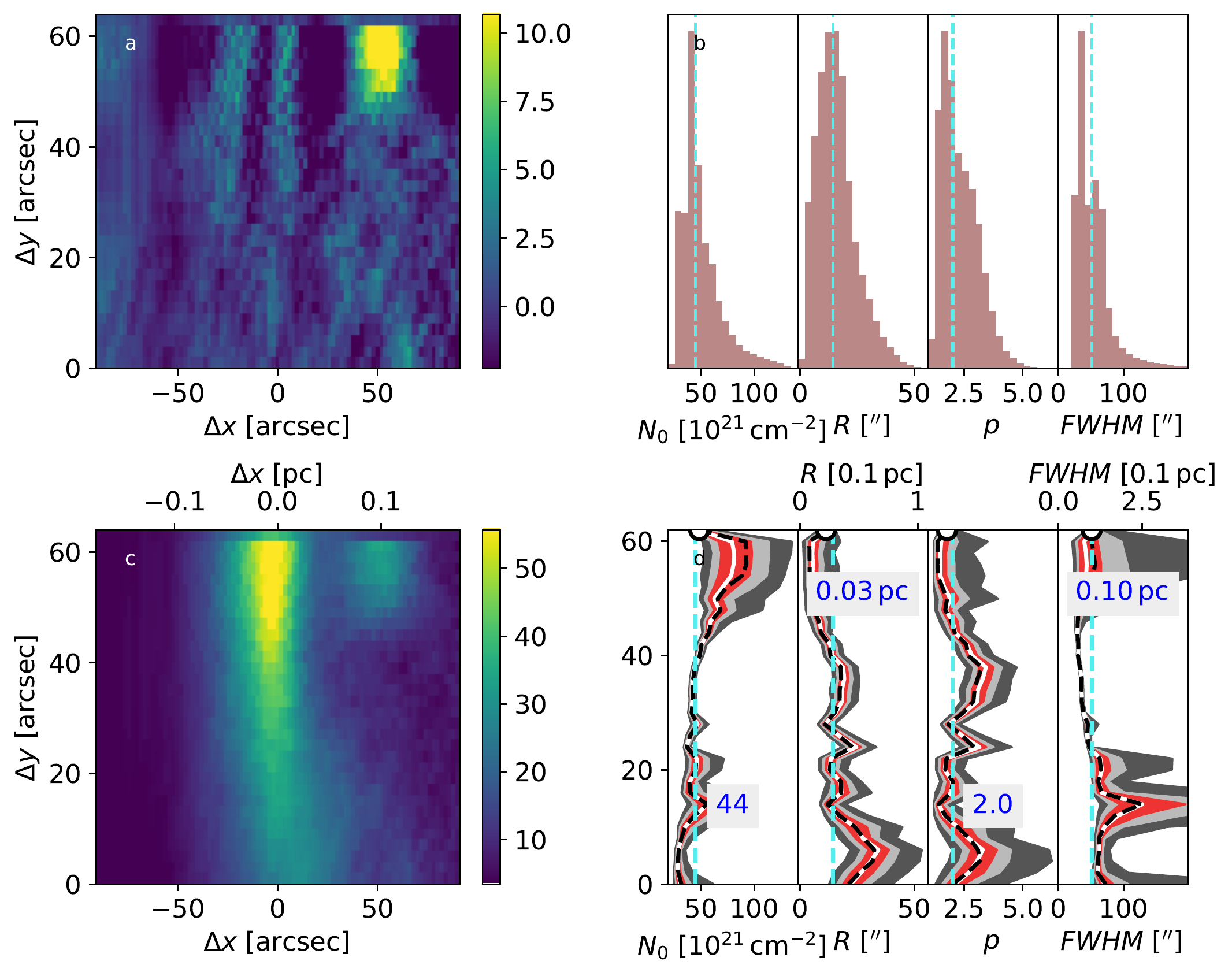}
\caption{
Plummer fits of combined \Herschel and \artemis data at 10$\arcsec$ resolution. As Fig.~\ref{fig:G208_NH2_feathered_MCMC_A0} but for OMC-3 filament segment B.
}
\label{fig:G208_NH2_feathered_MCMC_A1}
\end{figure}

\begin{figure}
\sidecaption
\centering
\includegraphics[width=9cm]{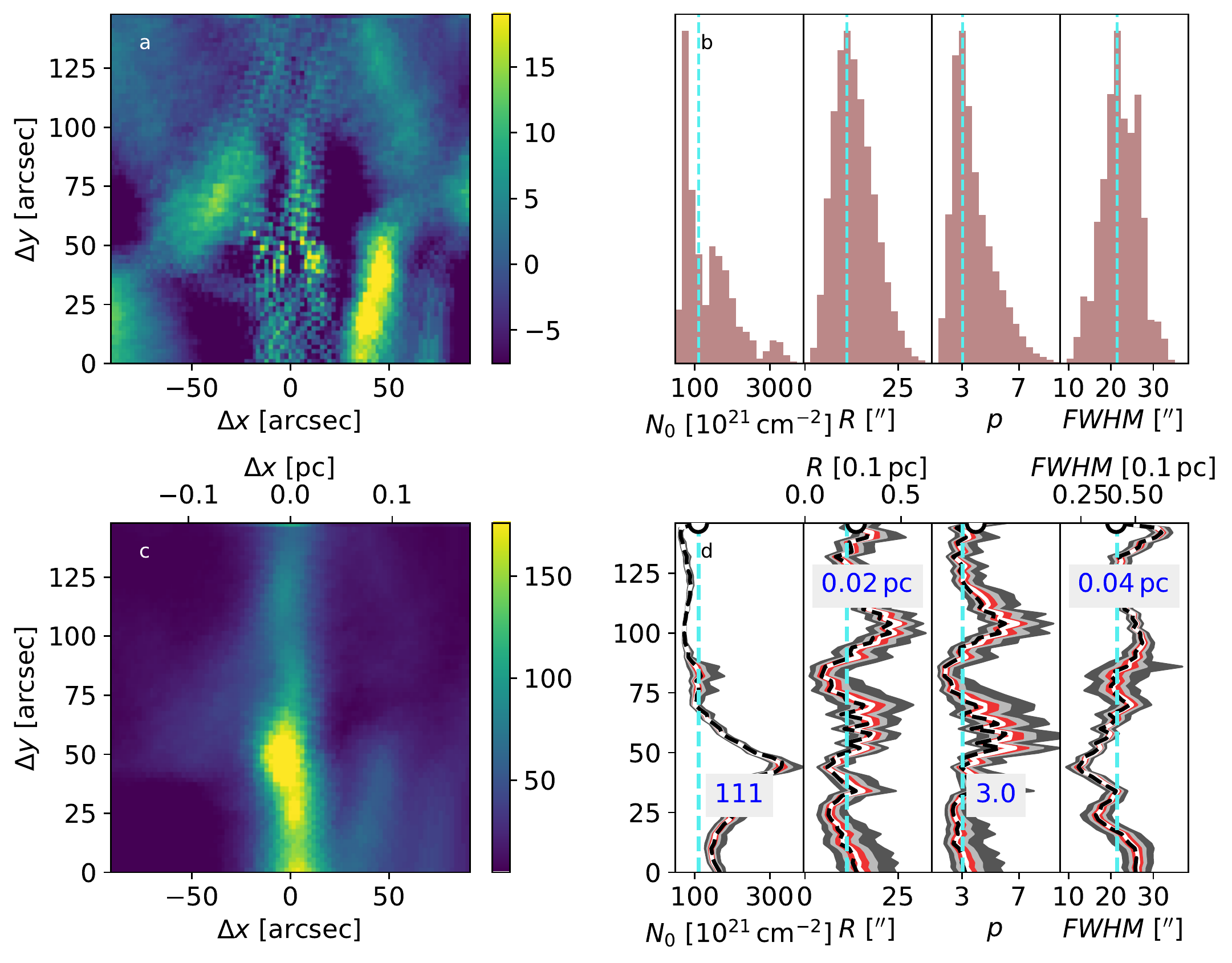}
\caption{
Plummer fits of combined \Herschel and \artemis data at 10$\arcsec$ resolution.
As Fig.~\ref{fig:G208_NH2_feathered_MCMC_A0} but for OMC-3 filament segment C.
}
\label{fig:G208_NH2_feathered_MCMC_A2}
\end{figure}

\begin{figure}
\sidecaption
\centering
\includegraphics[width=9cm]{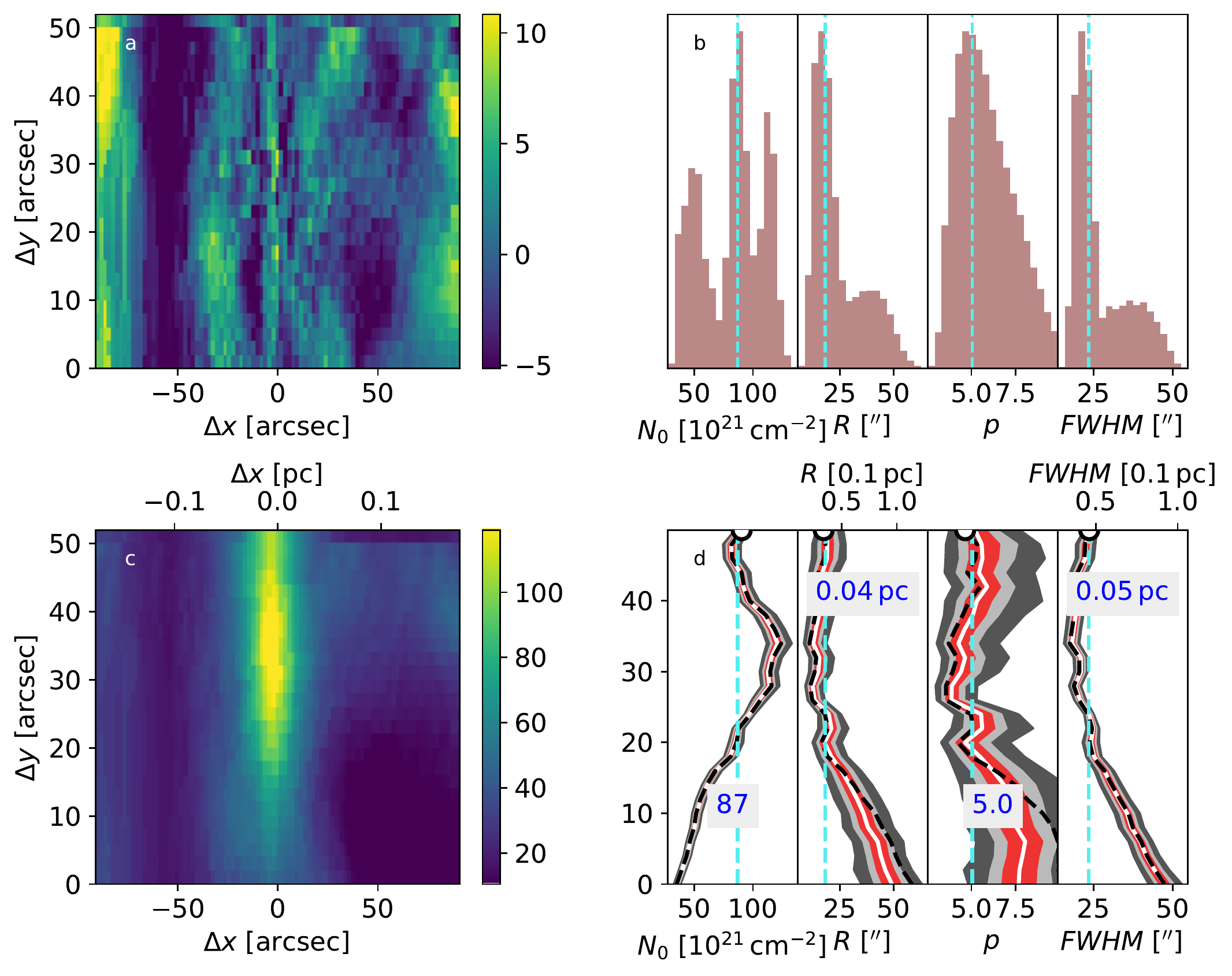}
\caption{
Plummer fits of combined \Herschel and \artemis data at 10$\arcsec$ resolution. As Fig.~\ref{fig:G208_NH2_feathered_MCMC_A0} but for OMC-3 filament segment D.
}
\label{fig:G208_NH2_feathered_MCMC_A3}
\end{figure}

Section~\ref{sect:results_FIR} presented results of Plummer fits to column-density data
that were obtained from \Herschel and \artemis FIR measurements, including figures for
the filament segment A. We present here the corresponding figures for segments B-D.
Figures~\ref{fig:G208_NH2_Palmeirim_NH2_MCMC_A1}-\ref{fig:G208_NH2_Palmeirim_NH2_MCMC_A3}
show the results based on \Herschel data ($20\arcsec$ map resolution and fits to an area
[-90$\arcsec$,+90$\arcsec$] around the filament centre). The corresponding results for
the combined \Herschel and \artemis data are plotted in
Figs.~\ref{fig:G208_NH2_feathered_MCMC_A1}-\ref{fig:G208_NH2_feathered_MCMC_A3}.

\section{Effect of filament inclination} \label{sect:app_inclination}

We tested the effect of different viewing angles using filament models with column
densities $N({\rm H}_2)=3\cdot 10^{22}$, $10^{23}$, and $3 \cdot
10^{23}$\,cm$^{-2}$ perpendicular to the filament axis. By increasing the inclination
from zero to 70.5 degrees, the line-of-sight column density increases by a factor of three.
Therefore, the comparison to the $N({\rm H}_2)=10^{23}- 10^{24}$\,cm$^{-2}$ models shows
the relative importance of the true versus line-of-sight column densities.

\begin{figure}
\sidecaption
\centering
\includegraphics[width=9cm]{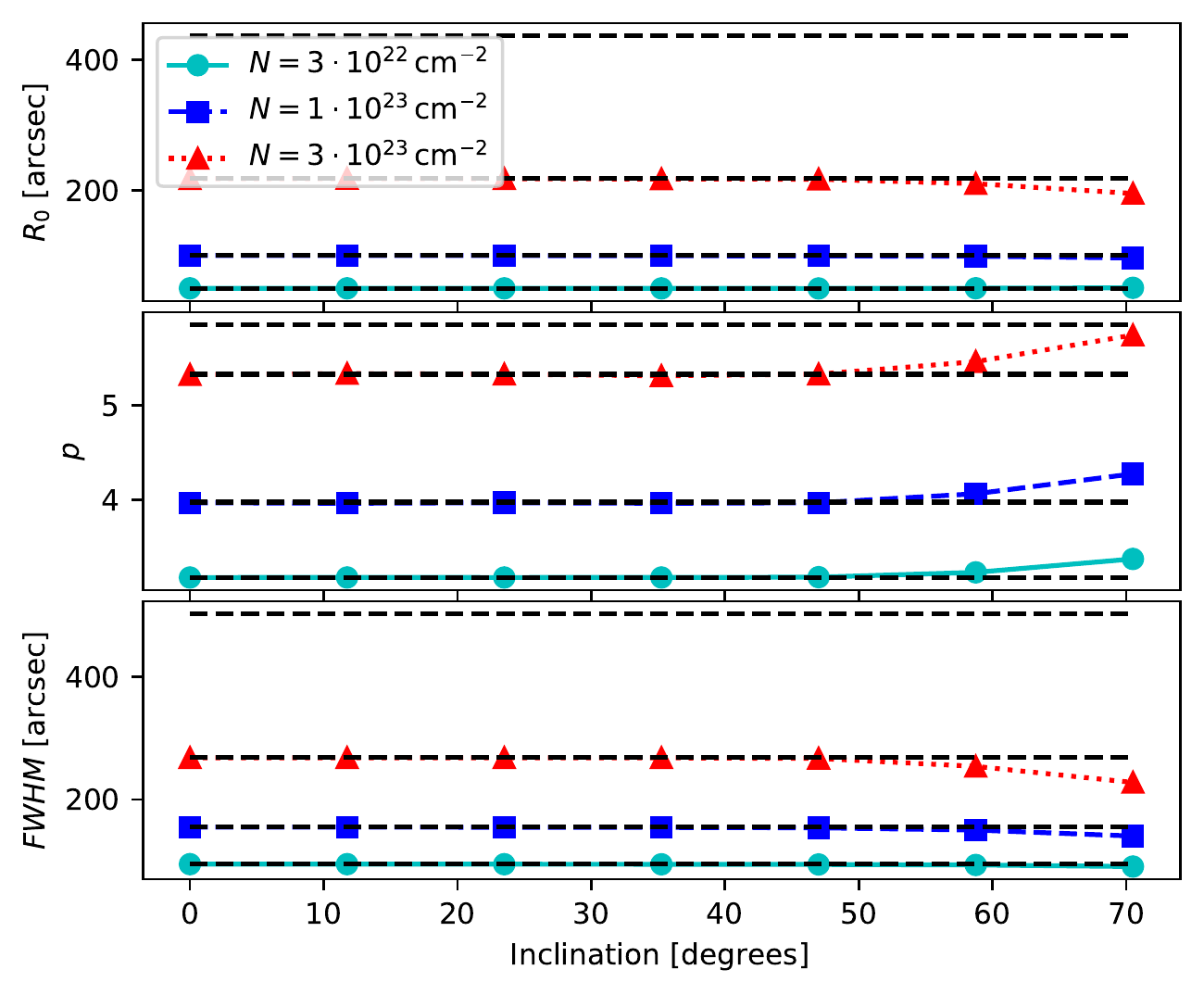}
\caption{
Dependence of filament parameters on the column density and inclination of model filaments. The
cyan, blue, and red curves and symbols show, respectively, parameter estimates at column densities
$N({\rm H}_2)=3\cdot 10^{22}$\,cm$^{-2}$, $N({\rm H}_2)=10^{23}$\,cm$^{-2}$, and $N({\rm H}_2)=3 \cdot
10^{23}$\,cm$^{-2}$ as functions of the filament inclination. Zero inclination corresponds to the
perpendicular view direction and 70.5 degrees to a factor of three higher line-of-sight column density. For
reference, the horizontal black dashed lines show the zero-inclination values for the three column
densities and additionally for $N({\rm H}_2)= 10^{24}$\,cm$^{-2}$ (the uppermost lines).
}
\label{fig:test_inclination}
\end{figure}

Figure~\ref{fig:test_inclination} shows that the effect of inclination is much smaller than that of the
actual filament column density.  At higher inclinations, $R_{\rm 0}$ and FWHM show no indication of
rising close to the values of the filaments with the same line-of-sight column density observed at zero
inclination. This is not surprising, given that the relative amounts of dust at different temperatures
is not affected by the inclination, and the 160\,$\mu$m emission remains optically thin ($\tau \sim
0.3$ at $N({\rm H}_2)=10^{24}$\,cm$^{-2}$). At the highest inclinations, the results may also be
affected by edge effects, as the line of sight enters and leaves the model volume (on the front and back border
of the model volume) relatively close to the filament axis, where the density is not negligible. 

\section{Tests of fitting mid-infrared filament profiles} \label{sect:app_MIR}

Errors in the determination of the foreground surface-brightness component $I^{\rm fg}$
will affect not only the level of optical-depth estimates but also cause uncertainty in
the filament profiles. We examined these effects with simulated observations, where the
optical depth follows a Plummer profile. The dimensionless parameters of the Plummer
model are $\tau_{\rm 0}$=4.0 for the peak optical depth, $R_{\rm 0}$=3.0 for the central
flat region, and $p_{\rm 0}$=2.5 for the asymptotic power law. The filament is seen
against extended background $I^{\rm bg}$=10.0 units and the foreground is set to $I^{\rm
fg}$=2.0 units. We simulate a piece of a filament where the ridge optical depth is
further modified by a Gaussian function, dropping to 40\% of the maximum at both ends.

The simulation consists of synthetic observations with two sources of error. The first is the
observational noise and the second an error in the estimated level of the extended emission component
$I^{\rm ext}$. For simplicity, we use the same noise value for both components. Although $I^{\rm ext}$
can often be estimated based on a large number of pixels, the interpolation across the extincted
regions is uncertain, and the uncertainty of $I^{\rm ext}$ could be larger or smaller than the
uncertainty of a single surface brightness measurement. The error of the $I^{\rm fg}$ estimate follows
from the observational noise, because it is here estimated from Eq.~(\ref{eq:BT2012}). To keep the
statistics similar to observations, $I^{\rm fg}$ is estimated using only 64 points along the filament
length. However, to gather more statistics for the following plots, the rest of the simulation is done
using 1024 independent profiles along the filament.

\begin{figure}
\sidecaption
\centering
\includegraphics[width=9cm]{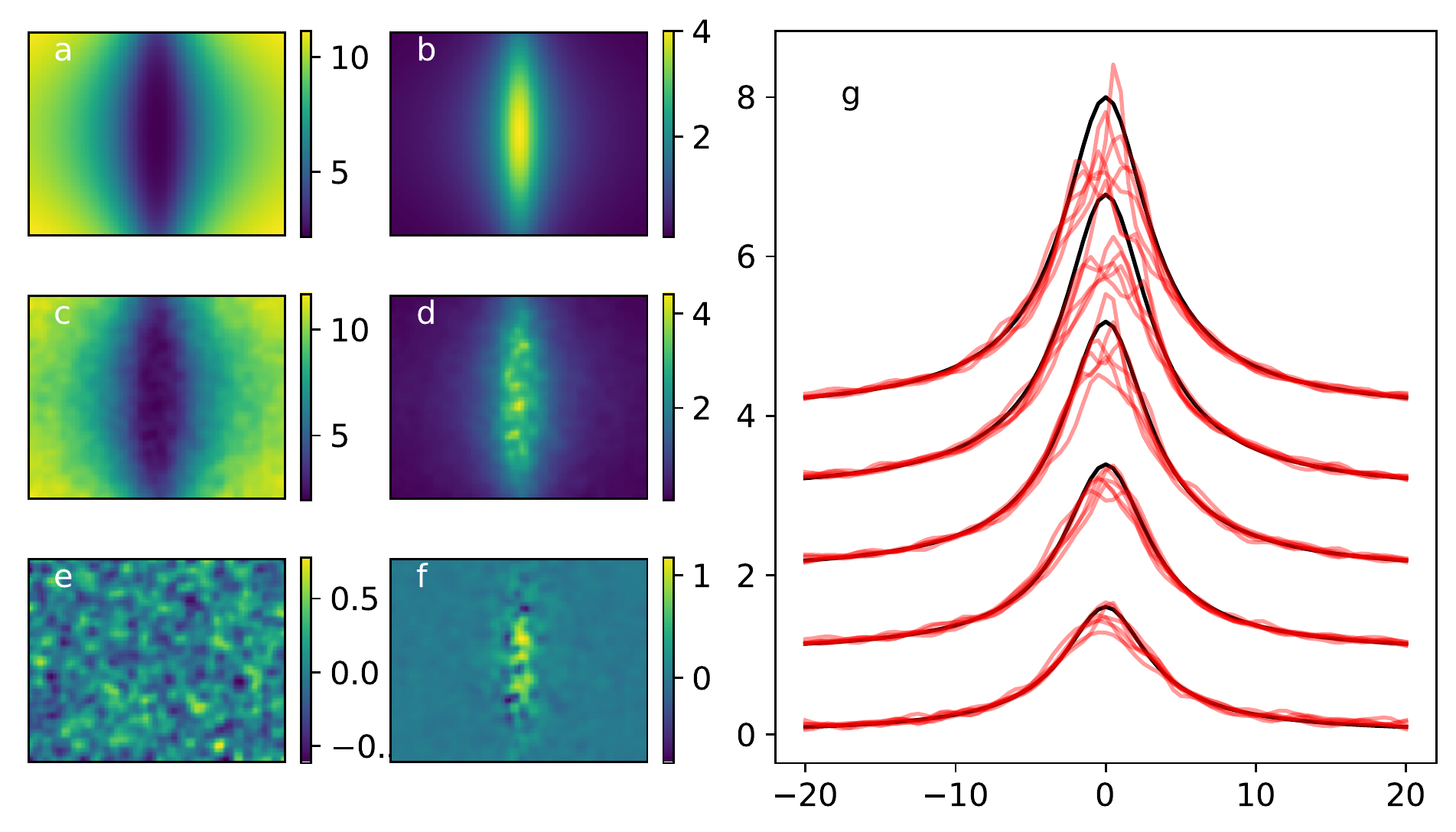}
\caption{
Example simulated MIR observations of a model filament. Frames (a) and (b) show the ideal absorption and the
corresponding true optical depth image. Frames (c) and (d) are the corresponding noisy realisations, with noise
0.2 units for both surface brightness observations and for the determination of the level of $I^{\rm
ext}$, and frames (e) and (f) show the differences between frames (a) and ()b and between frames (b) and (d). The black
curves in frame g show examples of profiles, from the end of the filament (bottom curve) to the centre
of the filament length (uppermost curve). The curves are offset along the y axis for better
readability. For each ideal profile, there are ten realisations of the profiles derived from noisy
observations.
}
\label{fig:Plummer_test_MIR_1}
\end{figure}

First, Fig.~\ref{fig:Plummer_test_MIR_1} shows examples of filament profiles that are
derived from noisy observations with 0.2 units of noise in both surface brightness and
$I^{\rm ext}$ values. Since Eq.~(\ref{eq:BT2012}) is used, the optical depth estimate is
always defined. On the other hand, there is a small bias whereby the optical depth at the
filament centre tends to be underestimated. Because this bias is largest when the
observed surface brightness approaches the assumed $I^{\rm fg}$ value, the recovered
profiles tend to be flatter in the centre and thus also have FWHM values that are
slightly larger than for the true filament profile.

\begin{figure}
\sidecaption
\centering
\includegraphics[width=9cm]{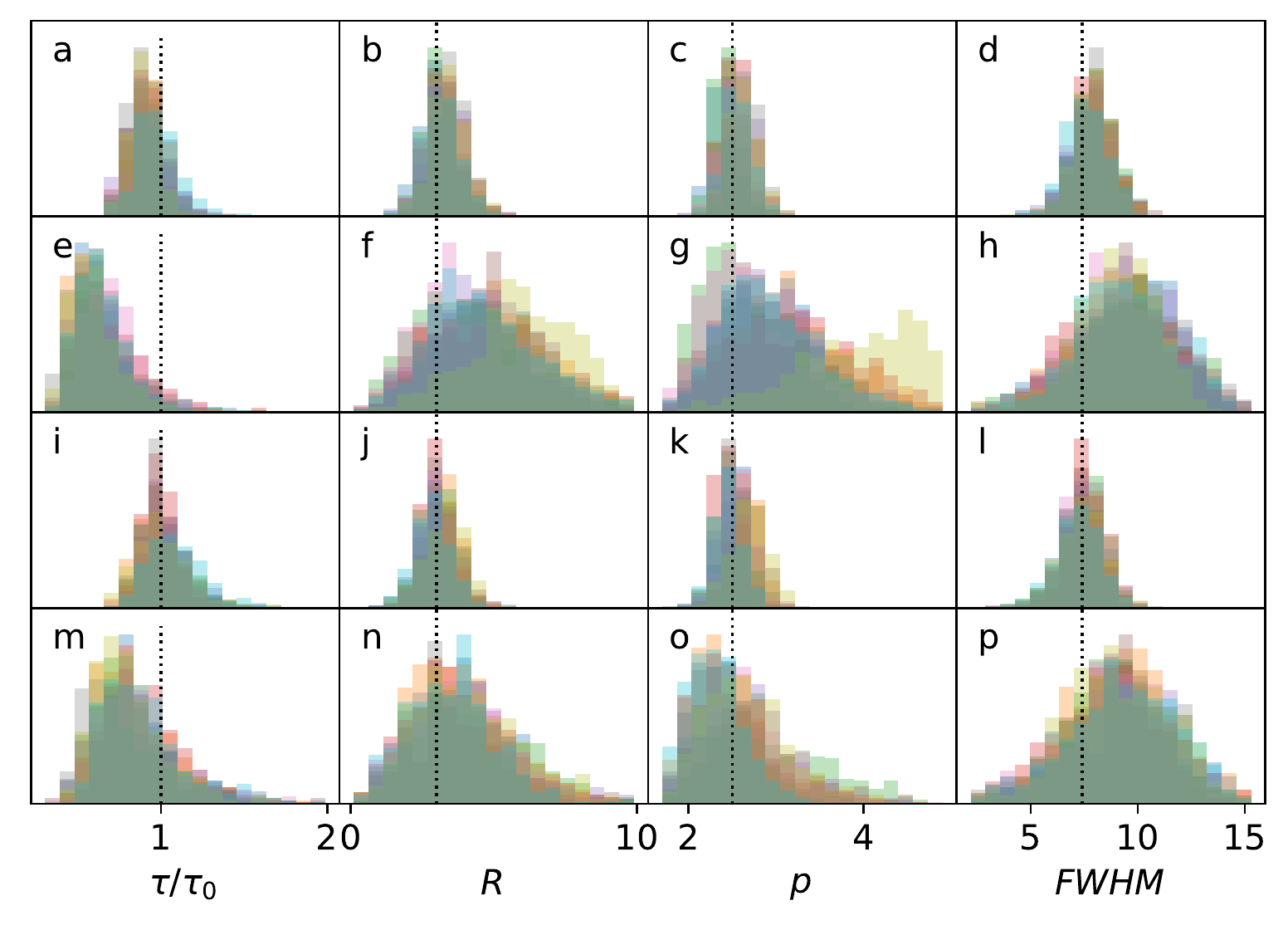}
\caption{
Histograms of extracted filament parameters for simulated observations. The first three columns show the
recovered central optical depth relative to the true value, $\tau/\tau_0$, and the parameters $R$ and
$p$. The last column contains the FWHM values computed from the Plummer parameters.  In frames (a)-(h),
$I^{\rm fg}$ is estimated with Eq.~(\ref{eq:BT2012}), in frames (i)-(p) simply based on the minimum
observed value. The noise values are 0.2 in frames (a)-(d) and (i)-(l) and 0.75 in frames (e)-(h) and m-p. In each
frame, we overplot ten histograms. Each of the individual histograms corresponds one random error in
the $I^{\rm ext}$ value and contains 640 realisations of filament profiles with random observational
errors.
}
\label{fig:Plummer_test_MIR_2}
\end{figure}

To study the accuracy of the recovered filament parameters, we simulated further four situations.
First, the noise values were set to either 0.2 or 0.55 units. Second, in addition to direct application
of Eq.~(\ref{eq:BT2012}), we also tested a case where $I^{\rm fg}$ is set based on the minimum observed
surface brightness value. The results in Fig.~\ref{fig:Plummer_test_MIR_2} suggest that the method used
for $I^{\rm ext}$ estimation has only little effect. On the other hand, with increasing noise the central
optical depth becomes significantly underestimated while the computed FWHM values are biased upwards.

\section{Further images related to dust emission} \label{sect:app_bias_emission}

\begin{figure}
\sidecaption
\centering
\includegraphics[width=9cm]{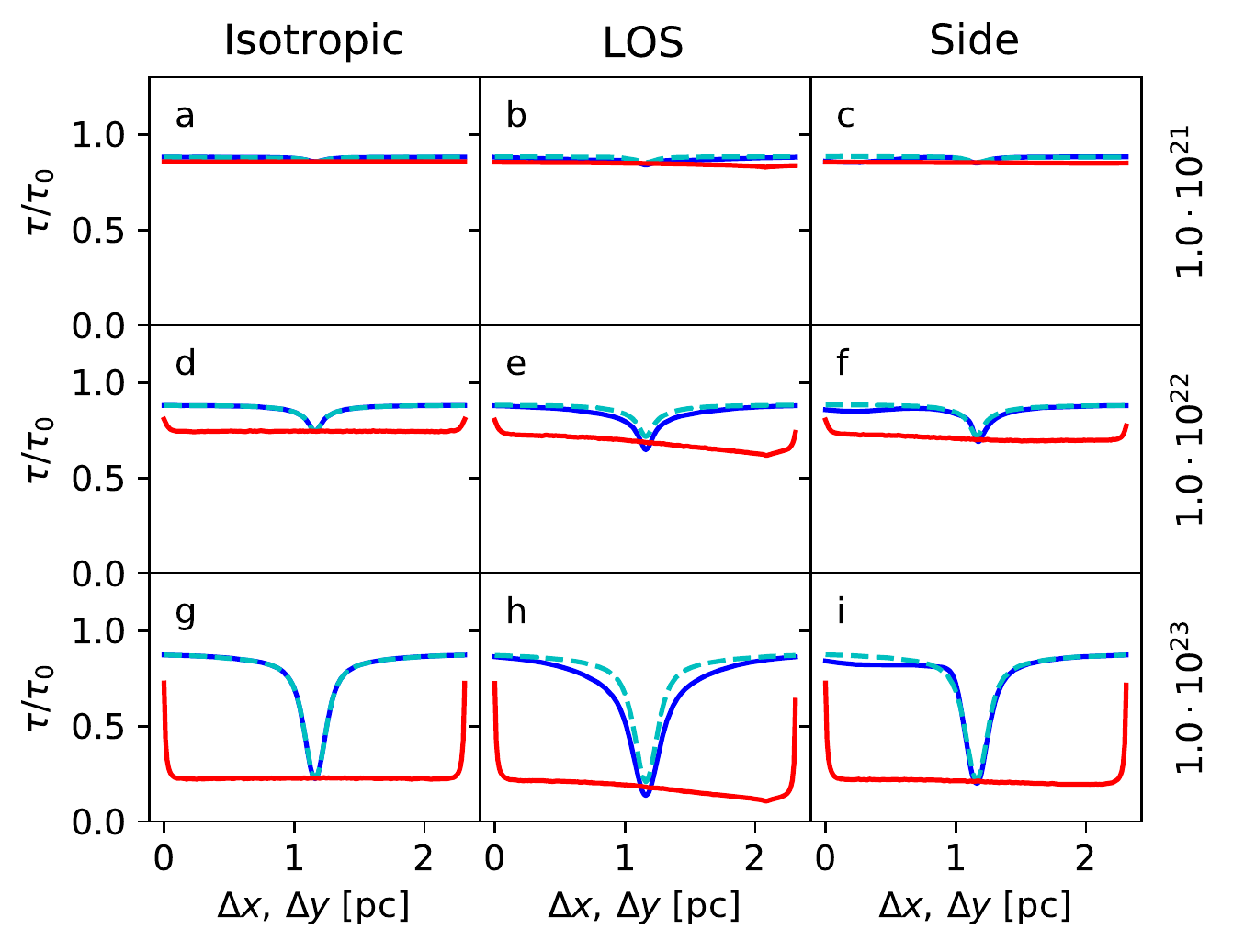}
\caption{
Ratio of estimated and true optical depths in case of simulated FIR observations of dust emission. The
rows correspond different filament column densities. The columns correspond to different cases of the
radiation field, with isotropic illumination or, alternatively, also including a point source at $\Delta
y=2.09$\,pc and at a distance of 0.93\,pc from the filament centre, either along the line of sight towards the
filament or to the left of the filament ($\Delta x$=-0.93\,pc). The ratios are plotted along the
filament centre line (red curves, as a function of $\Delta y$) or across the filament at $\Delta
y=1.4$\,pc (blue line) or $\Delta y=0.7$\,pc (dashed cyan line).
}
\label{fig:test_MBB_bias}
\end{figure}

Related to the discussion in Sect.~\ref{sect:bias_FIR}, Fig.~\ref{fig:test_MBB_bias} 
illustrates further the bias in the optical depths derived from FIR dust emission.
The figure shows the ratio of estimated and true optical depths for the $N({\rm
H}_2)=10^{21}-10^{23}$\,cm$^{-2}$ models and the alternative descriptions of the
radiation field.  The optical-depth ratios are shown along the filament centre line
and, at two positions, across the filament.

\begin{figure}
\sidecaption
\centering
\includegraphics[width=9cm]{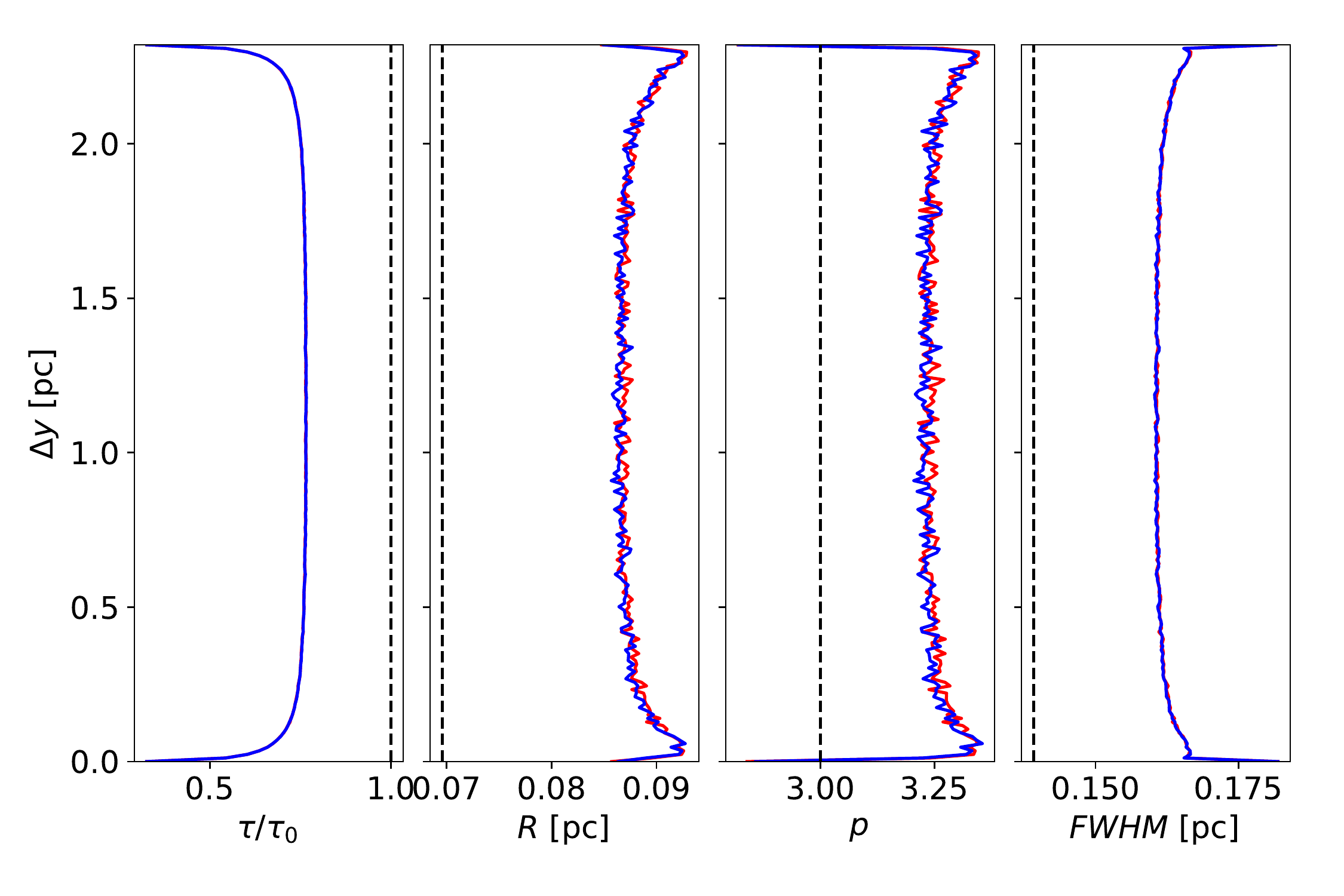}
\caption{
As Fig.~\ref{fig:SED_SHG_trace_PS3_HI1}, but showing the effect of dust emission on the MIR filament
parameters in the case of an isotropic external radiation field only ($\chi$=1). The simulated surface
brightness is assumed to consist of extincted background radiation and thermal emission from
stochastically heated grains.
}
\label{fig:SED_SHG_trace_PS0_HI1}
\end{figure}

\begin{figure}
\sidecaption
\centering
\includegraphics[width=9cm]{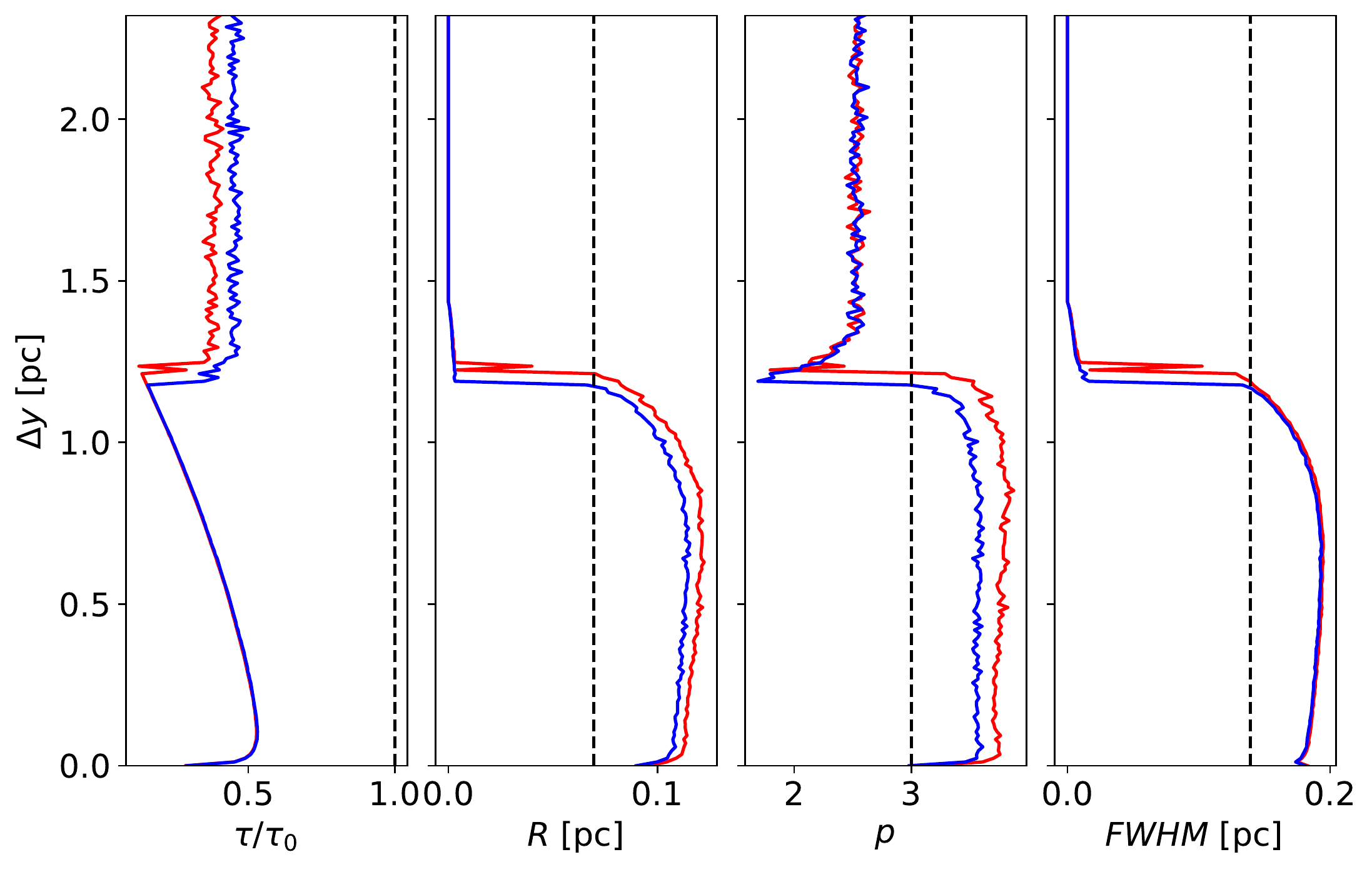}
\caption{
As Fig.~\ref{fig:SED_SHG_trace_PS0_HI1} but with the point source located 0.93\,pc in front of the
model filament. Because of thermal emission from stochastically heated grains, the filament is not visible as
an MIR absorption feature at $\Delta y>1.2$\,pc and the corresponding data should be ignored.
}
\label{fig:SED_SHG_trace_PS1_HI1}
\end{figure}

\begin{figure}
\sidecaption
\centering
\includegraphics[width=9cm]{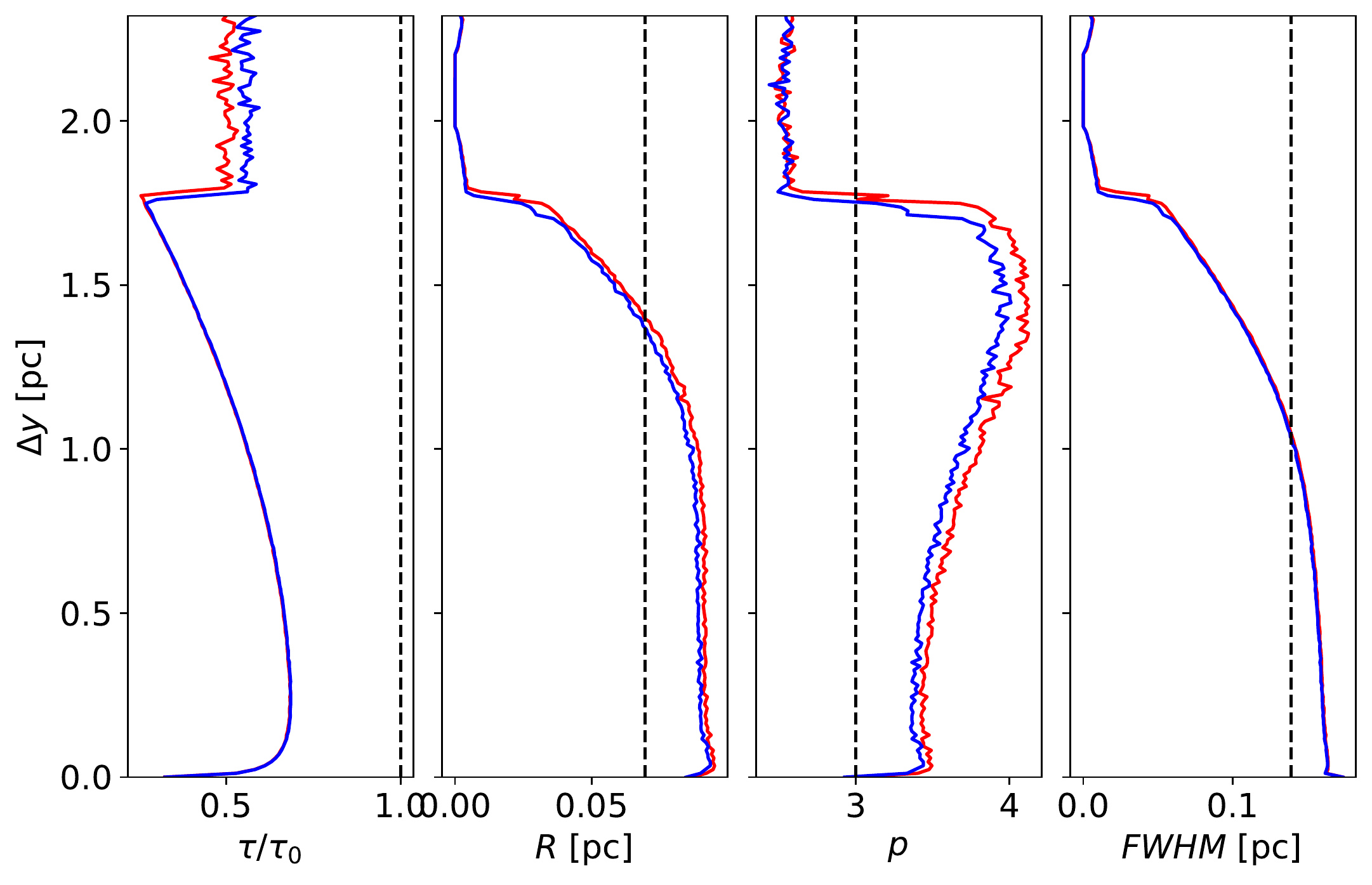}
\caption{
As Fig.~\ref{fig:SED_SHG_trace_PS0_HI1} but with the point source located 0.93\,pc
behind the model filament. The filament is not visible as an absorption feature at
$\Delta y>1.6$\,pc, and the corresponding data should be ignored.
}
\label{fig:SED_SHG_trace_PS2_HI1}
\end{figure}

Results on MIR dust emission were shown in Sect.~\ref{sect:bias_SHG}. Similar to
Fig.~\ref{fig:SED_SHG_trace_PS3_HI1}, we include below figures for the other radiation
field configurations. Figure~\ref{fig:SED_SHG_trace_PS0_HI1} shows parameter estimates
along a $N({\rm H}_2) = 3 \cdot 10^{23}$\,cm$^{-2}$ filament that is illuminated by an
isotropic radiation field with $\chi=1$. In Figs.~\ref{fig:SED_SHG_trace_PS1_HI1} and
\ref{fig:SED_SHG_trace_PS2_HI1}, the model also includes the 590\,$L_{\sun}$ radiation
source that is located 0.94\,pc in front of or behind the filament.

\begin{figure}
\sidecaption
\centering
\includegraphics[width=9cm]{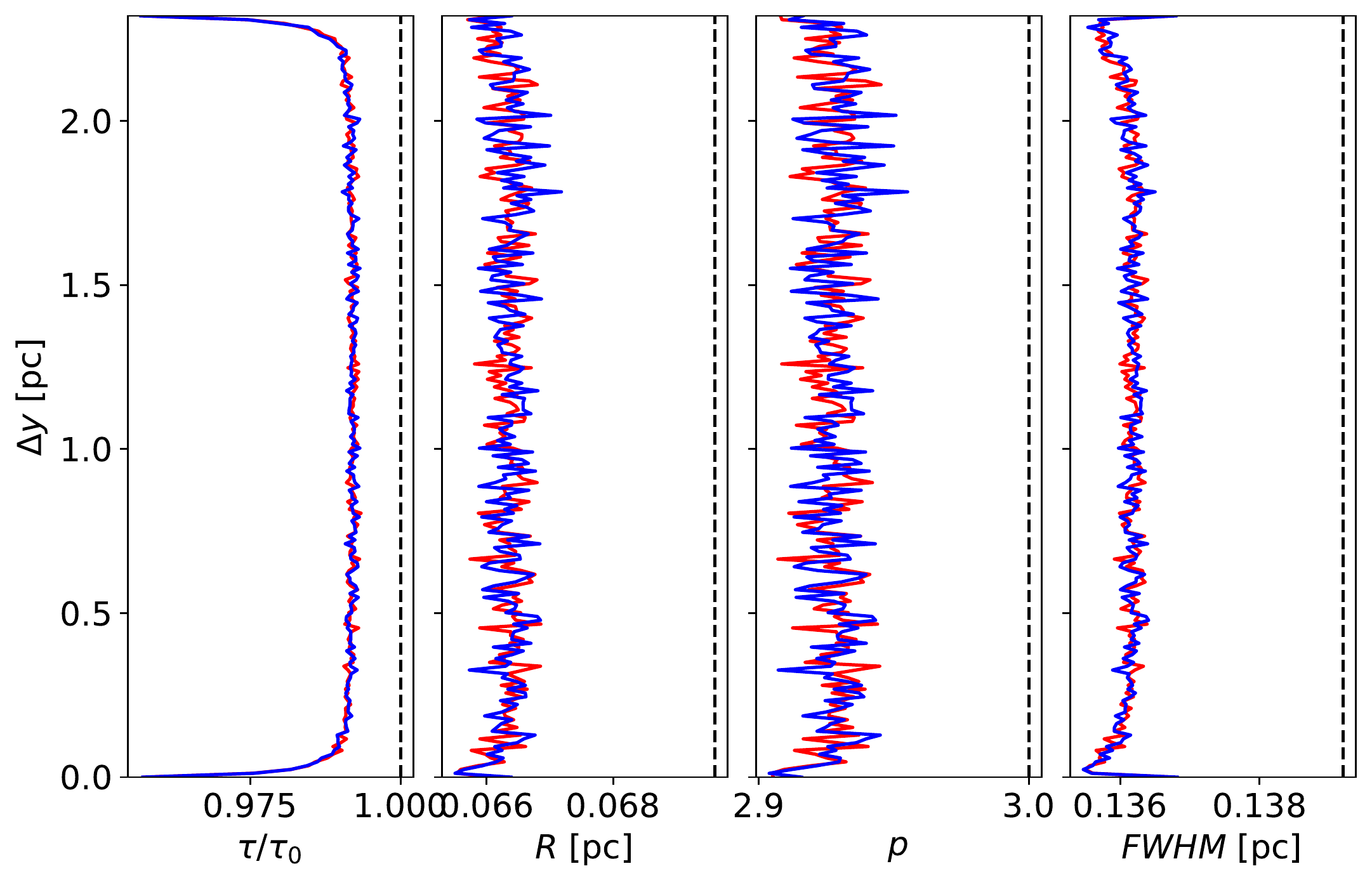}
\caption{
As Fig.~\ref{fig:SED_SHG_trace_PS3_HI1} but for the case of lower column density $N({\rm H}_2) = 3
\cdot 10^{22}$\,cm$^{-2}$, higher background surface brightness $I^{\rm bg}$=100\,MJy\,sr$^{-1}$, and
the model filament illumination by an isotropic radiation field only ($\chi$=1). The observed surface
brightness is assumed to consist only of the extincted background and thermal dust emission from
stochastically heated grains.
}
\label{fig:SED_SHG_trace_PS0_HI0}
\end{figure}

\begin{figure}
\sidecaption
\centering
\includegraphics[width=9cm]{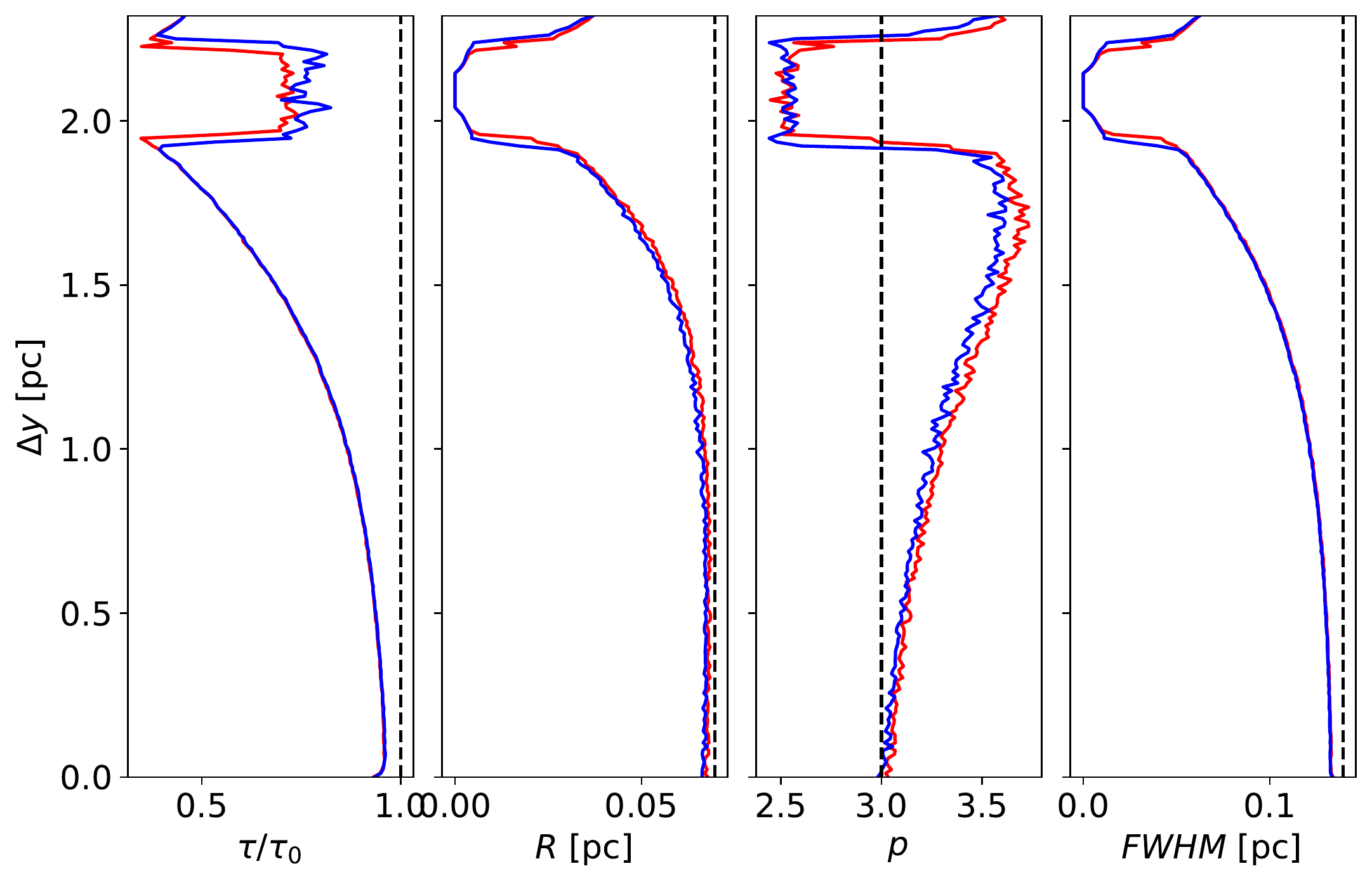}
\caption{
As Fig.~\ref{fig:SED_SHG_trace_PS0_HI0} but including a point source at $\Delta y=2.09$\,pc and
0.93\,pc in front of the model filament. Around $\Delta y \approx 2-2.2$\,pc, the filament is not seen in
absorption because of thermal emission from stochastically heated grains.
}
\label{fig:SED_SHG_trace_PS1_HI0}
\end{figure}

\begin{figure}
\sidecaption
\centering
\includegraphics[width=9cm]{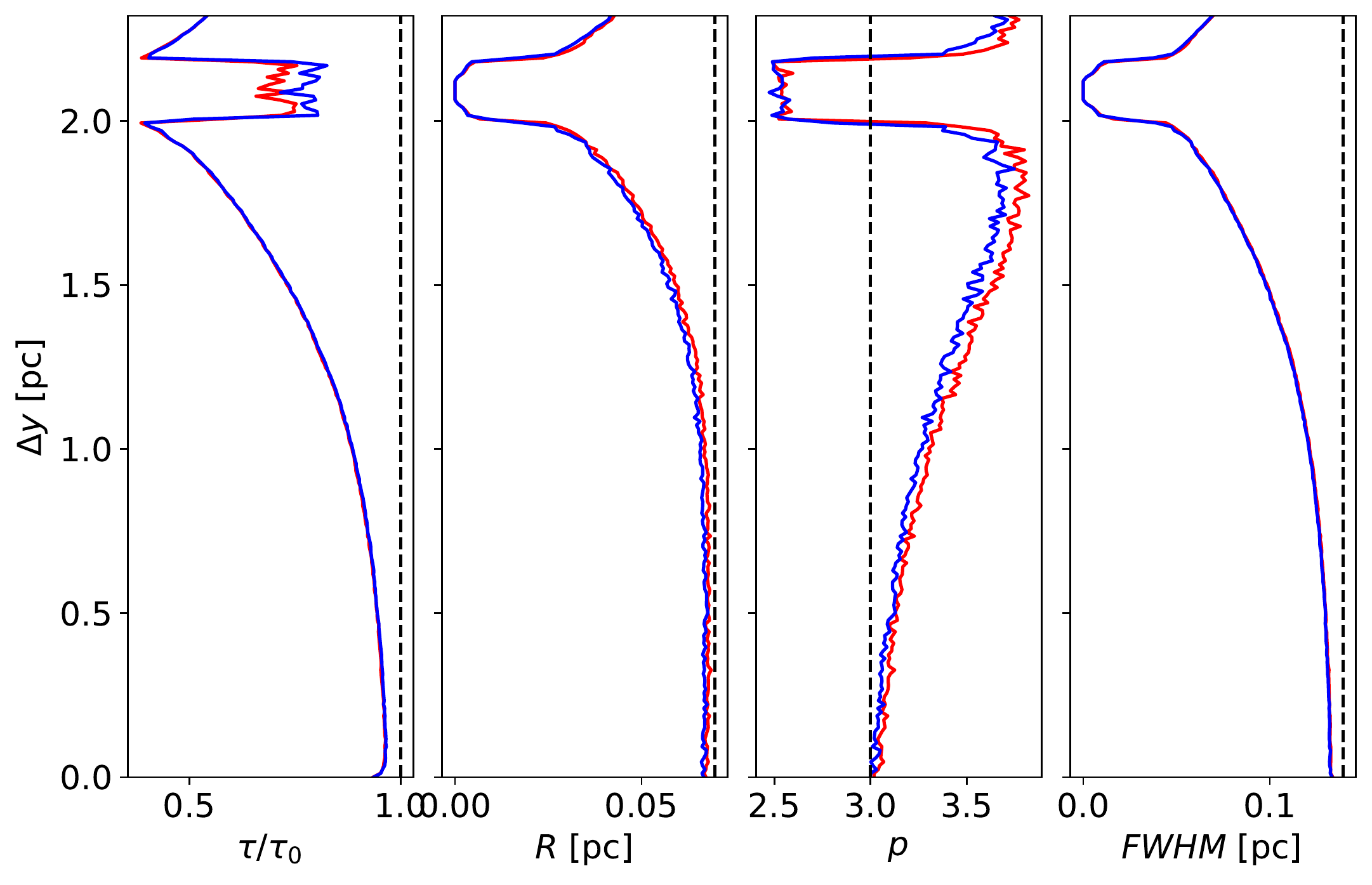}
\caption{
As Fig.~\ref{fig:SED_SHG_trace_PS0_HI0} but including a point source at $\Delta y=2.09$\,pc and
0.93\,pc behind the model filament. Around $\Delta y \approx 2-2.2$\,pc, the filament is not seen in
absorption because of thermal emission from stochastically heated grains.
}
\label{fig:SED_SHG_trace_PS2_HI0}
\end{figure}

\begin{figure}
\sidecaption
\centering
\includegraphics[width=9cm]{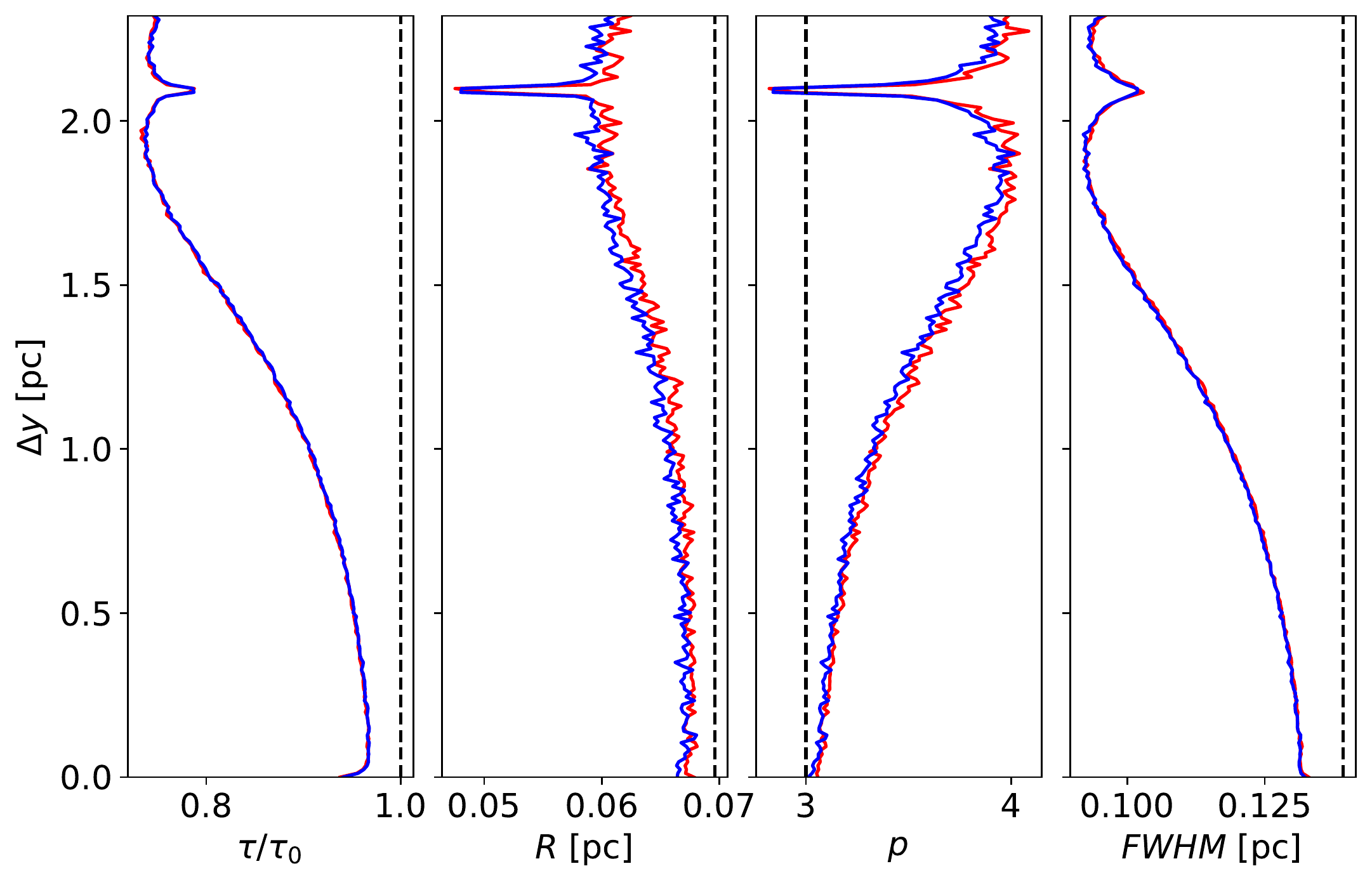}
\caption{
As Fig.~\ref{fig:SED_SHG_trace_PS0_HI0} but including a point source 
at $\Delta y=2.09$\,pc and 0.93\,pc to one side of the model filament.
}
\label{fig:SED_SHG_trace_PS3_HI0}
\end{figure}

Figures~\ref{fig:SED_SHG_trace_PS0_HI0}-\ref{fig:SED_SHG_trace_PS3_HI0} show the
results for a less massive filament. The filament has $N({\rm H}_2) = 3 \cdot
10^{22}$\,cm$^{-2}$ but the assumed background sky brightness is higher,
100\,MJy\,sr$^{-1}$.

\section{Toy model of a filament with two temperature components}
\label{sect:app_toy}

\begin{figure}
\sidecaption
\centering
\includegraphics[width=9cm]{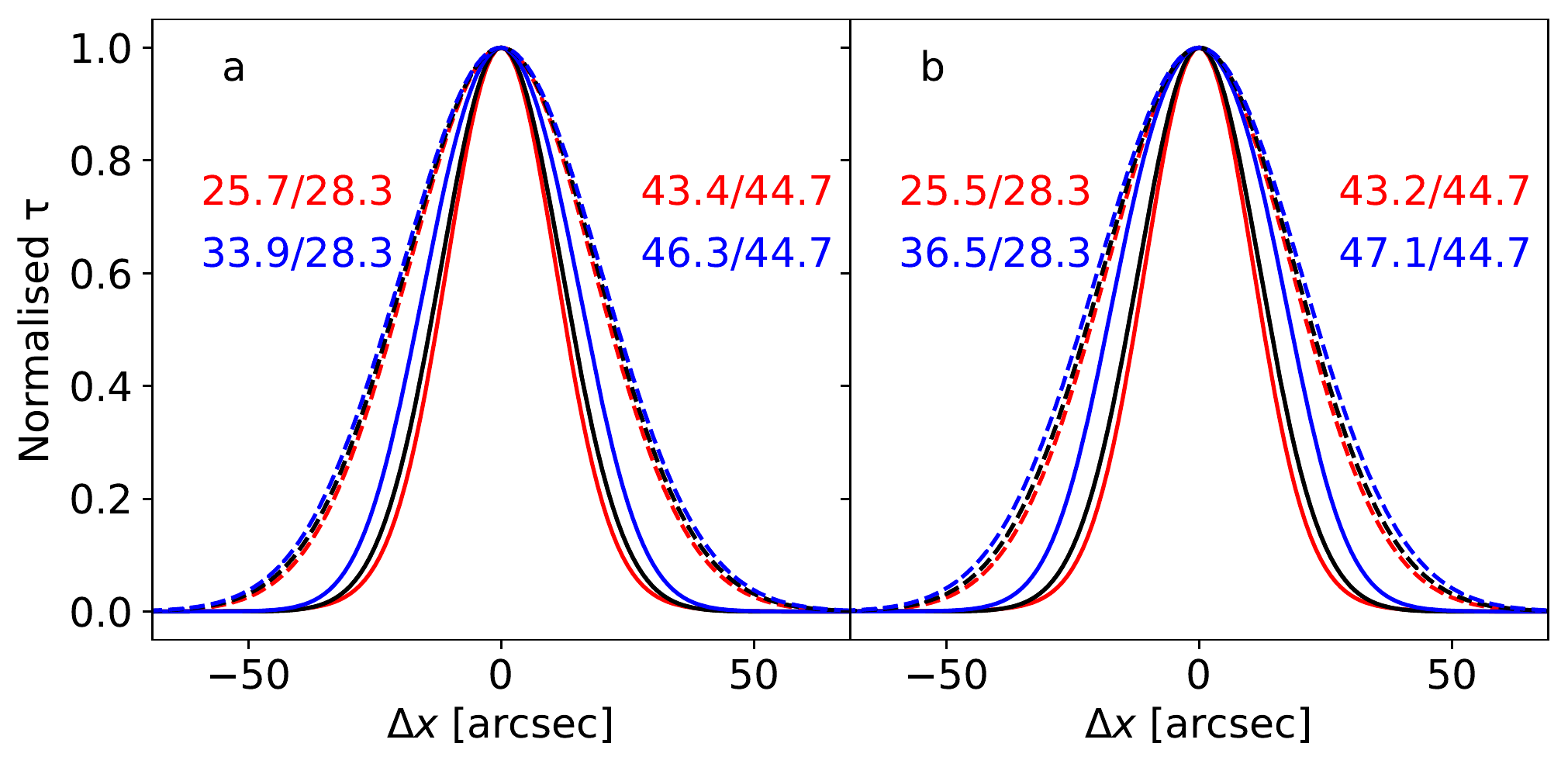}
\caption{
Optical-depth profiles of toy models consisting of 10\,K and 20\,K dust. The
model in the left frame is a Gaussian sphere, and the model on the right a
cylinder with the same Gaussian cross-section with $FWHM=20\arcsec$. The black
curves are the expected profiles observed at 20$\arcsec$ and 40$\arcsec$
resolution. The actually profiles (from HR and LR maps) are plotted in red
(models with warm centre) and blue (models with cold centre), where solid
lines show the HR and the dashed lines the LR results. The estimated and
expected FWHM values are quoted in the same colours (HR on left, and LR on
right side).
}
\label{fig:paltest}
\end{figure}

Figure~\ref{fig:paltest} show toy models of a core and a filament, both with a
Gaussian optical depth profile with $FWHM(\tau)=20\arcsec$. If the dust
properties and especially the dust temperature were constant, one could
precisely recover the expected convolved optical depth profiles with
$FWHM=\sqrt{FWHM(\tau)^2+FWHM({\rm beam})^2}$, where we now assume $FWHM({\rm
beam})= 20\arcsec$ and $40\arcsec$ for the HR and LR maps, respectively.

The model of Fig.~\ref{fig:paltest} consists however of two dust components at
10\,K and 20\,K temperatures. The fractional abundance of one component
increases proportionally to $\tau$, and the abundance of the other component
decreases, keeping their sum equal to $\tau$. Thus, at the distance from the
filament centre where the optical depth has dropped to half of its maximum
value, the optical depths of the warm and cold components are equal. Both
methods (HR and LR) now give biased estimates of the filament width. If the
filament centre is colder, the filament width is overestimated, and if the
centre is warmer, the width is underestimated. The relative errors are larger
for HR maps and, unlike in the isothermal cases, the errors are not identical
in spherical and cylindrical geometries. In the filament case of
Fig.~\ref{fig:paltest}b, the relative errors are +29\% and -10\% for the
HR map and -3\% and +5\% for the LR map. After
deconvolution\footnote{The deconvolved width is $\sqrt{FWHM({\rm
observed})^2-FWHM({\rm beam})^2}$. The formula is in this case exact because
the filament has a Gaussian profile.} the estimated width of a filament with a
cold centre is 30.6$\arcsec$ based on the HR map and
24.8$\arcsec$ based on the LR map. Relative differences are larger
for the HR map, which results in deconvolved sizes of 30.6$\arcsec$ and
15.8$\arcsec$ for the cold-core and the hot-core filaments, respectively.
Quantitatively, the errors depend on the average temperature, the temperature
contrast, and steepness and radial location of the temperature change. The
above example with 20\,K versus  10\,K components may exaggerate the typical
effects, but nevertheless suggests that the errors can be significant.

\end{document}